\documentclass[12pt,a4paper]{article}

\usepackage{bm}

\usepackage{amsmath}
\usepackage{amsfonts}
\usepackage{amssymb}

\usepackage[dvipdfmx]{graphicx}

\newcommand{\CC}{\mathbb{C}}
\newcommand{\ZZ}{\mathbb{Z}}
\newcommand{\RR}{\mathbb{R}}
\newcommand{\II}{\mathbb{I}}

\newcommand{\ol}{\overline}
\newcommand{\wt}{\widetilde}

\newcommand{\N}{\mathcal{N}}

\def\tr{\mathop{\rm tr}\nolimits}

\def\Pexp{\mathop{\rm Pexp}\nolimits}
\def\PE{\mathop{\rm Pexp}\nolimits}

\newcommand{\nn}{\nonumber}
\newcommand{\wh}{\widehat}

\newcommand{\bs}{\backslash}

\begin{document}
\begin{titlepage}
\title{
\vspace{-1.5cm}
\begin{flushright}
{\normalsize TIT/HEP-696\\ October 2023}
\end{flushright}
\vspace{1.5cm}
\LARGE{Simple-Sum Giant Graviton Expansions for Orbifolds and Orientifolds}}
\author{
Shota {\scshape Fujiwara$^1$\footnote{E-mail: shota.fujiwara@wits.ac.za}},
Yosuke {\scshape Imamura$^2$\footnote{E-mail: imamura@phys.titech.ac.jp}},\\
Tatsuya {\scshape Mori\footnote{T.M. left Tokyo Institute of Technology in March 2023.}},
Shuichi {\scshape Murayama$^2$\footnote{E-mail: s.murayama@th.phys.titech.ac.jp}},
and
Daisuke {\scshape Yokoyama$^3$\footnote{E-mail: ddyokoyama@meiji.ac.jp}}
\\
\\
{\itshape $^1$Mandelstam Institute for Theoretical Physics,} \\ {\itshape University of Witwatersrand, Johannesburg 2050, South Africa} \\
\textit{and} \\
{\itshape $^2$Department of Physics, Tokyo Institute of Technology,} \\ {\itshape Tokyo 152-8551, Japan} \\
\textit{and} \\ \textit{ $^3$Department of Physics, Meiji University,} \\ \textit{ Kanagawa 214-8571, Japan}}

\date{}
\maketitle
\thispagestyle{empty}

\begin{abstract}
We study giant graviton expansions of the superconformal index of 4d orbifold/orientifold theories.
In general, a giant graviton expansion is given as a multiple sum over wrapping numbers.
It has been known that the expansion can be reduced to a simple sum for the ${\cal N}=4$ $U(N)$ SYM
by choosing appropriate expansion variables.
We find such a reduction occurs
for a few examples of orbifold and orientifold theories: $\ZZ_k$ orbifold and orientifolds with $O3$ and $O7$.
We also argue that for a quiver gauge theory
associated with a toric Calabi-Yau $3$-fold
the simple-sum expansion works only if the toric diagram is a triangle,
that is,
the Calabi-Yau is an orbifold of $\CC^3$.
\end{abstract}
\end{titlepage}
\tableofcontents

\tableofcontents
\section{Introduction}\label{intro.sec}
AdS/CFT correspondence \cite{Maldacena:1997re,Gubser:1998bc,Witten:1998qj} provides a useful window to investigate quantum gravity.
Although we have not yet understood how to directly deal with quantum effects of gravity,
we can obtain information of such effects through the correspondence from analyses in the boundary theories.
The superconformal index \cite{Romelsberger:2005eg,Kinney:2005ej,Gadde:2011uv} is an important and useful quantity
for quantitative investigation of the correspondence.
The index can be calculated on the gauge theory side as long as the theory is Lagrangian,
and we are also able to calculate the index on the gravity side in an appropriate parameter region.
In the strict large $N$ limit, which means $N$ is much larger than the energy scale
(or the order in the Taylor expansion of the index) which we are focusing on,
the index obtained on the boundary side can be reproduced semi-classically
on the gravity side as the contribution from massless fields
living in the AdS background \cite{Kinney:2005ej}.
To access the quantum gravity effects via AdS/CFT correspondence, we should consider parameter regions out of the strict large $N$ limit.

One interesting region is the one with the energy scale of order $N^2$ with large $N$.
On the gravity side such a region is described by classical blackhole solutions,
and it was found that the superconformal index of large $N$ gauge theory
can correctly reproduces the Beckenstein Hawking entropy by taking appropriate limit of the index \cite{Hosseini:2017mds,Cabo-Bizet:2018ehj,Choi:2018hmj}.
This discovery is important because it indicates that the Boson-Fermion cancellation
does not occur for the majority of states in the Hilbert space and
the index can be used as the thermal partition function by taking appropriate values of fugacities.

Another important parameter region, which we focus on in this work, is the one with the energy comparable to $N$.
In the $q$ expansion of the index we find the deviation from the large $N$ limit around this order.
On the gauge theory side, this is related to the existence of additional operators or additional constraints due to the finiteness of the rank of the gauge group.
On the gravity side, this can be interpreted as the contribution of extended branes.
The first example of such a brane was found in the orientifold model,
in which D3-branes wrapped on the topologically non-trivial three-cycle in $\bm{S}^5/\ZZ_2$ correspond to
Pfaffian operators \cite{Witten:1998xy}.
Because Pfaffian operators are BPS operators contributing to the index,
the corresponding wrapped branes must also contribute to the index.
In such an example it is natural to expect the finite $N$ corrections
can be given in the form of expansion with respect to the wrapping numbers associated with nontrivial cycles.
Even if there are no such non-trivial cycles,
there exist stable extended brane configurations called giant gravitons \cite{McGreevy:2000cw,Grisaru:2000zn,Hashimoto:2000zp,Mikhailov:2000ya}.
They are BPS configurations, and should also contribute to the index.

Direct analyses of the contributions of wrapped branes to the superconformal index were carried out
for ${\cal N}=4$ SYM \cite{Arai:2019xmp,Arai:2020qaj,Imamura:2021ytr}
and many other examples \cite{Arai:2019wgv,Arai:2019aou,Arai:2020uwd,Fujiwara:2021xgu,Imamura:2021dya,Fujiwara:2023azx}.
Essentially the same expansions were also studied on the gauge theory side in \cite{Gaiotto:2021xce,Lee:2022vig}, and named giant graviton expansions.
There exists a similar expansion of superconformal index proposed by Murthy \cite{Murthy:2022ien}.
The relation between Murthy's expansion and the giant graviton expansion
was studied in \cite{Liu:2022olj,Eniceicu:2023uvd}.
The contribution of giant gravitons to
the black hole entropy were studied in \cite{Choi:2022ovw,Beccaria:2023hip}.
See also
\cite{Kinney:2005ej,Biswas:2006tj,Mandal:2006tk,Bourdier:2015wda,Bourdier:2015sga}
for earlier works for the giant graviton contribution to indices and supersymmetric partition functions.

In the analysis of finite $N$ corrections to the superconformal index
we need to include extended branes regardless of whether the brane wrapped on topologically non-trivial cycles.
Although the term ``giant gravitons'' originally means
extended branes without topological wrappings,
in this work we call general extended branes giant gravitons regardless of whether they have topological wrapping or not.

Let us consider the ${\cal N}=4$ $U(N)$ SYM, whose dual geometry is $AdS_5\times S^5$.
The superconformal index
is defined by
\begin{align}
I=\tr[(-1)^Fq^{J_1}p^{J_2}x^{R_x}y^{R_y}z^{R_z}],
\label{scidef}
\end{align}
where $J_1$ and $J_2$ are angular momenta and $R_x$, $R_y$, and $R_z$ are $R$-charges.
The fugacities for these generators, $q$, $p$, $x$, $y$, and $z$ are constrained by
\begin{align}
qp=xyz,
\label{qpxyz}
\end{align}
to respect one of the supercharges.
We can calculate the index $I_{U(N)}$ of the ${\cal N}=4$ $U(N)$ SYM by the localization formula
\begin{align}
I_{U(N)}=\int_{U(N)}dg \Pexp(f_{\rm vec}\chi_{\rm adj}^{U(N)}(g)),
\label{indexun}
\end{align}
where $\Pexp$ is the plethystic exponential,
$\int_{U(N)}dg$ is the gauge group integral with the Haar measure,
$\chi_{\rm adj}^{U(N)}(g)$ is the $U(N)$ adjoint
character\footnote{See Appendix \ref{haar.sec} for the explicit definitions of the character and the Haar measure.},
and $f_{\rm vec}$ is the letter index of the ${\cal N}=4$ $U(1)$ vector multiplet
\begin{align}
f_{\rm vec}(q,p,x,y,z)=i[{\cal B}^{\frac{1}{2},\frac{1}{2}}_{[0,1,0]}]=1-\frac{(1-x)(1-y)(1-z)}{(1-q)(1-p)}.
\label{n4vectorindex}
\end{align}
We use the notation $i[{\cal R}]$ for
the index of an irreducible superconformal representation ${\cal R}$,
and we adopt the notation in \cite{Dolan:2002zh} for ${\cal R}$.

In the large $N$ limit, the 
$U(N)$ integral in (\ref{indexun}) can be easily evaluated with the
saddle point method, and the result is \cite{Kinney:2005ej}
\begin{align}
I_{U(\infty)}=\Pexp f_{\rm sugra},
\end{align}
where $f_{\rm sugra}$ is
the letter index of the supergravity multiplet
in $AdS_5\times S^5$.
\begin{align}
f_{\rm sugra}
=\sum_{n=1}^\infty i[{\cal B}^{\frac{1}{2},\frac{1}{2}}_{[0,n,0]}]
=\frac{x}{1-x}+\frac{y}{1-y}+\frac{z}{1-z}-\frac{q}{1-q}-\frac{p}{1-p}.
\label{sugraindex}
\end{align}
This is obtained by summing up contributions from modes in $AdS_5\times S^5$ given in \cite{Kim:1985ez,Gunaydin:1984fk}.

If $N$ is finite, we have finite $N$ corrections, and are given by the giant graviton expansion.
Let us introduce three complex coordinates $X$, $Y$, and $Z$ such that the $S^5$
is given by $|X|^2+|Y|^2+|Z|^2=1$.
We take account of giant gravitons wrapped around three cycles $X=0$, $Y=0$, and $Z=0$, and
the giant graviton expansion of the index is given by the triple sum \cite{Arai:2019xmp,Imamura:2021ytr}
\begin{align}
\frac{I_{U(N)}}{I_{U(\infty)}}=\sum_{m_x,m_y,m_z=0}^\infty x^{m_xN}y^{m_yN}z^{m_zN}F_{m_x,m_y,m_z},
\label{triplesum}
\end{align}
where $m_x$, $m_y$, and $m_z$ are wrapping numbers associated with three three-cycles in
$S^5$: $X=0$, $Y=0$, and $Z=0$, respectively.
For each set of wrapping numbers $(m_x,m_y,m_z)$ the
function $F_{m_x,m_y,m_z}$ is the index of the theory realized on the system consisting of giant gravitons,
and is $N$-independent
because open strings on the giant gravitons do not couple with the background RR flux.
On each cycle $U(m)$ gauge group is realized, and
the gauge group of the theory on the giant graviton system is
$G=U(m_x)\times U(m_y)\times U(m_z)$.
We also have bi-fundamental fields coming from open strings attached on
two D-branes wrapped around different cycles.
The theory is a gauge theory with the triangle quiver diagram.
We can calculate the functions $F_{m_x,m_y,m_z}$ by the formula
similar to (\ref{indexun}):
\begin{align}
F_{m_x,m_y,m_z}
=\int_G dg \Pexp(i_x[m_x]+i_y[m_y]+i_z[m_z]+\cdots).
\label{fmxmymz}
\end{align}
$\int_G dg$ is the integral over the gauge group $G=U(m_x)\times U(m_y)\times U(m_z)$
with the Haar measure.
$i_x[m_x]$ ($i_y[m_y]$, $i_z[m_z]$)
 is the letter index of $U(m_x)$ ($U(m_y)$, $U(m_z)$) adjoint fields
living on the cycle $X=0$ ($Y=0$, $Z=0$),
and
the dots in the letter index
represent the contribution of bi-fundamental fields living on the intersections.

Because the fields living on the cycle $X=0$ belong to the $U(m_x)$ adjoint representation,
$i_x[m_x]$ is given by $i_x[m_x]=f_{X=0}\chi_{\rm adj}^{U(m_x)}$ with
the letter index $f_{X=0}$ of the $U(1)$ vector multiplet living on $X=0$.
Because the worldvolume of the giant graviton is $S^3\times\RR$, and is the same as the AdS boundary,
the theory on the worldvolume is essentially the same as the ${\cal N}=4$ SYM.
An important difference is the action of symmetry generators,
and the generators acting on the boundary and those acting on the cycle $X=0$
are related by the involution map \cite{Arai:2019xmp}:
\begin{align}
\sigma_x:&
H\rightarrow H-2R_x,\quad
A\rightarrow-A,\quad
J_1\rightarrow R_y,\quad
J_2\rightarrow R_z,\nonumber\\
&
R_x\rightarrow-R_x,\quad
R_y\rightarrow J_1,\quad
R_z\rightarrow J_2.
\label{generatormap}
\end{align}
where $A$ is the generator of $U(1)_R$ of type IIB supergravity normalized so that $A\in\ZZ/2$.%
\footnote{The $U(1)_R$ symmetry acts on the two three-form flux fields non-trivially,
and is broken to $\ZZ_4$
generated by $e^{\pi iA}$ (for a generic value of the axiodilaton field)
due to the flux quantization.
Similar to angular momenta $J_i$ and R-charges $R_a$, $A$ also related
to the fermion number $F$ by $e^{2\pi iA}=(-1)^F$.
We use a convention
with the quantum numbers
$(J_1,J_2,R_x,R_y,R_z,A)=(-\tfrac{1}{2},-\tfrac{1}{2},+\tfrac{1}{2},+\tfrac{1}{2},+\tfrac{1}{2},+\tfrac{1}{2})$
for the supercharge respected by the definition of the superconformal index.}
Correspondingly, we can obtain $f_{X=0}$ from $f_{\rm vec}$ by a simple variable change.
This is also the case for the other two-cycles,
and the variable changes to obtain the letter indices for three cycles are given by
\cite{Arai:2019xmp,Gaiotto:2021xce}.
\begin{align}
\sigma_x:(q,p,x,y,z)&\rightarrow(y,z,x^{-1},q,p),\nonumber\\
\sigma_y:(q,p,x,y,z)&\rightarrow(z,x,p,y^{-1},q),\nonumber\\
\sigma_z:(q,p,x,y,z)&\rightarrow(x,y,q,p,z^{-1}).
\label{sigmaxyz}
\end{align}
(We use $\sigma_I$ ($I=x,y,z$) for both the involutions acting on the generators
and variable changes for the fugacities.)
With these variable changes, we can give $i_I[m_I]$ ($I=x,y,z$) as follows.
\begin{align}
i_I[m_I]=\sigma_I f_{\rm vec}\chi_{\rm adj}^{U(m_I)}.
\label{ixyzn4}
\end{align}

The contribution from bi-fundamental fields denoted by dots in (\ref{fmxmymz})
can be obtained by directly analyzing the open string states.
For example, the contribution from the intersection of cycles
$X=0$ and $Y=0$ is $f_{xy}\chi^{(m_x,m_y)}$
with
\begin{align}
f_{xy}=\left(\frac{z}{qp}\right)^{\frac{1}{2}}\frac{(1-q)(1-p)}{1-z}
\label{hypers5}
\end{align}
and $\chi^{(m,m')}$ is the bi-fundamental character
\begin{align}
\chi^{(m,m')}=
\chi_{\rm fund}^{U(m)}\chi_{\ol{\rm fund}}^{U(m')}
+\chi_{\ol{\rm fund}}^{U(m)}\chi_{\rm fund}^{U(m')}.
\label{chbifund}
\end{align}

Although we can write down the integrand in (\ref{fmxmymz}),
there is a difficulty in carrying out the gauge integral.
To obtain the functions $F_{m_x,m_y,m_z}$ that correctly reproduce the known index
we have to carefully choose contours in the integrals and pick up correct poles.
Although a set of rules for the pole selection for ${\cal N}=4$ $U(N)$ SYM
was proposed in \cite{Imamura:2021ytr},
its derivation and rules for more general theories have not yet been known.
Although the rules for the functions associated
with a single cycle like $F_{m,0,0}$ are simple and natural,
treatment of bi-fundamental fields is involved and calculation of $F_{m_x,m_y,m_z}$ for
intersecting giant gravitons is complicated.
(See \cite{Lee:2022vig} for a proposal for integration contours and
\cite{Beccaria:2023zjw} for its application to
the Schur index of ${\cal N}=4$ SYM.)

We can avoid this problem if we can somehow remove the contributions from intersecting giant gravitons.
Surprisingly, this is possible.
Gaiotto and Lee \cite{Gaiotto:2021xce}
proposed a giant graviton expansion with simple-sum:
\begin{align}
\frac{I_{U(N)}}{I_{U(\infty)}}=\sum_{m_x=0}^\infty x^{m_xN}F_{m_x,0,0}.
\label{simplesum}
\end{align}
The reason of the reduction of the triple-sum expansion to the simple sum
is explained with a special behavior of the functions $F_{m_x,m_y,m_z}$,
which is referred to as ``the wall-crossing'' in \cite{Gaiotto:2021xce}.
Namely, functions $F_{m_x,m_y,m_z}$ are not analytic on some walls
in the space parametrized by the fugacities,
and by choosing an appropriate chamber,
some of the functions become identically zero, and they are decoupled from
the index calculation.
In other words,
by choosing appropriate expansion variables, we can decouple some contributions
and we can simplify the giant graviton expansion \cite{Imamura:2022aua}.

To clarify what is happening in functions $F_{m_x,m_y,m_z}$, let us first
consider a simple toy model.
\begin{align}
F(q)=\Pexp(q+q^2+q^3+\cdots).
\end{align}
If we Taylor expand this function around $q=0$,
this gives the following non-trivial expansion.
\begin{align}
F(q)=1+q+2q^2+3q^3+5q^4+7q^5+\cdots.
\label{nontrivexp}
\end{align}
However, if we regard $F(q)$ as a function of $s=q^{-1}$ and perform the $s$-expansion (expansion around $q=\infty$),
we obtain
\begin{align}
F(q)
=\prod_{k=1}^\infty\frac{1}{1-q^k}
=\prod_{k=1}^\infty\frac{-s^k}{1-s^k}=s^\infty+\cdots=0.
\label{toymodel}
\end{align}
Indeed, the function $F(q)$ has a singular wall
along the unit circle $|q|=1$, and
it is a non-trivial function inside the wall,
while it is trivial outside the wall.

Let us return to the functions $F_{m_x,m_y,m_z}$.
We can explain the relation between two expansions,
one with the simple sum and the other with the triple sum,
by different choices of the expansion variables.
Now we have five fugacities, $q_I=(q,p,x,y,z)$ constrained by
(\ref{qpxyz}).
To specify expansion variables we introduce
four independent auxiliary variables $t_i$ ($i=1,\ldots,4$)
and write five fugacities in terms of $t_i$ as follows
\begin{align}
q_I=\prod_{i=1}^4 t_i^{d_{I,i}}.
\end{align}
Then, we carry out $t_1$-expansion first, and
then sequentially perform $t_2$, $t_3$, and $t_4$-expansions in that order.
This multiple expansion is specified by the set of constants $d_{I,i}$.
Actually, we focus only on the first expansion specified by $d_{I,1}$.
Let us denote $t_1$ by $t$ and $d_{I,1}$ by $d_I$.
The first expansion with respect to $t(=t_1)$ is equivalently performed by
the $t$ expansion after the replacement
\begin{align}
q_I\rightarrow t^{d_I}q_I.
\end{align}
We can regard $t$ as a fugacity for the
operator
\begin{align}
\sum_Id_IQ_I,\quad
Q_I=(J_1,J_2,R_x,R_y,R_z).
\end{align}
In the following we call the constant $d_I$ ``the degree'' assigned to the fugacity $q_I$
and denote it by $d_I=\deg(q_I)$.
The consistency with the constraint (\ref{qpxyz}) requires the degrees
satisfy
\begin{align}
\deg(q)+\deg(p)=\deg(x)+\deg(y)+\deg(z).
\end{align}
In \cite{Imamura:2021ytr}, the triple-sum expansion (\ref{triplesum}) with the degrees
\begin{align}
\deg(q,p,x,y,z)=(\tfrac{3}{2},\tfrac{3}{2},1,1,1)
\end{align}
was studied.
Then the expansion variable $t$ is the fugacity for the operator
\begin{align}
\sum_Id_IQ_I=\frac{3}{2}(J_1+J_2)+R_x+R_y+R_z=H+\frac{1}{2}(J_1+J_2).
\end{align}
In this case all $F_{m_x,m_y,m_z}$ give non-trivial contributions.
The degrees adopted in the reference \cite{Gaiotto:2021xce}, which proposed the simple-sum expansion (\ref{simplesum}), are
\begin{align}
\deg(q,p,x,y,z)=(1,1,0,1,1)
\label{gldegrees}
\end{align}
corresponding to the charge
\begin{align}
\sum_Id_IQ_I=J_1+J_2+R_y+R_z=H-R_x.
\end{align}
With the degrees (\ref{gldegrees}) the contributions with $m_y+m_z\geq1$ decouple.

Let us see how the decoupling occurs with the degrees in (\ref{gldegrees}).
In the toy model with (\ref{toymodel}), the expansion with $\deg(q)=+1$ gives
the non-trivial expansion (\ref{nontrivexp}),
while $\deg(q)=-1$ gives the trivial one.
This occurs as follows.
Each term $q^k$ in the letter index
gives the factor $1/(1-q^k)$ in the plethystic exponential, and if $d=\deg(q)$
is negative, the $t$-expansion of this factor starts with
$-t^{|d|}q^{-1}$.
Namely, each negative-degree term in the letter index gives positive power of $t$
in the plethystic exponential,
and if we have infinitely many such terms, the result becomes $t^{+\infty}=0$.
Based on this, we obtain the following simple criterion for the decoupling:
\begin{itemize}
\item{}{\bf The decoupling criterion:}\\
If the letter index includes infinitely many negative-degree terms with positive coefficients,
its plethystic exponential is trivial and the contribution decouples.\footnote{We have also to confirm that
negative-degree terms with negative coefficients does not give the factor $t^{-\infty}$ canceling the $t^{+\infty}$.
In the following examples we can easily confirm it.}
\end{itemize}

Let us apply the criterion to $F_{m_x,m_y,m_z}$ for $\mathcal N=4$ $U(N)$ SYM and show the decoupling for $m_y+m_z\geq1$.
As we explained, $F_{m_x,m_y,m_z}$ is given in (\ref{fmxmymz})
with the adjoint contributions (\ref{ixyzn4}).
In particular,
if $m_y\geq1$, $i_y[m_y]$ includes
\begin{align}
\sigma_y f_{\rm vec}=1-\frac{(1-y^{-1})(1-q)(1-p)}{(1-z)(1-x)},
\end{align}
corresponding to the constant term (the Cartan part) in $\chi^{U(m_y)}_{\rm adj}$.
If we adopt the degrees in (\ref{gldegrees}),
this letter index contains infinitely many negative-degree terms of the form $x^ky^{-1}$
($k=0,1,2,\ldots$).
Therefore, the contributions with $m_y\geq1$ decouple.
This is also the case for $i_z[m_z]$ with $m_z\geq1$.
This does not happen to $i_x[m_x]$
because
\begin{align}
\sigma_x f_{\rm vec}=1-\frac{(1-x^{-1})(1-q)(1-p)}{(1-y)(1-z)}
\end{align}
does not contain negative-degree terms and
$F_{m_x,0,0}$ with $m_x\geq1$ give non-trivial contributions.

An advantage of the simple-sum expansion is that we can calculate $F_{m_x,0,0}$
much more easily than general contributions from intersecting branes.
By using the relation $f_{X=0}=\sigma_Xf_{\rm vec}$ we can relate $F_{m,0,0}$
and $I_{U(m)}$ by
\begin{align}
F_{m,0,0}
=\int_{U(m)}dg \Pexp(\sigma_x f_{\rm vec}\chi^{U(m)}_{\rm adj})
=\sigma_xI_{U(m)}.
\end{align}
Therefore,
the expansion (\ref{simplesum}) can be written as
\begin{align}
\frac{I_{U(N)}}{I_{U(\infty)}}=\sum_{m=0}^\infty x^{mN}\sigma_x I_{U(m)}.
\label{n4unexp}
\end{align}

A purpose of this paper is to discuss generalization of the simple-sum expansions to orbifold and orientifold theories.
We will not give comprehensive analysis.
We demonstrate in a few examples that the decoupling occurs and the triple-sum expansion
reduces to the simple-sum expansion.
We use the decoupling criterion above as a main tool to check the decoupling
and we numerically test that the simple-sum expansion actually gives the correct index.

An interesting point of the simple-sum
GG expansion is that not only
the LHS in (\ref{n4unexp})
 but also the RHS is given in terms of the superconformal index of
four-dimensional theories
labeled by the rank $m$ of the gauge group.
We can thus consider the large $m$ limit.
In fact, the theory also has the holographic dual,
and we can apply the giant graviton expansion to the theory again.
We will show in some examples
that the expansion
of the ``dual'' theory gives the original theory.
Namely, the relation is mutual
and invertible.
For ${\cal N}=4$ $U(N)$ SYM, this relation
is ``self-dual'', but in general two theories may be different.
In the following sections
we demonstrate how we can obtain
the ``dual'' theory from the original one.

Another interesting point of the orbifold and orientifold theories
is that three fugacities $x$, $y$, and $z$ may not be symmetric.
As we mentioned above, the degrees (\ref{gldegrees}) give
simple-sum expansion associated with the cycle $X=0$.
Let us call such an expansion ``the $X$-expansion''.
Similarly, we can also define the $Y$-expansion and the $Z$-expansion associated with $Y=0$ and $Z=0$, respectively.
For ${\cal N}=4$ SYM this does not give anything new because of the symmetry among fugacities.
However, in more general cases with less supersymmetries, the three simple-sum expansions may give
different expansions for a single theory.

It is natural to ask if the simple-sum expansion works for more general
examples like $AdS_5\times{\rm SE}_5$, where ${\rm SE}_5$ is a Sasaki-Einstein
fivefold.
It is hard to believe that the simple-sum expansion works
for such a case
because the expansion is based on the analysis of fluctuation modes on branes
wrapped around a specific supersymmetric three-cycle in ${\rm SE}_5$,
and the global structure
of the manifold cannot be captured.
We will discuss the decoupling of supersymmetric cycles for toric ${\rm SE}_5$ based on the decoupling criterion above,
and show that the simple-sum expansion works only for ${\rm SE}_5$
whose toric diagram is a triangle.
This means that the ${\rm SE}_5$ needs to be an orbifold of $S^5$ for
the simple-sum expansion to work.

This paper is organized as follows.
In section \ref{orb.sec}, we study giant graviton expansions for $\ZZ_k$ orbifold
with ${\cal N}=2$ supersymmetry.
In section \ref{o3.sec}, we discuss O3 orientifold models with ${\cal N}=4$ supersymmetry.
In section \ref{o7.sec}, we consider another orientifold projection with O7-plane.
In section \ref{toric.sec}, we discuss extension to toric quiver gauge theories.
Section \ref{disc.sec} is devoted for discussion.

\section{$S^5/\ZZ_k$}\label{orb.sec}
In this section we consider ${\cal N}=2$ quiver gauge theory
realized on probe D3-branes in $\CC^2/\ZZ_k$ background, which we call $T_{N_1,\ldots,N_k}$.
The AdS/CFT correspondence of these theories was studied in
\cite{Kachru:1998ys,Lawrence:1998ja,Nakayama:2005mf}.
See \cite{Bourdier:2015sga} for analytic results for the
Schur limit of the index.

\begin{figure}[htb]
\centering
\includegraphics{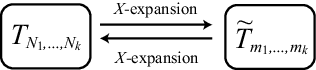}
\caption{$T_{N_1,\ldots,N_k}$ and $\wt T_{m_1,\ldots,m_k}$ are mutually related by the $X$-expansions.}\label{tntm.eps}
\end{figure}

We will find that the decoupling works for $X$-expansion.
By the $X$-expansion we obtain a dual theory as the theory on GG, which we call $\wt T_{m_1,\ldots,m_k}$.
We also find that the $X$-expansion of $\wt T_{m_1,\ldots,m_k}$
gives the original theory $T_{N_1,\ldots,N_k}$ (Figure \ref{tntm.eps}).

\subsection{Boundary theories: $T_{N_1,\ldots,N_k}$}\label{zkbdr.sec}
\subsubsection{Projection}
\begin{figure}[htb]
\centering
\includegraphics{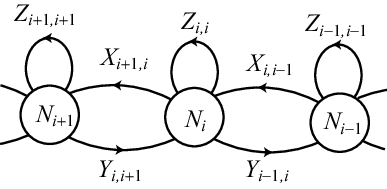}
\caption{A part of the circular quiver diagram of the $\ZZ_k$ orbifold theory}\label{zkfig01.eps}
\end{figure}

We consider the boundary theory obtained
from the ${\cal N}=4$ $U(N)$ SYM by the $\ZZ_k$ orbifold
projection with the generator
\begin{align}
U_k=\exp\left(\frac{2\pi i}{k}(R_x-R_y)\right).
\label{s5zk}
\end{align}
The field contents are obtained
from that of the ${\cal N}=4$ $U(N)$ SYM
by picking up the $\ZZ_k$ invariant degrees of freedom \cite{Douglas:1996sw}.
The insertion of $U_k$ in the trace of the superconformal index
is realized by the following $\ZZ_k$ action on the fugacities.
\begin{align}
(q,p,x,y,z)\rightarrow
(q,p;\omega_kx,\omega_k^{-1}y,z),\quad
\omega_k=e^{\frac{2\pi i}{k}}.
\label{zkaction1}
\end{align}
In addition, the action on the Chan-Paton factor is realized by the following
$\ZZ_k$ action on the gauge fugacities:
\begin{align}
\zeta_a\rightarrow \omega_k^{h_a}\zeta_a\quad(a=1,\ldots,N)
\label{zkaction2}
\end{align}
$h_a$ are holonomy variables and each component takes value in $\ZZ/(k\ZZ)=\{1,\ldots,k-1,k\}$,
and without loosing generality
we can assume $h_1\leq h_2\leq\cdots\leq h_N$
(by using the Weyl group).
In other words, holonomy variables
are given by
\begin{align}
\{h_a\}=\{1^{N_1},2^{N_2},\ldots,k^{N_k}\},
\end{align}
where $N_i$ are non-negative integers constrained by $N_1+N_2+\cdots+N_k=N$.
The gauge group $U(N)$ is broken by the orbifolding to $G_{\rm UV}=U(N_1)\times\cdots\times U(N_k)$,
and flows in the IR to $G_{\rm IR}=SU(N_1)\times\cdots\times SU(N_k)$.
The resulting theory is the quiver gauge theory
shown in Figure \ref{zkfig01.eps},
which we denote by $T_{N_1,N_2,\ldots,N_k}$.

In the IR the diagonal subgroups $U(1)_i\subset U(N_i)$ become global baryonic symmetries.
We will put off the discussion of the baryonic charges,
and we here focus on the sector with vanishing baryonic charges.
In the index calculation
this is equivalent to carrying out the gauge fugacity integral
not for $G_{\rm IR}$ but for $G_{\rm UV}$ including $U(1)_i$.

In the following we first consider the case with
$N_i=\ol N$ (and so $N=k\ol N$).
The superconformal index of the non-baryonic sector of $T_{\ol N,\ldots,\ol N}$ is
given by
\begin{align}
I^0_{T_{\ol N,\ldots,\ol N}}=\int_{G_{\rm UV}} dg \Pexp\left(P_k\left[f_{\rm vec}\chi_{\rm adj}^{U(N)}\right]\right),
\label{itn0}
\end{align}
where the superscript `0' indicates this is the index for the non-baryonic sector
and
$P_k$ denotes the projection associated with the $\ZZ_k$ action
(\ref{zkaction1}) and
(\ref{zkaction2}) on the fugacities:
\begin{align}
P_k[f_{\rm vec}(x,y,z,q,p)\chi^{U(N)}_{\text{adj}}(\zeta_a)]
=\frac{1}{k}\sum_{\omega\in\ZZ_k}
f_{\rm vec}(\omega x,\omega^{-1}y,z;q,p)\chi^{U(N)}_{\text{adj}}(\omega^{h_a}\zeta_a).
\label{pk0}
\end{align}
Note that
we do not remove the contribution from the $k$ IR-free $U(1)$ vector multiplets
in (\ref{itn0}).

The large $N$ limit of the theory $T_\infty=\lim_{\ol N\rightarrow\infty}T_{\ol N,\ldots,\ol N}$ is dual to
$AdS_5\times(S^5/\ZZ_k)$ where the $\ZZ_k$ action on $S^5$ is given by (\ref{s5zk}).
The fixed locus of the orbifold is $AdS_5\times S^1$.
The index is
\begin{align}
I_{T_{\infty}}
&=\Pexp\left[P_k(f_{\rm sugra})+(k-1)f_{\rm tensor}^{AdS_5\times S^1}\right]
\nonumber\\
&=\Pexp\left(\frac{1}{1-x^k}
+\frac{1}{1-y^k}
+k\frac{1}{1-z}
-k\frac{q}{1-q}
-k\frac{p}{1-p}
\right).
\label{largenorb1}
\end{align}
The letter index consists of two parts,
the contribution from
the supergravity multiplet in the ten-dimensional bulk
and the contribution from $k-1$ tensor multiplets
living on the six-dimensional fixed locus.
$f_{\rm tensor}^{AdS_5\times S^1}$ is the letter index of a
single tensor multiplet.
See \ref{tensor51} for a derivation of $f_{\rm tensor}^{AdS_5\times S^1}$.

\subsubsection{Decoupling}\label{orbdec.sec}
As in the $S^5$ case, the index of the theory on a system of giant gravitons
with wrapping numbers $m_x$, $m_y$, and $m_z$ is given by (\ref{fmxmymz})
with different $i_I[m_I]$ (and different terms represented by dots).
In the orbifold case $i_I[m_I]$ are given by
\begin{align}
i_I[m_I]=P_k(\sigma_If_{\rm vec}\chi^{U(m_I)}_{\text{adj}}),
\label{impk}
\end{align}
where $P_k$ is defined by (\ref{pk0}).
Let us focus on the constant term in the adjoint characters.
For each cycle we obtain
\begin{align}
&P_k[\sigma_xf_{\rm vec}]
=1-\frac{(1-x^{-1}y^{k-1})(1-q)(1-p)}{(1-y^k)(1-z)},\nonumber\\
&P_k[\sigma_yf_{\rm vec}]
=1-\frac{(1-y^{-1}x^{k-1})(1-q)(1-p)}{(1-x^k)(1-z)},\nonumber\\
&P_k[\sigma_zf_{\rm vec}]
=1-\frac{(1-z^{-1})(1-x^ky^k)(1-q)(1-p)}{(1-x^k)(1-y^k)(1-xy)}.
\label{pksigmaf}
\end{align}

Let us first consider the $Z$-expansion defined with the degrees $\deg(q,p,x,y,z)=(1,1,1,1,0)$.
We can easily see that none of three functions in
(\ref{pksigmaf}) contains negative-degree terms for $k\ge 2$.
This means all three cycles give non-trivial contributions,
and the triple-sum expansion does not reduce to the simple-sum expansion.

Next, let us consider $X$-expansion defined with the degrees $\deg(q,p,x,y,z)=(1,1,0,1,1)$.
In this case, the functions $P_k[\sigma_yf_{\rm vec}]$ and $P_k[\sigma_zf_{\rm vec}]$ contain infinitely many negative-degree terms.
Then, the contributions with $m_y+m_z\geq1$ become trivial, and we obtain the simple-sum expansion.

As we will discuss in detail in the next subsection,
the gauge group $U(m)$ on $m$ coincident giant gravitons
is broken down to $\wt G_{\rm UV}=U(m_1)\times\cdots\times U(m_k)$ due to
non-trivial holonomies on the giant gravitons.
We denote the theory by $\wt T_{m_1,m_2,\ldots,m_k}$.
The baryonic charges in $T_{\ol N,\ldots,\ol N}$ are related to the ranks $m_k$ of the unbroken gauge groups,
and vanishing baryonic charges correspond to the gauge groups with equal ranks: $m_1=m_2=\cdots=m_k=:\ol m$.
We denote such a theory by $\wt T_{\ol m,\ldots,\ol m}$.
This means that only the terms with wrapping number $m=k\ol m$ contribute to the index of non-baryonic sector.
Therefore, the simple-sum expansion takes the form
\begin{align}
\frac{I^0_{T_{\ol N,\ldots,\ol N}}}{I_{T_\infty}}=\sum_{\ol m=0}^\infty x^{k\ol m\ol N}\sigma_xI^0_{{\wt T}_{\ol m,\ldots,\ol m}},
\label{itnexpansion}
\end{align}
where the index appearing on the RHS is that for non-baryonic sector
because the baryonic charges in $\wt T_{\ol m,\ldots,\ol m}$ are related to
the ranks $N_i$, and we consider the special case with all $N_i$ being the same.

\subsection{Theories on giant gravitons: $\wt T_{m_1,\ldots,m_k}$}
As we discussed in
\ref{orbdec.sec}
the $X$-expansion of the index $I_{T_{\ol N,\ldots,\ol N}}$
gives $I_{{\wt T}_{\ol m,\ldots,\ol m}}$, which is the index of the theory
realized on the worldvolume of $m=k\ol m$ coincident giant gravitons
with unbroken gauge group $\wt G_{\rm UV}=U(\ol m)^k$.

\subsubsection{Projection}
Theory on giant gravitons on the cycle $X=0$ is also $\ZZ_k$ orbifold theory,
and $\ZZ_k$ action is obtained from the action in the original
theory by the operator map (\ref{generatormap}):
\begin{align}
\wt U_k=\sigma_xU_k\sigma_x=\exp\left(-\frac{2\pi i}{k}J_1-\frac{2\pi i}{k}R_x\right),
\label{ggtwist}
\end{align}
where $U_k$ is the $\ZZ_k$ generator in (\ref{s5zk}).
This $\ZZ_k$ generator contains $J_1$ nontrivially acting on the boundary
coordinates, and the boundary becomes $\RR_t\times S^3/\ZZ_k$,
which has a fixed locus $S^1\times \RR_t$.
The theory is locally the
${\cal N}=4$ $U(m)$ SYM and $\ZZ_k$ identification
breaks the gauge symmetry
to $\wt G_{\rm UV}=U(m_1)\times\cdots U(m_k)$.

The insertion of the operator $\wt U_k$ in the trace of the index
is realized by the following $\ZZ_k$ action on the fugacities:
\begin{align}
(q,p;x,y,z)\rightarrow (\omega_k^{-1}q,p;\omega_k^{-1}x,y,z).
\label{zkw1}
\end{align}
The gauge fugacities are also transformed by
\begin{align}
\zeta_a\rightarrow \omega_k^{\wt h_a}\zeta_a\quad (a=1,\ldots,m),
\label{zkw2}
\end{align}
where $\wt h_a$
are holonomy variables
which take values in $\ZZ/(k\ZZ)=\{1,\ldots,k\}$.
Without loosing generality we assume
\begin{align}
\{\wt h_a\}=\{1^{m_1},2^{m_2},\ldots,k^{m_k}\},
\end{align}
where $m_i$ are non-negative integers constrained by
$m_1+\cdots+m_k=m$.
The holonomy breaks the $U(m)$ gauge symmetry to $\wt G_{\rm UV}=U(m_1)\times\cdots\times U(m_k)$.
As we mentioned at the end of the previous subsection
the vanishing baryonic charges in $T_{\ol N,\ldots,\ol N}$ correspond to holonomies with $m_1=\cdots=m_k=:\ol m$.
With such a choice of the holonomy, the index is given by
\begin{align}
I^0_{{\wt T}_{\ol m,\ldots,\ol m}}=\int_{{\wt G}_{\rm UV}} dg \Pexp\left[\wt P_k(f_{\rm vec}\chi_{\rm adj}^{U(m)})\right],
\end{align}
where $\wt P_k$ is the projection associated with the $\ZZ_k$ actions
(\ref{zkw1}) and (\ref{zkw2}),
and is explicitly defined by
\begin{align}
\wt P_kf_{\rm vec}\chi_{\text{adj}}^{U(m)}
=\frac{1}{k}\sum_{\omega\in\ZZ_k}f_{\rm vec}(\omega^{-1}q,p,\omega^{-1}x,y,z)\chi^{U(m)}_{\text{adj}}(\omega^{\wt h_a}\zeta_a).
\end{align}

The index of the large $\ol m$ limit $\wt T_\infty=\lim_{\ol m\rightarrow\infty}\wt T_{\ol m,\ldots,\ol m}$ is given by
\begin{align}
I_{\wt T_\infty}
&=\Pexp(\wt P_k(f_{\rm sugra})+(k-1)f_{\rm tensor}^{AdS_3\times S^3})
\nonumber\\
&=\Pexp\left(\frac{x^k}{1-x^k}
+k\frac{y}{1-y}+k\frac{z}{1-z}
-\frac{q^k}{1-q^k}
-k\frac{p}{1-p}\right)
\end{align}
On the gravity side this is reproduced as the
contributions from the gravity multiplet in the ten-dimensional bulk and
$k-1$ tensor multiplets on the six-dimensional fixed locus $AdS_3\times S^3$.
See \ref{tensor33} for the explicit form of $f_{\rm tensor}^{AdS_3\times S^3}$.

\subsubsection{Decoupling}
Let us consider the triple-sum expansion of $I_{\wt T_{\ol m,\ldots,\ol m}}$.
The functions $F_{m_x,m_y,m_z}$ appearing in the expansion take the form
(\ref{fmxmymz}) with
$i_I[m_I]$ given by
\begin{align}
i_I[m_I]=\wt P_k(\sigma_If_{\rm vec}\chi_{\rm adj}^{U(m_I)}),
\end{align}
which are similar to 
(\ref{impk}) but $P_k$ is replaced by $\wt P_k$.
Let us focus on the constant terms in the adjoint characters $\chi_{\rm adj}^{U(m_I)}$.
For each cycle we have
\begin{align}
\wt P_k[\sigma_x f_{\rm vec}]
&=1-\frac{(1+x^{-1}q)(1-p)}{(1-y)(1-z)},\nonumber\\
\wt P_k[\sigma_y f_{\rm vec}]
&=1-\frac{(1-y^{-1})(1-qx^{k-1})(1-p)}{(1-x^k)(1-z)},\nonumber\\
\wt P_k[\sigma_zf_{\rm vec}]
&=1-\frac{(1-z^{-1})(1-qx^{k-1})(1-p)}{(1-x^k)(1-y)}.
\end{align}

Let us first consider whether simple-sum $Y$-expansion works.
With the degrees $\deg(q,p,x,y,z)=(1,1,1,0,1)$,
only $\wt P_k\sigma_z f_{\rm vec}$ contains infinitely many negative-degree
terms, and the decoupling works only partially.
Therefore, we cannot obtain simple-sum expansion associated with
the cycle $Y=0$
(and it is also the case for $Z=0$).

On the other hand,
with the degrees $\deg(q,p,x,y,z)=(1,1,0,1,1)$
$\wt P_k[\sigma_y f_{\rm vec}]$ and $\wt P_k[\sigma_z f_{\rm vec}]$ contain
infinitely many negative-degree terms,
and the corresponding cycles decouple.
Therefore, the triple sum reduces to the simple sum in the form
\begin{align}
\frac{I_{\wt T_{\ol m,\ldots,\ol m}}^0}{I_{\wt T_\infty}}
=\sum_{\ol N=0}^\infty x^{k\ol m\ol N}\sigma_xI^0_{T_{\ol N,\ldots,\ol N}.}
\label{nonbwtexp}
\end{align}
On the right hand side the index of the original theory $T_{\ol N,\ldots,\ol N}$ appears
because $\sigma_x$ is an involution and applying $\sigma_x$ on $\wt U_k$ gives the
original $\ZZ_k$ generator $U_k$.

\subsection{Baryonic charges}
In the analysis in the previous subsections
we focused on the non-baryonic sector,
and obtained (\ref{itnexpansion}) and (\ref{nonbwtexp}).
Let us generalize the relations by
including states with non-vanishing baryonic charges.

We can introduce baryonic charges on the both sides of the relation.
We use $B_i$ and $\wt B_i$ to denote the charges in $T_{\ol N,\ldots,\ol N}$ and those in $\wt T_{\ol m,\ldots,\ol m}$,
respectively.
As we will explain shortly, baryonic charges on one side are related to the ranks of the
unbroken gauge symmetries on the other side.
Namely, $B_i$ are identified with $m_i$ and $\wt B_i$ are identified with $N_i$
up to certain equivalence relations.
In general, if $SU$ factors had different ranks, the spectrum of
gauge invariant baryonic operators would be complicated.
Therefore, in the following we turn on only one of $B_i$ and $\wt B_i$
for simplicity.

Let us first discuss the baryonic charges $B_i$ in $T_{\ol N,\ldots,\ol N}$.
The index of the baryonic sector with charges $B_i$ can be obtained by the insertion
of the background charges
\begin{align}
\prod_{i=1}^k\left(\frac{\det g_i}{\det g_{i+1}}\right)^{B_i}
=\prod_{i=1}^k\left(\frac{\zeta_{i,1}\cdots\zeta_{i,\ol N}}{\zeta_{i+1,1}\cdots\zeta_{i+1,\ol N}}\right)^{B_i}
\label{bkgcharge}
\end{align}
in (\ref{itn0}).
These are defined so that the baryonic operators $\det X_{i+1,i}$ and $\det Y_{i,i+1}$
carry $B_i=+1$ and $-1$, respectively.
Because (\ref{bkgcharge}) is invariant under the shift $B_i\rightarrow B_i+c$
with $i$-independent constant $c$,
the $k$ charges $B_i$ are redundant,
and the baryonic sectors are labeled by the equivalence classes $[B_1,\ldots,B_k]$
defined with the equivalence relation
\begin{align}
B_i\sim B_i+c.
\label{bequiv}
\end{align}
This means that only the differences $B_i-B_j$ are physical quantities.

To extend the GG expansion (\ref{itnexpansion}),
we should also modify $I^0_{\wt T_{\ol m,\ldots,\ol m}}$ on the right hand side.
The ranks $m_i$ appearing in $\wt T_{m_1,\ldots,m_k}$ are related to the baryonic charges $B_i$ in $T_{\ol N,\ldots,\ol N}$.
This is because giant gravitons correspond to baryonic operators and baryonic charges correspond
to the numbers of giant gravitons with different holonomy variables on them.
In the single wrapping sector with $m=1$ there are $k$ contributions with only one of $m_i$ being $1$.
They correspond to $k$ baryonic operators $\det X_{i+1,i}$ ($i=1,\ldots,k$).
With this relation we identify $m_i$ with the baryonic charges $B_i$
up to the equivalence relation (\ref{bequiv}).

The giant graviton expansion for the baryonic sector is given by
\begin{align}
\frac{I_{T_{\ol N,\ldots,\ol N}}^{[B_1,\ldots,B_k]}}
{I_{T_\infty}}
=\sum_{(m)\in [B]}x^{(m_1+\cdots+m_k)\ol N}\sigma_x I_{\wt T_{m_1,\ldots,m_k}}^0,
\label{ggexpz20}
\end{align}
where $\sum_{(m)\in [B]}$ is the summation over
non-negative integers $m_i$ belonging to the equivalence class $[B_i]$.
The large $N$ limit appearing in the denominator on the left hand side
in (\ref{ggexpz20})
is the same as before.
The inverse expansion of (\ref{ggexpz20}) is given by
\begin{align}
\frac{I^0_{\wt T_{m_1,\ldots,m_k}}}{I_{\wt T_\infty}}
=\sum_{\ol N=0}^\infty x^{\ol N(m_1+\cdots+m_k)}
\sigma_x I^{[m_1,\ldots,m_k]}_{T_{\ol N,\ldots,\ol N}} .
\end{align}

We can also consider the baryonic charges $\wt B_i$ in $\wt T_{\ol m,\ldots,\ol m}$.
Then the corresponding ranks $N_i$ change.
For arbitrary ranks $m_i$ the spectrum
of gauge invariant baryonic operators is complicated.
So, we consider the simple case with equal ranks $m_1=m_2=\cdots=m_k=:\ol m$.
On the $T_{N_1,\ldots,N_k}$ side this corresponds to restricting the states
to the non-baryonic sector.
The GG expansion (\ref{nonbwtexp}) becomes
\begin{align}
\frac{I^{[\wt B_1,\ldots,\wt B_k]}_{\wt T_{\ol m,\ldots,\ol m}}}{I_{\wt T_\infty}}=\sum_{(N)\in[\wt B]}x^{\ol m(N_1+\cdots+N_k)}
\sigma_xI^0_{T_{N_1,\ldots,N_k}},
\label{invbaryonic}
\end{align}
and its inverse expansion is
\begin{align}
\frac{I_{T_{N_1,\ldots,N_k}}^0}
{I_{T_\infty}}
=\sum_{\ol m=0}^\infty
x^{\ol m(N_1+\cdots+N_k)}\sigma_x I_{\wt T_{\ol m,\ldots,\ol m}}^{[N_1,\ldots,N_k]}.
\label{ggexpni}
\end{align}

\subsection{Numerical tests}
\subsubsection{$X$-expansion of $I^0_{T_{\ol N,\ldots,\ol N}}$}
First let us test the expansion (\ref{itnexpansion}),
whose right hand side
is an infinite sum over the wrapping number $\ol m$.
We introduce a cutoff $\ol m_{\rm max}$, and
see how the difference
\begin{align}
\Delta_{T_{\ol N,\ldots,\ol N}}(\ol m_{\rm max})
=\frac{I_{T_{\ol N,\ldots,\ol N}}^0}{I_{T_\infty}}
-\sum_{\ol m=0}^{\ol m_{\rm max}} x^{k\ol m\ol N}\sigma_xI_{\wt T_{\ol m,\ldots,\ol m}}^0
\end{align}
changes as we increase $\ol m_{\rm max}$.

\begin{figure}[htb]
\centering
\includegraphics[scale=0.5]{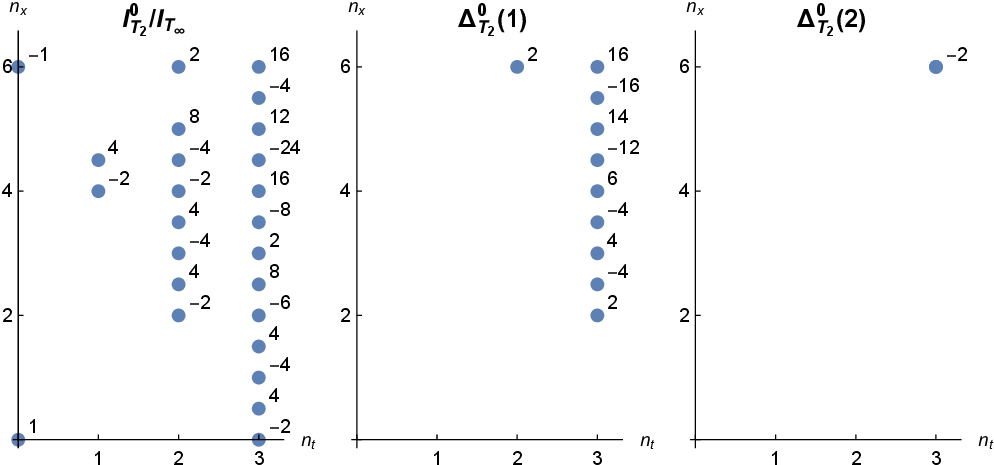}
\caption{$I_{T_{2,2}}/I_{T_\infty}$ and $\Delta_{T_{2,2}}(\ol m_{\rm max})$ with $\ol m_{\rm max}=1,2$ are shown as
two-dimensional plots.
A term $t^{n_t}x^{n_x}$ with non-vanishing coefficient
in the Taylor expansion is expressed as a dot at the coordinates $(n_t,n_x)$,
and the coefficient of the term is shown beside the dot.}\label{orbn2.eps}
\end{figure}

To realize the $X$-expansion
we take the degrees $\deg(q,p,x,y,z)=(1,1,0,1,1)$.
To reduce the computational cost, we take the unrefined parametrization
\begin{align}
(q,p,x,y,z)=(tx^{\frac{1}{2}},tx^{\frac{1}{2}},x,t,t),
\label{param1}
\end{align}
and treat the index as a function of two variables $t$ and $x$.
We first expand with respect to $t$, and then we perform the $x$-expansion.
For example, the ratio $I_{T_{2,2}}/I_{T_\infty}$ is
\begin{align}
I_{T_{2,2}}^0/I_{T_\infty}
&=(1-x^6+o(x^6))t^0\nonumber\\
&+(-2x^4+4x^{\frac{9}{2}}+o(x^6))t^1\nonumber\\
&+(-2x^2+4x^{\frac{5}{2}}-4x^3+4x^{\frac{7}{2}}-2x^4-4x^{\frac{9}{2}}+8x^5+2x^6+o(x^6))t^2\nonumber\\
&+(-2+4x^{\frac{1}{2}}-4x+4x^{\frac{3}{2}}-6x^2+8x^{\frac{5}{2}}+2x^3-8x^{\frac{7}{2}}+16x^4-24x^{\frac{9}{2}}
\nonumber\\
&\hspace{6cm}+12x^5-4x^{\frac{11}{2}}+16x^6+o(x^6))t^3\nonumber\\
&+o(t^3).
\label{ratio2}
\end{align}
If we subtract $\ol m=0$ and $\ol m=1$ contributions
many terms are canceled,
and further subtraction of the $\ol m=2$ contribution removes more terms
appearing in (\ref{ratio2}).
\begin{align}
\Delta_{T_{2,2}}(1)
&=o(x^6)t^0
+o(x^6)t^1
+(2x^6+o(x^6))t^2\nonumber\\
&
+(2x^2-4x^{\frac{5}{2}}+4x^3-4x^{\frac{7}{2}}+6x^4-12x^{\frac{9}{2}}+14x^5-16x^{\frac{11}{2}}+16x^6+o(x^6))t^3\nonumber\\
&+o(t^3),\nonumber\\
\Delta_{T_{2,2}}(2)
&=o(x^8)t^0
+o(x^8)t^1
+o(x^8)t^2
+(-2x^6+o(x^8))t^3
+o(t^3).
\end{align}
We show the cancellation graphically by using two-dimensional plots
in Figure \ref{orbn2.eps}.

\subsubsection{$X$-expansion of $I^0_{\wt T_{\ol m,\ldots,\ol m}}$}
\begin{figure}[htb]
\centering
\includegraphics[scale=0.5]{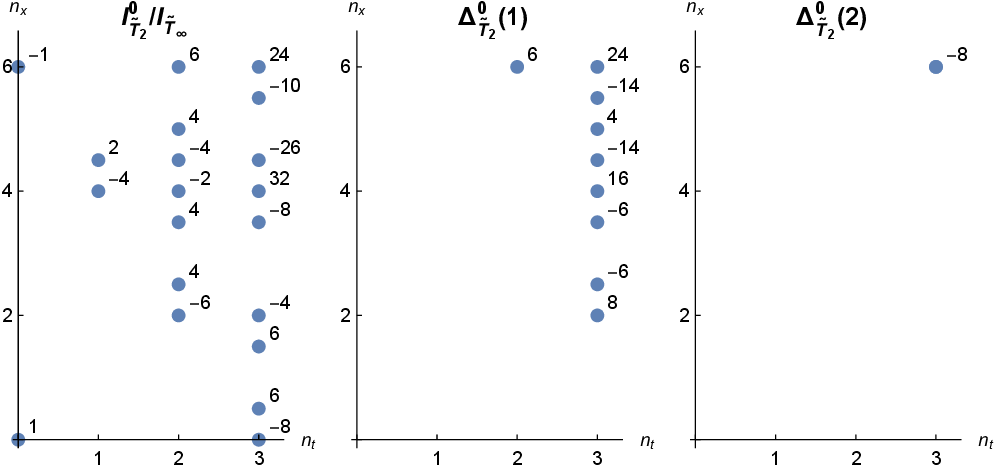}
\caption{
The ratio $I_{\wt T_{2,2}}^0/I_{\wt T_\infty}$ and the errors
$\Delta_{\wt T_{2,2}}(\ol N_{\rm max})$ with $\ol N_{\rm max}=1$ and $2$ are shown.}\label{z2inv.eps}
\end{figure}

Let us numerically test the expansion (\ref{nonbwtexp}) for $\wt T_{\ol m,\ldots,\ol m}$.
We introduce a cutoff $\ol N_{\rm max}$ and calculate the error
\begin{align}
\Delta_{\wt T_{\ol m,\ldots,\ol m}}(\ol N_{\rm max})
=\frac{I_{\wt T_{\ol m}}^0}{I_{\wt T_\infty}}
-\sum_{\ol N=0}^{\ol N_{\rm max}} x^{k\ol m\ol N}\sigma_xI_{T_{\ol N,\ldots,\ol N}}^0.
\end{align}
We show the ratio $I_{\wt T_{2,2}}^0/I_{\wt T_\infty}$
and the errors $\Delta_{\wt T_{2,2}}(\ol N_{\rm max})$ with $\ol N_{\rm max}=1$ and $2$
as two-dimensional plots in Figure \ref{z2inv.eps}.

\subsubsection{Baryonic sector}

\begin{figure}[htb]
\centering
\includegraphics[scale=0.5]{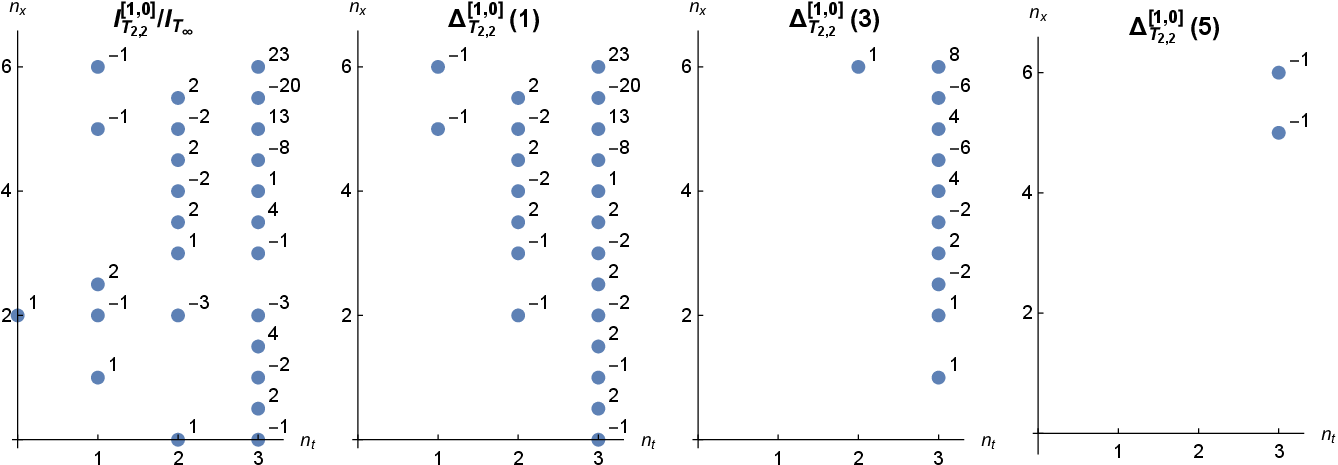}
\caption{The ratio $I_{T_{\ol N,\ol N}}^{[1,0]}/I_{T_\infty}$
and the error function $\Delta_{T_{\ol N,\ol N}}^{[1,0]}(c_{\rm max})$
for $c_{\rm max}=1$, $2$, and $3$ are shown as two-dimensional plots.}\label{orbb1n2.eps}
\end{figure}

Let us numerically test (\ref{ggexpz20}).
As a simple case we consider $k=2$ and $[B_1,B_2]=[1,0]$.
We define the error function
\begin{align}
\Delta_{T_{\ol N,\ol N}}^{[1,0]}(c_{\rm max})=
\frac{I_{T_{\ol N,\ol N}}^{[1,0]}}
{I_{T_\infty}}
-\sum_{c=0}^{c_{\rm max}}x^{(2c+1)\ol N}\sigma_x I_{\wt T_{c+1,c}}^0.
\end{align}
Figure \ref{orbb1n2.eps} shows the ratio $I_{T_{\ol N,\ol N}}^{[1,0]}/I_{T_\infty}$
and the error function
$\Delta_{T_{\ol N,\ol N}}^{[1,0]}(c_{\rm max})$
for $c_{\rm max}=1$, $2$, and $3$.

\begin{figure}[htb]
\centering
\includegraphics[scale=0.5]{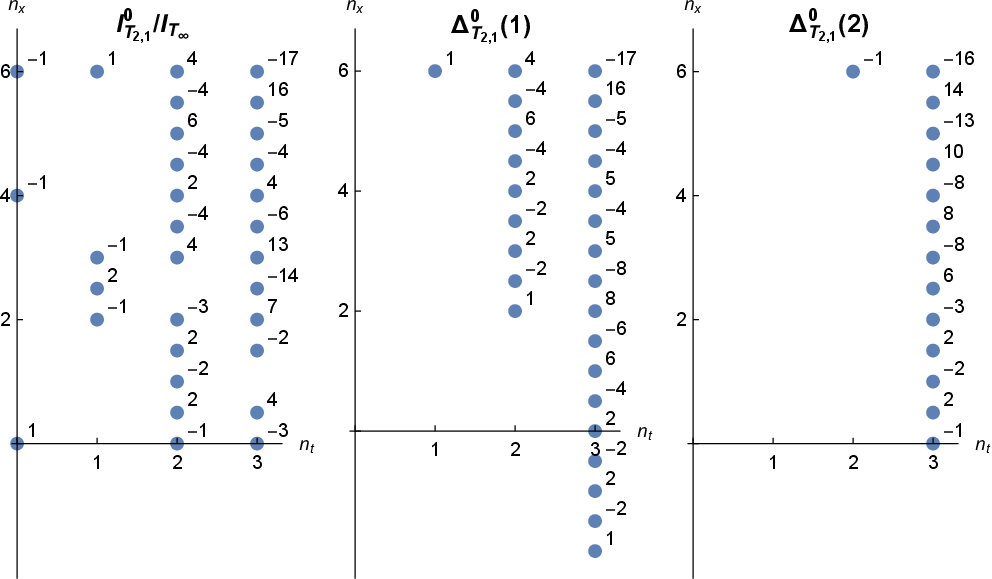}
\caption{
The ratio $I_{T_{3,2}}^0/I_{T_\infty}$ and the errors
$\Delta_{T_{3,2}}^0(\ol m_{\rm max})$ with $\ol m_{\rm max}=1$ and $2$ are
shown.
}\label{orbn21.eps}
\end{figure}

We test the expansion (\ref{ggexpni}) for $k=2$ and $(N_1,N_2)=(3,2)$.
We define the error function
\begin{align}
\Delta_{T_{3,2}}^0(\ol m_{\rm max})=
\frac{I_{T_{3,2}}^0}
{I_{T_\infty}}
-\sum_{\ol m=0}^{\ol m_{\rm max}}
x^{5\ol m}\sigma_x I_{\wt T_{\ol m,\ldots,\ol m}}^{[1,0]}.
\end{align}
Numerical results for $I_{T_{3,2}}^0/I_{T_\infty}$
and $\Delta_{T_{3,2}}^0(\ol m_{\rm max})$ with $\ol m_{\rm max}=1$ and $2$ are
shown in Figure \ref{orbn21.eps}.

\section{O3-D3 system}\label{o3.sec}
\begin{figure}[htb]
\centering
\includegraphics{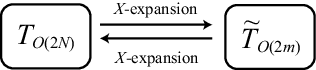}
\caption{$T_{O(2N)}$ and $\wt T_{O(2m)}$ are related by X-expansions.}\label{o3summary.eps}
\end{figure}

In this section we discuss ${\cal N}=4$ SYM realized by O3-D3 systems.
The AdS/CFT correspondence of the model was first studied in \cite{Witten:1998xy}.
We will mainly discuss the case with $O3^-$ plane,
and we denote the corresponding $O(2N)$ SYM by $T_{O(2N)}$.
We will find the $X$-expansion works for $T_{O(2N)}$,
and the theory on giant gravitons, which we denote by $\wt T_{O(2m)}$,
is another orientifold theory.
The inverse $X$-expansion also works, and it gives the original theory $T_{O(2N)}$
as the theory on giant gravitons (Figure \ref{o3summary.eps}).

\subsection{Boundary theories: $T_{O(2N)}$}
\subsubsection{Projection}
Let us consider ${\cal N}=4$ SYM with orthogonal and symplectic
gauge groups realized by orientifolds.
We first consider the case with $O3^-$-plane, which gives the orthogonal gauge groups.

The orientifold is defined with the O3 flip operator
\begin{align}
U_{O3}=e^{\pi iS},\quad
S=R_x+R_y+R_z+A,
\label{o3flip}
\end{align}
commuting with all generators in the $\mathcal N =4$ superconformal algebra.

Table \ref{d3o3setup.tbl} shows the directions of D3-branes,
O3-planes, and the worldvolume of giant gravitons.
The gauge group ($O$ or $Sp$) should be chosen
according to the relative positions of D-branes and O3-plane.
\begin{table}[htb]
\caption{Extended directions of branes are shown.
Directions of D3 and O3 are shown by using the coordinates in the flat ten-dimensional spacetime.
In the description of the directions for giant gravitons
the time directions are treated as the radial direction in the $XYZ$ space.
The column $O/Sp$ shows the gauge group realized on D-branes when the orientifold plane is $O3^-$.
For $O3^+$ all the gauge groups become $Sp$.
}\label{d3o3setup.tbl}
\centering
\begin{tabular}{cccccccccccc}
\hline
\hline
& $1$ & $2$ & $3$ & $4$ & \multicolumn{2}{c}{$X$} & \multicolumn{2}{c}{$Y$} & \multicolumn{2}{c}{$Z$} & $O/Sp$ \\
\hline
D3 & $\checkmark$ & $\checkmark$ & $\checkmark$ & $\checkmark$ &&&&&&& $O$ \\
O3$^-$ & $\checkmark$ & $\checkmark$ & $\checkmark$ & $\checkmark$ \\
GG$_{X=0}$ &&&&&&& $\checkmark$ & $\checkmark$ & $\checkmark$ & $\checkmark$ & $O$ \\
GG$_{Y=0}$ &&&&& $\checkmark$ & $\checkmark$ &&& $\checkmark$ & $\checkmark$ & $O$ \\
GG$_{Z=0}$ &&&&& $\checkmark$ & $\checkmark$ & $\checkmark$ & $\checkmark$ &&& $O$ \\
\hline
\end{tabular}
\end{table}

For the following analysis it is convenient to refine the superconformal index
by introducing a $\ZZ_2$ fugacity $\eta=\pm1$ for the operator $S$.
\begin{align}
 \wh I(q,p,x,y,z,\eta) = \tr[(-1)^Fq^{J_1}p^{J_2}x^{R_x}y^{R_y}z^{R_z} \eta^S].
 \label{twistedIndex}
\end{align}
Then, the orientifold projection operator $P_{O3}$ acting on letter indices is defined by
\begin{align}
P_{O3}[\cdots]=[\cdots]_+,
\end{align}
where we introduced the notation
\begin{align}
[\cdots]_\pm=\frac{1}{2}\left[(\cdots)|_{\eta=+1}\pm(\cdots)|_{\eta=-1}\right].
\end{align}

The introduction of the $\ZZ_2$ fugacity $\eta$ modifies
the variable changes in (\ref{sigmaxyz}) as follows \cite{Arai:2019xmp}:
\begin{align}
\sigma_x(q,p,x,y,z,\eta)&=(y\eta,z\eta,x^{-1},q\eta,p\eta,\eta),\nonumber\\
\sigma_y(q,p,x,y,z,\eta)&=(z\eta,x\eta,p\eta,y^{-1},q\eta,\eta),\nonumber\\
\sigma_z(q,p,x,y,z,\eta)&=(x\eta,y\eta,q\eta,p\eta,z^{-1},\eta).
\end{align}
To describe the orientifold action on the Chan-Paton factor
it is convenient to define the refined character $\wh\chi^{U(N)}_{\text{adj}}$ by
\begin{align}
\wh\chi^{U(N)}_{\text{adj}}
=\chi_+^{(N)}+\eta\chi_-^{(N)},
\end{align}
where $\chi_\pm^{(N)}$ for $g\in O(N)\subset U(N)$ are
\begin{align}
\chi_+^{(N)}=\chi^{O(N)}_{\rm adj},\quad
\chi_-^{(N)}=\chi^{O(N)}_{\rm sym} ,
\label{chiorth}
\end{align}
and for $g\in Sp(N/2)\subset U(N)$
\begin{align}
\chi_+^{(N)}=\chi^{Sp(N)}_{\text{anti-sym}},\quad
\chi_-^{(N)}=\chi^{Sp(N)}_{\rm adj}.
\label{chisp}
\end{align}
(\ref{chisp}) make sense only for even $N$.
The letter index of the orientifold theory with $O3^-$
is the $\ZZ_2$-invariant part of the refined $U(N)$ letter index
\begin{align}
P_{O3}[f_{\rm vec}\wh\chi^{U(N)}_{\text{adj}}]
=f_{\rm vec}\chi^{O(N)}_{\rm adj}.
\label{letterorth}
\end{align}
When we consider orientifold with $O3^+$ we should
replace $\wh\chi^{U(N)}_{\rm adj}$
by $\eta\wh\chi^{U(N)}_{\rm adj}$ and then
the orientifold projection gives
$f_{\rm vec}\chi^{Sp(N/2)}_{\rm adj}$.
With the letter index (\ref{letterorth}), the full index is given by
\begin{align}
I_{T_{O(2N)}}=
\int_{O(2N)}dg
\PE\left[f_{\rm vec} \chi^{O(2N)}_\mathrm{adj}\right] .
\label{o2nlocalization}
\end{align}
See Appendix \ref{haar.sec} for the explicit forms of the Haar measure and the character.

The large $N$ limit of the index $I_{T_{O(N)}}$ is given by
\begin{align}
I_{T_{O(\infty)}}
=\Pexp[P_{O3}\wh f_{\text{sugra}}].
\end{align}
$\wh f_{\text{sugra}}$ is the $\ZZ_2$-refined index of supergravity Kaluza-Klein modes in $AdS_5\times S^5$
defined by
\begin{align}
\wh f_{\text{sugra}}|_{\eta=+1}=f_{\rm sugra},\quad
\wh f_{\text{sugra}}|_{\eta=-1}=\wt f_{\rm sugra},
\end{align}
where the $\ZZ_2$ twisted index 
$\wt f_{\text{sugra}}$ is given by
\begin{align}
\wt f_{\text{sugra}}
&=\sum_{n=1}^\infty(-1)^ni[{\cal B}^{\frac{1}{2},\frac{1}{2}}_{[0,n,0](0,0)}]
\nonumber\\
&=\frac{1}{2}\left(
\frac{(1-x)(1-y)(1-z)(1+q)(1+p)}{(1+x)(1+y)(1+z)(1-q)(1-p)}-1\right).
\label{ikktwisted}
\end{align}
((\ref{ikktwisted}) is obtained by directly calculating the alternating sum over $n$.)

\subsubsection{Decoupling}
Let us discuss whether we can decouple some cycles by assigning appropriate degrees.
The giant graviton contribution again takes the form (\ref{fmxmymz}) with
\begin{align}
i_I[m_I]
&=P_{O3}[\sigma_If_{\rm vec}\wh\chi^{U(m_I)}_{\text{adj}}]
=\sum_{\pm}[\sigma_If_{\rm vec}]_\pm\chi_\pm^{(m_I)}.
\label{ixyzo3}
\end{align}
Let us assign the degrees $\deg(q,p,x,y,z)=(1,1,0,1,1)$.
The letter index $i_y[m_y]$ with $m_y\geq1$ includes
\begin{align}
[\sigma_yf_{\rm vec}]_\pm
=\sum_\pm\left[1-\frac{(1-y^{-1})(1-\eta q)(1-\eta p)}{(1-\eta y)(1-\eta z)}\right]_\pm.
\end{align}
We can easily check that for both signs $[\sigma_y f_{\rm vec}]_\pm$ contains infinitely many negative-degree terms.
It is the case also for $[\sigma_zf_{\rm vec}]_\pm$ associated with the cycle $Z=0$.
Therefore, $F_{m_x,m_y,m_z}$ with $m_y+m_z\geq1$ decouple, and the GG expansion reduces to the simple sum.

If the gauge group is not $SO(2N)$ but $O(2N)$, Pfaffian operators are not gauge invariant.
This means only contributions with even wrapping number should be included.
The simple-sum giant graviton expansion is given by
\begin{align}
\frac{I_{T_{O(2N)}}}{I_{T_{O(\infty)}}}
=\sum_{m=0}^\infty x^{2mN}\sigma_x I_{\wt T_{O(2m)}},
\label{expo2n}
\end{align}
where we denote the theory on $m$ coincident giant gravitons by $\wt T_{O(m)}$.

\subsection{Theories on giant gravitions: $\wt T_{O(2m)}$}
\subsubsection{Projection}
The $X$-expansion of $I_{T_{O(2N)}}$ is the
sum of the contributions from the theories
realized on giant gravitons wrapped around $X=0$.
Let $\wt T_{O(m)}$ be the theory realized on $m$ giant gravitons and $I_{\wt T_{O(m)}}$ be its index.
$\wt T_{O(m)}$ is defined as the orientifold of ${\cal N}=4$ $U(m)$ SYM
with the orientifold flip operator
\begin{align}
\wt U_{O3}=\sigma_x U_{O3}\sigma_x
= e^{\pi i (-R_x + J_1 + J_2 - A)}
=e^{\pi i(J_1+J_2+R_y+R_z-S)}
\label{omegawto3}
\end{align}
where $S$ is defined in (\ref{o3flip}).
The directions of the worldvolumes of D-branes and the O-plane are shown in Table \ref{d3o3dualsetup.tbl}.
\begin{table}[htb]
\caption{Extended directions of branes are shown.}\label{d3o3dualsetup.tbl}
\centering
\begin{tabular}{cccccccccccc}
\hline
\hline
& $1$ & $2$ & $3$ & $4$ & \multicolumn{2}{c}{$X$} & \multicolumn{2}{c}{$Y$} & \multicolumn{2}{c}{$Z$} & $O/Sp$ \\
\hline
D3 & $\checkmark$ & $\checkmark$ & $\checkmark$ & $\checkmark$ &&&&&&& $O$ \\
O3$^-$ &&&&&&& $\checkmark$ & $\checkmark$ & $\checkmark$ & $\checkmark$ \\
GG$_{X=0}$ &&&&&&& $\checkmark$ & $\checkmark$ & $\checkmark$ & $\checkmark$ & $O$ \\
GG$_{Y=0}$ &&&&& $\checkmark$ & $\checkmark$ &&& $\checkmark$ & $\checkmark$ & $Sp$ \\
GG$_{Z=0}$ &&&&& $\checkmark$ & $\checkmark$ & $\checkmark$ & $\checkmark$ &&& $Sp$ \\
\hline
\end{tabular}
\end{table}

(\ref{omegawto3}) non-trivially acts on the AdS boundary,
and the orientifold gives a theory in $\RR_t\times S^3/\ZZ_2$.
It is locally the same as the ${\cal N}=4$ $U(m)$ SYM,
but non-trivial holonomy breaks the gauge symmetry down to $O(m)$.
The fixed locus is an O3-plane, which is localized at the center of $AdS_5$
and wrapped around $S^3\subset S^5$ given by $X=0$.

The insertion of $\wt U_{O3}$ in the trace of the $\ZZ_2$-refined index (\ref{twistedIndex})
is equivalent to
the sign flips for some fugacities
\begin{align}
(q,p,x,y,z,\eta)
\rightarrow(-q,-p,x,-y,-z,-\eta).
\end{align}
Therefore, the orientifold projection operator acting on letter indices is defined by
\begin{align}
\wt P_{O3}F(q,p,x,y,z,\eta)
&=\frac{1}{2}(F(q,p,x,y,z,+1)+F(-q,-p,x,-y,-z,-1))
\nonumber\\
&=[F(\eta q,\eta p,x,\eta y,\eta z,\eta)]_+.
\end{align}
With this projection, the index $I_{\wt T_{O(m)}}$ is given by
\begin{align}
I_{\wt T_{O(m)}}=\int_{O(m)} dg \Pexp [\wt P_{O3}(f_{\rm vec}\wh\chi_{\text{adj}}^{U(m)})].
\label{iwttom}
\end{align}

The large $m$ limit of the index $I_{\wt T_{O(m)}}$ is given by
\begin{align}
I_{\wt T_{O(\infty)}}
=\Pexp[\wt P_{O3}\wh f_{\text{sugra}}]
=\Pexp\frac{f_{\text{sugra}}(q,p,x,y,z)+\wt f_{\text{sugra}}(-q,-p,x,-y,-z)}{2}
\end{align}

\subsubsection{Decoupling}
The index for giant gravitons can be easily obtained by combining the variable changes and the $\wt U_{O3}$-projection.
The functions $F_{m_x,m_y,m_z}$
are given by (\ref{fmxmymz}) with
\begin{align}
i_x[m_x]=\wt P_{O3}[\sigma_xf_{\rm vec}\wh\chi_{\rm adj}^{U(m_x)}],\quad
i_I[m_I]=\wt P_{O3}[\sigma_If_{\rm vec}\eta\wh\chi_{\rm adj}^{U(m_I)}]\quad(\mbox{for $I=y,z$}).
\end{align}
These are explicitly given by
\begin{align}
i_x[m_x]&=\left(1-\frac{(1-x^{-1})(1-q)(1-p)}{(1-y)(1-z)}\right)\chi^{O(m_x)}_{\rm adj},\nonumber\\
i_y[m_y]&=\sum_\pm\left[1-\frac{(1-\eta y^{-1})(1-q)(1-p)}{(1-z)(1-\eta x)}\right]_\pm\chi_\mp^{(m_y)},\nonumber\\
i_z[m_z]&=\sum_\pm\left[1-\frac{(1-\eta z^{-1})(1-q)(1-p)}{(1-y)(1-\eta x)}\right]_\pm\chi_\mp^{(m_z)}.
\end{align}
We can easily check that both $i_y[m_y]$ with $m_y\geq1$ and $i_z[m_z]$ with $m_z\geq1$
contain infinitely many negative-degree terms provided we assign the degrees $\deg(q,p,x,y,z)=(1,1,0,1,1)$.
Hence, the X-expansion reduces to the simple-sum expansion of the form
\begin{align}
\frac{I_{\wt T_{O(2m)}}}{I_{\wt T_{O(\infty)}}}
=\sum_{N=0}^\infty x^{2mN}\sigma_xI_{T_{O(2N)}}.
\label{io2mexp}
\end{align}

\subsection{$Sp(N)$, $SO(2N+1)$, and $SO(2N)$}
If we replace $O3^-$ with $O3^+$, we obtain the theory with symplectic gauge group.
In the index calculation, this is realized by replacing
$\wh\chi^{U(N)}_{\rm adj}$ in (\ref{o2nlocalization}) and 
(\ref{iwttom}) with
$\eta\wh\chi^{U(N)}_{\rm adj}$.
The decoupling again works, and instead of (\ref{expo2n}) and
(\ref{io2mexp}) we obtain the expansion
\begin{align}
\frac{I_{T_{Sp(N)}}}{I_{T_{Sp(\infty)}}}
=\sum_{m=0}^\infty
x^{2mN}\sigma_x I_{\wt T_{Sp(m)}},
\label{spexpansion}
\end{align}
and the inverse expansion
\begin{align}
\frac{I_{\wt T_{Sp(m)}}}{I_{\wt T_{Sp(\infty)}}}
=\sum_{N=0}^\infty
x^{2mN}\sigma_x I_{T_{Sp(N)}}.
\label{spinvexpansion}
\end{align}

The gauge group $O(2N)$ consists of two disconnected components.
One of them is $SO(2N)$, and we denote the other component by $\ol{SO}(2N)$.
Correspondingly, the index for $O(2N)$ splits into two parts:
\begin{align}
I_{T_{O(2N)}}=
\frac{1}{2}(I_{T_{SO(2N)}}+I_{T_{\ol{SO}(2N)}}).
\end{align}
Two contributions correspond to the two elements of $\ZZ_2=O(2N)/SO(2N)$.
This $\ZZ_2$ symmetry couples to Pfaffian operators.
In $O(2N)$ gauge theory
this $\ZZ_2$ is gauged,
and Pfaffian operators do not contribute to the index.
This is analogous to the baryonic $U(1)$ symmetries
studied in Section \ref{orb.sec}.
For the latter we can extract the contribution
of states
with non-vanishing baryonic charges by
inserting the factor (\ref{bkgcharge}) in the gauge fugacity integral in (\ref{o2nlocalization}).
This is also possible for the $\ZZ_2$ charge.
By inserting the factor $\det g$ ($g\in O(2N)$)
in the integral (\ref{o2nlocalization}) we obtain
the index of the $\ZZ_2$-odd states.
\begin{align}
I_{T_{O(2N)}}^{(-)}=
\frac{1}{2}(I_{T_{SO(2N)}}- I_{T_{\ol{SO}(2N)}}).
\label{z2oddindex}
\end{align}

Because a giant graviton, which correspond to a Pfaffian operator,
is $\ZZ_2$ odd,
only configurations with odd wrapping numbers contribute to (\ref{z2oddindex}),
and the expansion (\ref{expo2n}) is changed to
\begin{align}
\frac{I_{T_{O(2N)}}^{(-)}}{I_{T_{O(\infty)}}}=\sum_{m=0}^\infty x^{(2m+1)N}\sigma_x I_{\wt T_{O(2m+1)}}.
\label{onminus}
\end{align}
The sum of (\ref{expo2n}) and (\ref{onminus}) gives the expansion for $SO(2N)$ gauge group
\begin{align}
\frac{I_{T_{SO(2N)}}}{I_{T_{SO(\infty)}}}=\sum_{m=0}^\infty x^{mN}\sigma_x I_{\wt T_{O(m)}}.
\end{align}

Two expansions (\ref{expo2n}) and (\ref{onminus}) are
similar to the expansion (\ref{ggexpz20}) for the baryonic sector
in the sense that
the number of giant gravitons is constrained according to the value of the
$\ZZ_2$ charge.

We can interchange the roles of the $\ZZ_2$ charge and the number of branes,
and obtain the following expansion similar to (\ref{ggexpni}).
\begin{align}
\frac{I_{T_{O(2N+1)}}}{I_{T_{O(\infty)}}}=\sum_{m=0}^\infty x^{(2N+1)m}\sigma_x I_{\wt T_{O(2m)}}^{(-)} ,
\label{o2np1}
\end{align}
where $I_{\wt T_{O(2m)}}^{(-)}$ is defined with the $\det g$ insertion into (\ref{iwttom}),
and given in a similar way to (\ref{z2oddindex}).

The left hand side of (\ref{o2np1})
should be identical with the left hand side of (\ref{spexpansion})
due to the Montonen-Olive duality.
The consistency requires non-trivial relation
between $I_{{\wt T}_{Sp(m)}}$ and $I_{\wt T_{O(2m)}}^{(-)}$ appearing in the expansions.
In fact, we can numerically confirm the following relation holds:
\begin{align}
I_{\wt T_{Sp(m)}}=x^{-m}I_{\wt T_{O(2m)}}^{(-)}.
\end{align}
This is the counterpart of the Montonen-Olive duality on the giant graviton side.

\subsection{Numerical tests}
\subsubsection{X-expansion of $I_{T_{O(2N)}}$}
\begin{figure}[htb]
 \centering
 \includegraphics[scale=0.6]{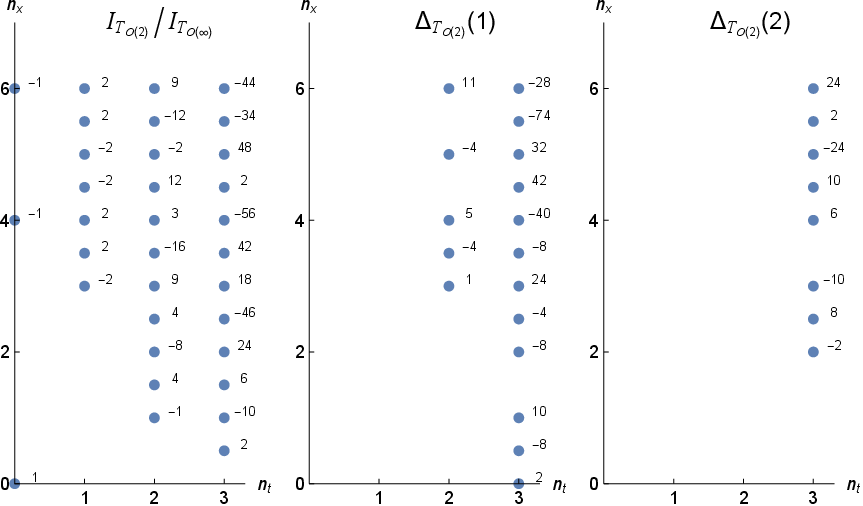}
 \caption{The ratio $I_{T_{O(2)}}/I_{T_{O(\infty)}}$ and the errors $\Delta_{T_{O(2)}}(m_\mathrm{max})$
 with $m_\mathrm{max}=1$ and $2$ are shown.}
 \label{o2.eps}
\end{figure}

Let us check the expansion of (\ref{expo2n}) for $T_{O(2N)}$.
Here and in following numerical tests, we again employ the unrefined parametrization (\ref{param1}).
We introduce a cutoff $m_\mathrm{max}$, 
and calculate the error function below for $N=1$,
\begin{align}
\Delta_{T_{O(2N)}}(m_\mathrm{max})=\frac{I_{T_{O(2N)}}}{I_{T_{O(\infty)}}}
-\sum_{m=0}^{m_\mathrm{max}} x^{2mN}\sigma_x I_{\wt T_{O(2m)}}.
\end{align}
The results are shown in Figure~\ref{o2.eps} as two-dimensional plots.

\subsubsection{X-expansion of $I_{\wt T_{O(2N)}}$}

\begin{figure}[htb]
 \centering
 \includegraphics[scale=0.6]{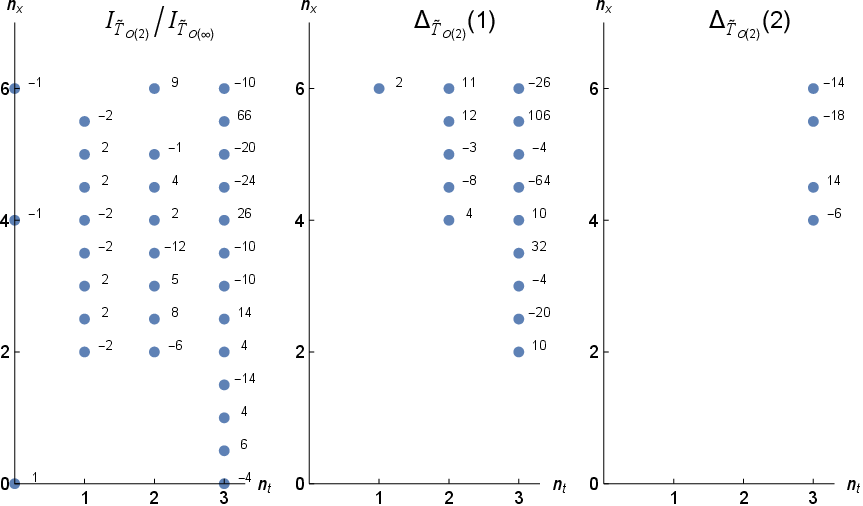}
 \caption{The ratio $I_{\wt T_{O(2)}}/I_{\wt T_{O(\infty)}}$ and the errors $\Delta_{\wt T_{O(2)}}(m_\mathrm{max})$
 with $m_\mathrm{max}=1$ and $2$ are shown.}
 \label{oTil2.eps}
\end{figure}

Let us check the inverse expansion for $T_{O(2N)}$, namely, the expansion of $I_{\wt T_{O(2N)}}$ (\ref{io2mexp}).
We calculate the error function
\begin{align}
\Delta_{\wt T_{O(2N)}}(m_\mathrm{max})=\frac{I_{\wt T_{O(2N)}}}{I_{\wt T_{O(\infty)}}}
-\sum_{m=0}^{m_\mathrm{max}} x^{2mN}\sigma_x I_{T_{O(2m)}}
\end{align}
for $N=1$ with cutoff $m_\mathrm{max} = 1$ and $2$. The results are shown in Figure~\ref{oTil2.eps}.

\subsubsection{X-expansion of $I_{T_{Sp(N)}}$, $I_{T_{SO(2N+1)}}$, and $I_{T_{SO(2N)}}$}

\begin{figure}[htb]
 \centering
 \includegraphics[scale=0.6]{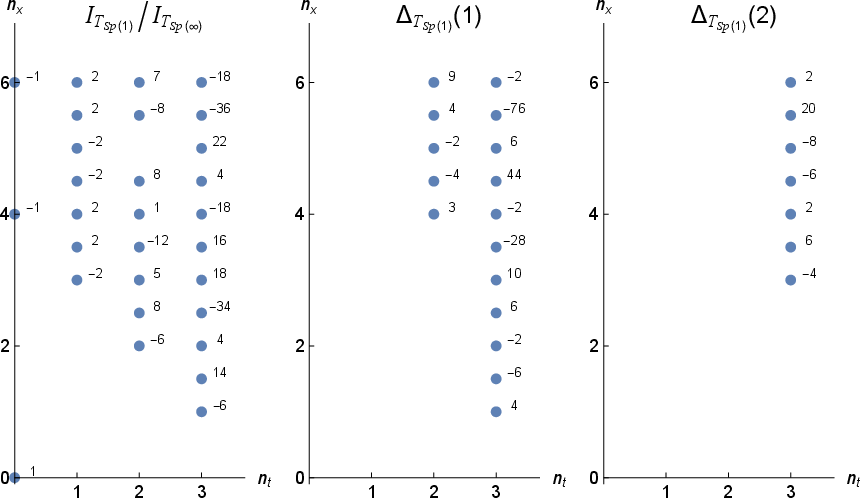}
 \caption{The ratio $I_{T_{Sp(1)}}/I_{T_{Sp(\infty)}}$ and the errors $\Delta_{T_{Sp(1)}}(m_\mathrm{max})$
 with $m_\mathrm{max}=1$ and $2$ are shown.}
 \label{sp1.eps}
\end{figure}

\begin{figure}[htb]
 \centering
 \includegraphics[scale=0.6]{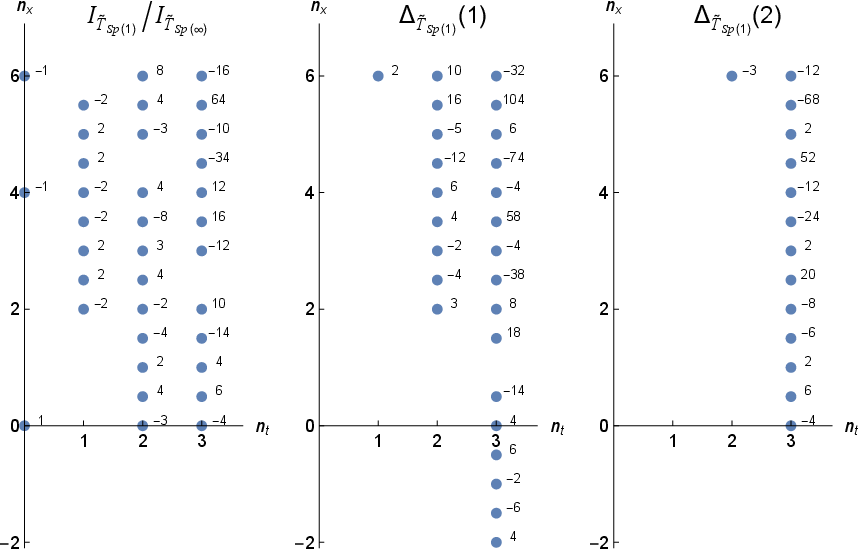}
 \caption{The ratio $I_{\wt T_{Sp(1)}}/I_{\wt T_{Sp(\infty)}}$ and the errors $\Delta_{\wt T_{Sp(1)}}(m_\mathrm{max})$
 with $m_\mathrm{max}=1$ and $2$ are shown.}
 \label{sp1Til.eps}
\end{figure}

Firstly, let us check the expansion of $I_{T_{Sp(N)}}$ (\ref{spexpansion}) and its inverse expansion (\ref{spinvexpansion}).
We introduce cutoff $m_\mathrm{max}$ and calculate the error functions
\begin{align}
\Delta_{T_{Sp(N)}}(m_\mathrm{max}) &=\frac{I_{T_{Sp(N)}}}{I_{T_{Sp(\infty)}}}
-\sum_{m=0}^{m_\mathrm{max}} x^{2mN}\sigma_x I_{\wt T_{Sp(m)}} ,  \\
\Delta_{\wt T_{Sp(N)}}(m_\mathrm{max}) &=\frac{I_{\wt T_{Sp(N)}}}{I_{\wt T_{Sp(\infty)}}}
-\sum_{m=0}^{m_\mathrm{max}} x^{2mN}\sigma_x I_{T_{Sp(m)}} ,
\end{align}
for $N=1$.
The results are shown in Figure \ref{sp1.eps} and Figure \ref{sp1Til.eps}.

\begin{figure}[htb]
 \centering
 \includegraphics[scale=0.7]{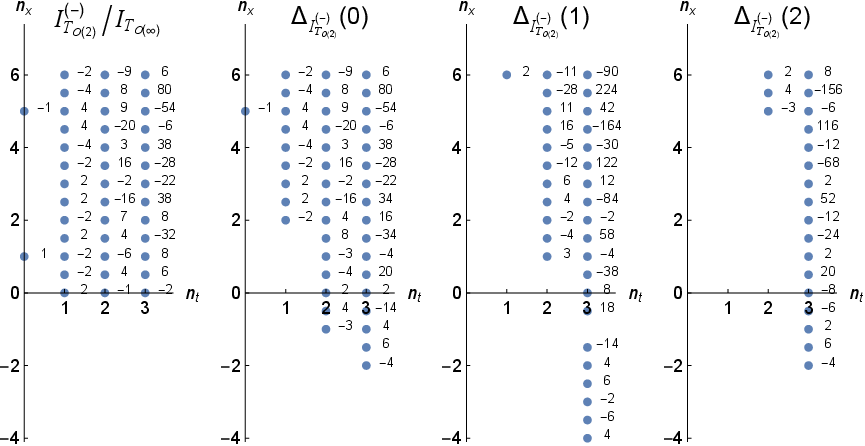}
 \caption{The ratio $I_{T_{O(2)}}^{(-)}/I_{T_{O(\infty)}}$ and the errors $\Delta_{I^{(-)}_{T_{O(2)}}}(m_\mathrm{max})$
 with $m_\mathrm{max}=0,\ 1,$ and $2$ are shown.}
 \label{om2.eps}
\end{figure}

Secondary, we check the expansion of (\ref{onminus}).
We calculate the error function
\begin{align}
\Delta_{I_{T_{O(2N)}}^{(-)}}(m_\mathrm{max}) = 
\frac{I_{T_{O(2N)}}^{(-)}}{I_{T_{O(\infty)}}} -\sum_{m=0}^{m_\mathrm{max}} x^{(2m+1)N}\sigma_x I_{\wt T_{O(2m+1)}}
\end{align}
for $N=1$.
The results are shown in Figure \ref{om2.eps}.

\begin{figure}[htb]
 \centering
 \includegraphics[scale=0.6]{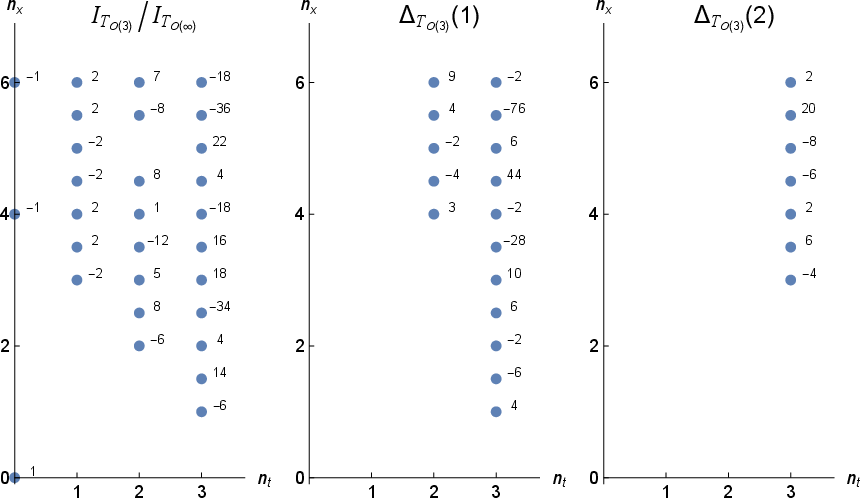}
 \caption{The ratio $I_{T_{O(3)}}/I_{T_{O(\infty)}}$ and the errors $\Delta_{T_{O(3)}}(m_\mathrm{max})$
 with $m_\mathrm{max}=1$ and $2$ are shown.}
 \label{o3.eps}
\end{figure}

Finally, we check the expansion of (\ref{o2np1}),
by calculating the error function
\begin{align}
 \Delta_{T_{O(2N+1)}}(m_\mathrm{max}) =\frac{I_{T_{O(2N+1)}}}{I_{T_{O(\infty)}}} -\sum_{m=0}^{m_\mathrm{max}} x^{(2N+1)m}\sigma_x I_{\wt T_{O(2m)}}^{(-)}
\end{align}
with $m_\mathrm{max} =1$ and $2$ for $N=1$.
The results are shown in Figure \ref{o3.eps}.

\section{O7-D3 system}\label{o7.sec}
\begin{figure}[htb]
\centering
\includegraphics{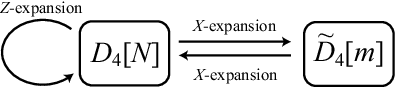}
\caption{GG expansions for $D_4[N]$ and $\wt D_4[m]$.}\label{o7d3summary.eps}
\end{figure}

In this section we discuss ${\cal N}=2$ SCFT realized on D3-branes
probing the O7-plane background, which we denote by $D_4[N]$.
We will find all $X$-, $Y$-, and $Z$-expansions work.
$X$-expansion (which is essentially the same as the $Y$-expansion up to 
the exchange $x\leftrightarrow y$) gives another orientifold theory denoted by
$\wt D_4[m]$, while $Z$-expansion gives the original theory.
The $X$-expansion works for $\wt D_4[m]$ and it gives the original theory
$D_4[N]$ (Figure \ref{o7d3summary.eps}).

\subsection{Boundary theories: $D_4[N]$}\label{o7bdr.sec}
\subsubsection{Projection}

Let us consider the $\mathcal{N}=2$ SCFT realized on $N$ D3-branes in the
background of an $O7^-$-plane
defined with the orientifold flip operator
\begin{align}
U_{\rm O7}=e^{i\pi(R_z-A)}
=e^{i\pi(2R_z+R_x+R_y-S)}
=e^{i\pi(R_x-R_y-S)}.
\label{o7flip}
\end{align}
The $O7^-$ worldvolume is space-filling in $AdS_5$ and wraps around $S^3\subset S^5$ given by $Z=0$.
To keep the conformal invariance we need
to introduce four D7-branes (and their mirror images) coincident with the O7-plane.

\begin{table}[htb]
\caption{Extended directions of branes are shown.}\label{}
\centering
\begin{tabular}{cccccccccccc}
\hline
\hline
& $1$ & $2$ & $3$ & $4$ & \multicolumn{2}{c}{$X$} & \multicolumn{2}{c}{$Y$} & \multicolumn{2}{c}{$Z$} & $O/Sp$ \\
\hline
D3 & $\checkmark$ & $\checkmark$ & $\checkmark$ & $\checkmark$ &&&&&&& $Sp$ \\
O7$^-$ & $\checkmark$ & $\checkmark$ & $\checkmark$ & $\checkmark$ & $\checkmark$ & $\checkmark$ & $\checkmark$ & $\checkmark$ \\
GG$_{X=0}$ &&&&&&& $\checkmark$ & $\checkmark$ & $\checkmark$ & $\checkmark$ & $O$ \\
GG$_{Y=0}$ &&&&& $\checkmark$ & $\checkmark$ &&& $\checkmark$ & $\checkmark$ & $O$ \\
GG$_{Z=0}$ &&&&& $\checkmark$ & $\checkmark$ & $\checkmark$ & $\checkmark$ &&& $Sp$ \\
\hline
\end{tabular}
\end{table}

The orientifold projection with (\ref{o7flip}) breaks
the $SO(6)_R$ symmetry down to $SU(2)_R\times U(1)_R\times SU(2)_F$,
with Cartan operators $R_x+R_y$, $R_z$, and $R_x-R_y$.
The theory realized on the worldvolume of $N$ D3-branes
is an $\mathcal N=2$ $Sp(N)$ superconformal field theory \cite{Sen:1996vd},
which we call $D_4[N]$.
It has $SO(8)$ flavor symmetry realized on the 7-branes,
and we can refine the index by introducing $SO(8)$ fugacities.

The insertion of $U_{O7}$ in the $\ZZ_2$ refined index (\ref{twistedIndex})
is equivalent to the variable change
\begin{align}
(q,p,x,y,z,\eta)\rightarrow(q,p,-x,-y,z,-\eta).
\end{align}
Correspondingly, we define the projection
operator $P_{O7}$ acting on letter indices by
\begin{align}
P_{O7}F(q,p,x,y,z,\eta)
&=\frac{1}{2}[F(q,p,x,y,z,+1)+F(q,p,-x,-y,z,-1)]
\nonumber\\
&=[F(q,p,\eta x,\eta y,z,\eta)]_+.
\end{align}
With this projection operator, the index of $D_4[N]$ is given by
\begin{align}
I_{D_4[N]}
=\int_{Sp(N)} dg \Pexp(P_{O7}[f_{\rm vec}\eta\wh\chi^{U(2N)}_{\rm adj}]
+f_{\rm hyp}\chi^{SO(8)}_{\bm{8}_v}\chi_{\rm fund}^{Sp(N)}).
\end{align}
The first term in the letter index is the contribution from D3-D3 open strings,
and is explicitly given by
\begin{align}
P_{O7}[f_{\rm vec}\eta\wh\chi^{U(2N)}_{\rm adj}]
&=\sum_\pm[f_{\rm vec}(q,p,\eta x,\eta y,z)]_\pm\chi_\mp^{U(2N)}
\nonumber\\
&=\left(1-\frac{(1+xy)(1-z)}{(1-q)(1-p)}\right)\chi^{Sp(N)}_{\rm adj}
+\frac{(x+y)(1-z)}{(1-q)(1-p)}\chi^{Sp(N)}_{\rm anti-sym}.
\end{align}
The second term in the letter index is the contribution of
D3-D7 open strings.
$f_{\rm hyp}$ is the letter index of the $\mathcal N=2$ hypermultiplet.
\begin{align}
f_{\rm hyp}=\frac{(xy)^{\frac12}(1-z)}{(1-q)(1-p)}.
\end{align}
See Appendix \ref{haar.sec} for the explicit forms of the Haar measure and the characters.\footnote{
Note that the anti-symmetric tensor representation
of $Sp(N)$ is reducible and contains the singlet representation, which
corresponds to the center of mass degrees of freedom
in the O7-D3 system.
In the following analysis we will not remove the singlet contribution,
and $\chi^{Sp(N)}_{\rm anti-sym}$ includes the contribution from the singlet.}

The holographic dual of $D_4[N]$ is $AdS_5\times (S^5/\ZZ_2)$,
where $\ZZ_2$ action on $S^5$ is given by (\ref{o7flip}),
and the worldvolume of the O7-plane and the four coincident D7-branes
is $AdS_5\times S^3$ \cite{Fayyazuddin:1998fb,Aharony:1998xz}.
The large $N$ index can be calculated on the gravity side
as the contribution from massless fields.
There are two contributions:
\begin{align}
I_{D_4[\infty]}=\Pexp[P_{O7}\wh f_{\rm sugra}+f_{\rm vector}^{AdS_5\times S^3}\chi^{SO(8)}_{\bm{28}}] .
\end{align}
One is the contribution from the gravity multiplet
\begin{align}
P_{O7}\wh f_{\rm sugra}
=&[\wh f_{\rm sugra}(q,p,\eta x,\eta y,z,\eta)]_+
\nonumber\\
=&\frac{1}{2}(f_{\rm sugra}(x,y)+\wt f_{\rm sugra}(-x,-y))
\nonumber\\
=&\frac12\left(\frac{x}{1-x}+\frac{y}{1-y}+\frac{z}{1-z}
-\frac{q}{1-q}-\frac{p}{1-p}\right)\nn\\
&+\frac14\left(\frac{(1+x)(1+y)(1-z)(1+q)(1+p)}{(1-x)(1-y)(1+z)(1-q)(1-p)}-1\right).
\end{align}
The other is the contribution from D7-D3 open strings.
$f_{\rm vector}^{AdS_5\times S^3}$
is the letter index of the eight-dimensional vector multiplet in $AdS_5\times S^3$.
See \ref{vector53} for a brief derivation of $f_{\rm vector}^{AdS_5\times S^3}$.

\subsubsection{Decoupling}
Again, the contribution from the system of giant gravitons
with wrapping numbers $m_x$, $m_y$, and $m_z$
is given by (\ref{fmxmymz}).
We again focus only on the three terms $i_I[m_I]$.
They are given by
\begin{align}
i_x[m_x]
&=P_{O7}[\sigma_x f_{\rm vec}\wh\chi^{U(m_x)}_{\rm adj}]
=\sum_\pm\left[1-\frac{(1-\eta x^{-1})(1-\eta q)(1-\eta p)}{(1-y)(1-\eta z)}\right]_\pm\chi_\pm^{(m_x)},
\nonumber\\
i_y[m_y]
&=P_{O7}[\sigma_y f_{\rm vec}\wh\chi^{U(m_y)}_{\rm adj}]
=\sum_\pm\left[1-\frac{(1-\eta y^{-1})(1-\eta q)(1-\eta p)}{(1-x)(1-\eta z)}\right]_\pm\chi_\pm^{(m_y)},
\nonumber\\
i_z[m_z]
&=P_{O7}[\sigma_z f_{\rm vec}\eta\wh\chi^{U(2m_z)}_{\rm adj}]
=\sum_\pm\left[1-\frac{(1-z^{-1})(1-\eta q)(1-\eta p)}{(1-x)(1-y)}\right]_\pm\chi_\mp^{(2m_z)},
\label{ixyzo7}
\end{align}
With these letter indices, we can check that all the $X$-, $Y$-, and $Z$-expansions work.

To obtain $Z$-expansion, we adopt the degrees $\deg(q,p,x,y,z)=(1,1,1,1,0)$.
Then, we can show that both $i_x[m_x]$ with $m_x\geq1$ and $i_y[m_y]$ with $m_y\geq1$ contain
infinitely many negative-degree terms, and they decouple.
We obtain the simple-sum GG expansion
\begin{align}
\frac{I_{D_4[N]}}{I_{D_4[\infty]}}
=\sum_{m=0}^\infty z^{2mN}\sigma_z I_{D_4[m]}.
\label{d4zexp}
\end{align}
In this case, the theory on the $m$ giant gravitons is again $D_4[m]$.
This is because $\sigma_z$ maps the orientifold operator $U_{O7}$ to itself.

The $X$-expansion and the $Y$-expansion are essentially the same with each other and they are related by the
Weyl reflection $x\leftrightarrow y$ of the $SU(2)_R$ symmetry.
Let us consider the $X$-expansion for concreteness.
To obtain $X$-expansion, we adopt the degrees $\deg(q,p,x,y,z)=(1,1,0,1,1)$.
Then, we can easily confirm that both $i_y[m_y]$ with $m_y\geq1$ and $i_z[m_z]$ with $m_z\geq1$ contain
infinitely many negative-degree terms, and the cycles $Y=0$ and $Z=0$ decouple.
As the result, we obtain the simple-sum giant graviton expansion
\begin{align}
\frac{I_{D_4[N]}}{I_{D_4[\infty]}}
=\sum_{m=0}^\infty x^{mN}\sigma_x I_{\wt D_4[m]},
\label{d7xexp}
\end{align}
where we denote the theory realized on giant gravitons by $\wt D_4[m]$.
This will be defined in the next subsection.

\subsection{Theories on giant gravitons: $\wt D_4[m]$}
\subsubsection{Projection}
$I_{\wt D_4[m]}$ appearing on the right hand side in (\ref{d7xexp})
is the index of the orientifold theory $\wt D_4[m]$ defined with the orientifold flip operator
\begin{align}
\wt U_{O7}=\sigma_x U_{O7}\sigma_x=e^{\pi i (J_2+A)}
=e^{i\pi(J_2-R_x-R_y-R_z+S)}.
\label{o7wt}
\end{align}
The fixed locus of $\wt U_{O7}$ is the
O7-plane (together with four coincident D7-branes) along different directions
from the one for $U_{O7}$.
See Table \ref{oconst}.
\begin{table}[htb]
\caption{Extended directions of branes are shown.}\label{oconst}
\centering
\begin{tabular}{cccccccccccc}
\hline
\hline
& $1$ & $2$ & $3$ & $4$ & \multicolumn{2}{c}{$X$} & \multicolumn{2}{c}{$Y$} & \multicolumn{2}{c}{$Z$} & $O/Sp$ \\
\hline
D3&\checkmark&\checkmark&\checkmark&\checkmark&&&&&&& $O$ \\
O7$^-$ &\checkmark&\checkmark&&&\checkmark&\checkmark&\checkmark&\checkmark&\checkmark&\checkmark\\
GG$_{X=0}$ &&&&&&& $\checkmark$ & $\checkmark$ & $\checkmark$ & $\checkmark$ & $Sp$ \\
GG$_{Y=0}$ &&&&& $\checkmark$ & $\checkmark$ &&& $\checkmark$ & $\checkmark$ & $Sp$ \\
GG$_{Z=0}$ &&&&& $\checkmark$ & $\checkmark$ & $\checkmark$ & $\checkmark$ &&& $Sp$ \\
\hline
\end{tabular}
\end{table}
The worldvolume theory on D3-branes probing the $7$-brane
background is locally the ${\cal N}=4$ $U(m)$ SYM,
and the $\ZZ_2$ identification breaks the $U(m)$ gauge symmetry to $O(m)$.

The insertion of $\wt U_{O7}$ in the $\ZZ_2$ refined index (\ref{twistedIndex}) is equivalent to
the variable change
\begin{align}
(q,p,x,y,z,\eta)\rightarrow(q,-p,-x,-y,-z,-\eta) .
\end{align}
Correspondingly, the projection operator $P_{\wt{O7}}$ acting on letter indices
is defined by
\begin{align}
P_{\wt{O7}}F(q,p,x,y,z,\eta)
&=\frac{1}{2}[F(q,p,x,y,z,+1)+F(q,-p,-x,-y,-z,-1)]
\nonumber\\
&=[F(q,\eta p,\eta x,\eta y,\eta z,\eta)]_+.
\end{align}
With this projection operator,
the index is given by
\begin{align}
I_{\wt{D}_4[m]} =
\int_{O(m)} dg \Pexp(P_{\wt{O7}}[f_{\rm vec}\wh\chi^{U(m)}_{\rm adj}]
+f_\psi\chi_{\rm vec}^{O(m)}\chi^{D_4}_{\bm{8}_v}),
\end{align}
where the first term in the letter index is the contribution of
D3-D3 open strings, and is rewritten as
\begin{align}
P_{\wt{O7}}[f_{\rm vec}\wh\chi^{U(m)}_{\rm adj}]
&=\sum_\pm[f_{\rm vec}(q,\eta p,\eta x,\eta y,\eta z)]_\pm\chi_\pm^{(m)}.
\end{align}
The second term is the contribution of chiral fermions arising from D3-D7 open strings.
They are living on the $J_2$-fixed locus in the boundary $S^3$,
and quantum numbers carried by them are $J_1$ and
$SO(8)$ flavor charges.
$f_\psi$ is given by
\begin{align}
f_\psi=-\frac{q^{\frac{1}{2}}}{1-q}.
\end{align}

The index of the  theory $\wt D_4[\infty]$ in the large $N$ limit is
\begin{align}
I_{\wt D_4[\infty]}=\Pexp(P_{O7}[\wh f_{\rm sugra}]+f_{\rm vector}^{AdS_3\times S^5}\chi^{SO(8)}_{\bm{28}}) ,
\end{align}
where
\begin{align}
P_{\wt{O7}}[\wh f_{\rm sugra}]&=\frac{1}{2}[f_{\rm sugra}+\wt f_{\rm sugra}(q,-p,-x,-y,-z)]
\end{align}
is the contribution from the supergravity multiplet in the ten-dimensional bulk,
and the second term is the contribution from D7-D7 open strings.
$f_{\rm vector}^{AdS_3\times S^5}$ is the letter index of the eight-dimensional vector multiplet
in $AdS_3\times S^5$.
See \ref{vector35}.

\subsubsection{Decoupling}
In the triple-sum GG expansion
each contribution with a specific set of wrapping numbers $(m_x,m_y,m_z)$
takes the form (\ref{fmxmymz}).
The contribution from each cycle $i_I[m_I]$ is given by
\begin{align}
i_I[m_I]=P_{\wt{O7}}[\sigma_If_{\rm vec}\eta\wh\chi^{U(2m_I)_{\rm adj}}] .
\end{align}
Because $P_{\wt{O7}}$ is symmetric under the permutations among $x$, $y$, and $z$,
$X$-, $Y$-, and $Z$-expansions are essentially the same.
Let us consider $X$-expansion with the degrees $\deg(q,p,x,y,z)=(1,1,0,1,1)$.
We need to show the decoupling of the cycles $Y=0$ and $Z=0$.
The letter index associated with $Y=0$ is
\begin{align}
i_y[m_y]=\sum_\pm\left[1-\frac{(1-\eta y^{-1})(1-\eta q)(1-p)}{(1-x)(1-z)}\right]_\pm\chi_\mp^{(2m_y)} .
\end{align}
We can easily confirm that this contains infinitely many negative-degree terms if $m_y\geq 1$,
and hence the cycle $Y=0$ decouples.
It is the case for $Z=0$ cycle, too,
and the triple-sum expansion reduces to the simple-sum expansion
of the form
\begin{align}
\frac{I_{\wt D_4(m)}}{I_{\wt D_4(\infty)}}
=\sum_{N=0}^\infty x^{mN}\sigma_xI_{D_4[N]}.
\label{d4tildeexp}
\end{align}

\subsection{Numerical tests}
\subsubsection{$X$-expansion of $I_{D_4[N]}$}
\begin{figure}[htb]
\centering
\includegraphics[scale=0.5]{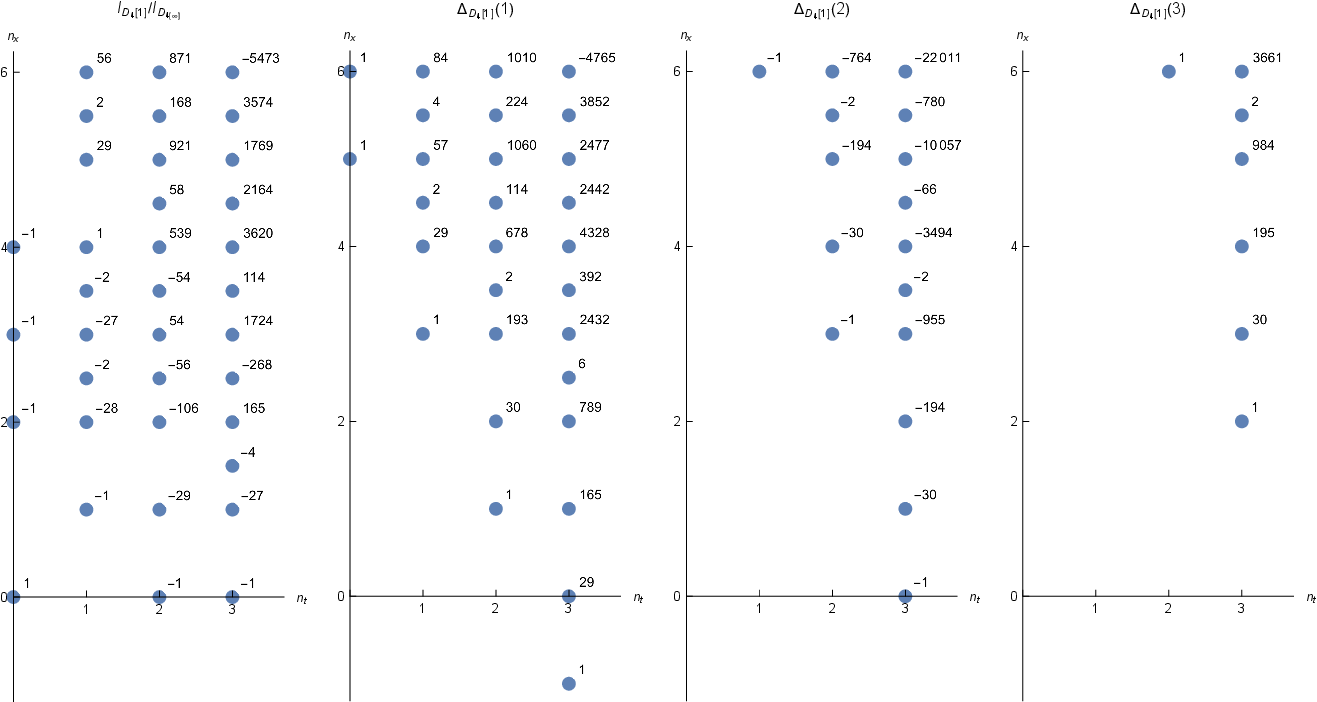}
\caption{The ratio $I_{D_4[1]}/I_{D_4[\infty]}$ and the errors $\Delta_{D_4[1]}(m_{\rm max})$
with $m_{\rm max}=1$, $2$, and $3$ are shown as two-dimensional plots.}
\label{ploto7}
\end{figure}

Let us numerically test the expansion (\ref{d7xexp}).
We introduce a cutoff $m_{\rm max}$ and define the error function
\begin{align}
\Delta_{D_4[N]}(m_{\rm max})=\frac{I_{D_4[N]}}{I_{D_4[\infty]}}-\sum_{m=0}^{m_{\rm max}} x^{mN}\sigma_x I_{\wt D_4[m]}.
\end{align}
We calculate the ratio
$I_{D_4[1]}/I_{D_4[\infty]}$
and the error $\Delta_{D_4[1]}(m_{\rm max})$
for the unrefined fugacities (\ref{param1})
and the results with $m_{\rm max}=1$, $2$, and $3$ are shown in Figure \ref{ploto7}.

\subsubsection{$X$-expansion of $I_{\wt D_4[m]}$}

\begin{figure}[htb]
\centering
\includegraphics[scale=0.5]{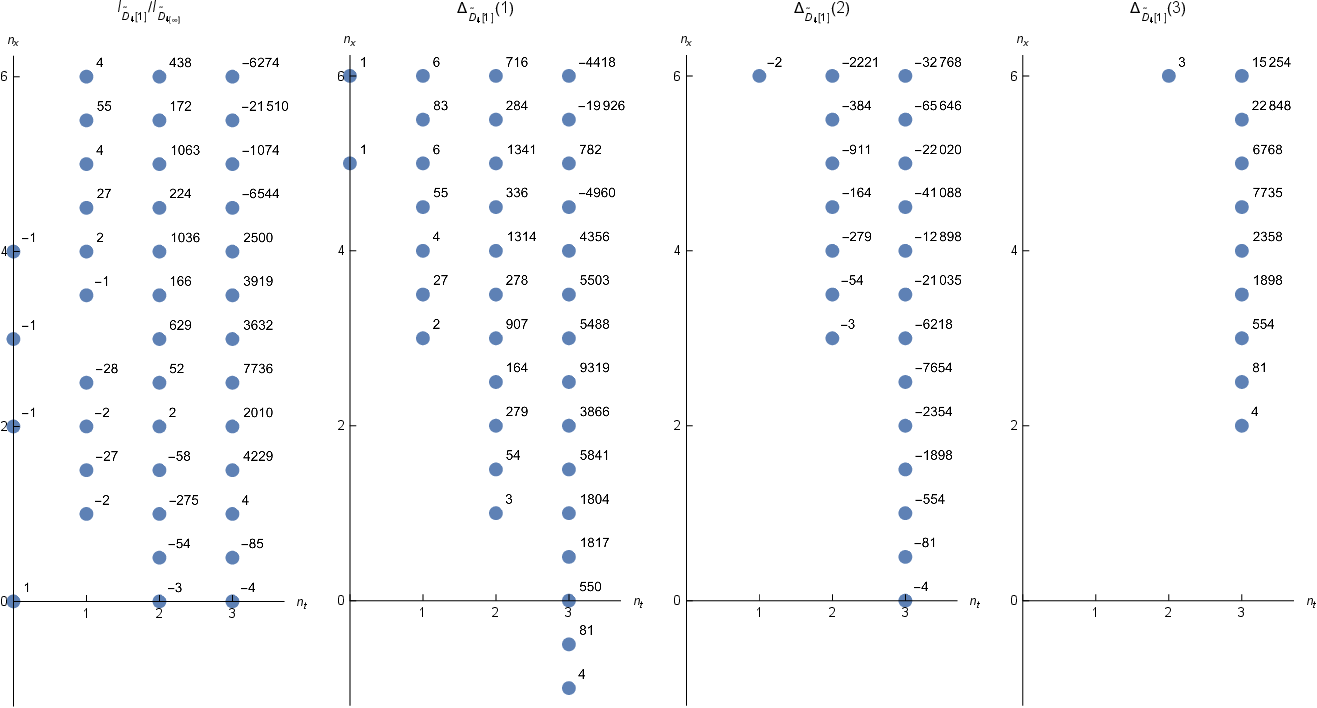}
\caption{The ratio $I_{\wt D_4(1)}/I_{\wt D_4(\infty)}$ and the errors $\Delta_{\wt D_4[1]}(N_{\rm max})$ with $N_{\rm max}=1$, $2$, and $3$ are shown.}\label{ploto7_dual}
\end{figure}

Let us test the expansion
of $I_{\wt D_4[m]}$ in (\ref{d4tildeexp}).
We introduce a cutoff $N_{\rm max}$ and
define the error function
\begin{align}
\Delta_{\wt D_4[m]}(N_{\rm max})
&=\frac{I_{\wt D_4(m)}}{I_{\wt D_4(\infty)}}
-\sum_{N=0}^{N_{\rm max}} x^{mN}\sigma_xI_{D_4[N]}.
\end{align}
The ratio
$I_{\wt D_4(1)}/I_{\wt D_4(\infty)}$ and the errors
$\Delta_{\wt D_4[1]}(N_{\rm max})$ with $N_{\rm max}=1$, 2 and 3 are calculated,
and the results are shown in Figure \ref{ploto7_dual}.

\subsubsection{$Z$-expansion of $I_{D_4[N]}$}
\begin{figure}[htb]
\centering
\includegraphics[width=0.9\textwidth]{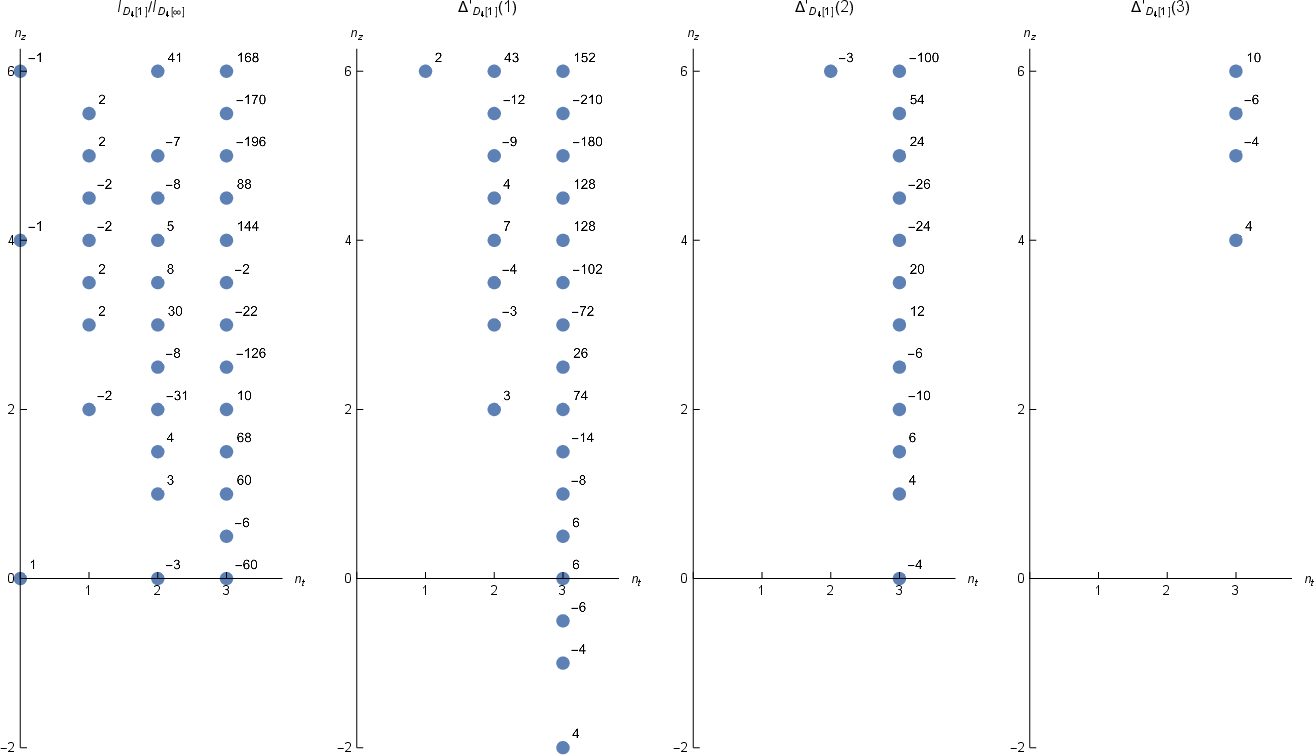}
\caption{The ratio $I_{D_4[1]}/I_{D_4[\infty]}$ and the errors $\Delta_{D_4[1]}'(m_{\rm max})$ 
with $m_{\rm max}=1$, $2$, and $3$ are shown.}
\label{ploto7_z}
\end{figure}

Let us numerically test the $Z$-expansion of $I_{D_4[N]}$ in (\ref{d4zexp}).
We introduce a cutoff $m_{\rm max}$ and define the error function
\begin{align}
\Delta_{D_4[N]}'(m_{\rm max})
&=\frac{I_{D_4[N]}}{I_{D_4[\infty]}}
-\sum_{m=0}^{m_{\rm max}} z^{2mN}\sigma_z I_{D_4[m]}.
\end{align}
The ratio
$I_{D_4[N]}/I_{D_4[\infty]}$ and the errors
$\Delta_{D_4[N]}'(m_{\rm max})$ are calculated for unrefined
fugacities
\begin{align}
(q,p,x,y,z)=(tz^{\frac{1}{2}},tz^{\frac{1}{2}},t,t,z),
\end{align}
and the results are shown in Figure \ref{ploto7_z}.

\section{Toric quiver gauge theories}\label{toric.sec}
\subsection{General rules}
Let us consider a toric quiver gauge theory associated with
a toric Calabi-Yau cone.
There is a systematic prescription to determine the gauge theory from the toric data
of the Calabi-Yau \cite{Feng:2000mi,Feng:2002zw,Hanany:2005ve,Franco:2005rj,Franco:2005sm,Hanany:2005ss,Feng:2005gw,Franco:2006gc}.
Let $(x_I,y_I)$ ($I=1,\ldots,d$) be the lattice
points on the boundary of the corresponding toric diagram
labeled in the counter-clockwise order.
The gauge theory has $d$ $U(1)$ symmetries corresponding to the $d$ boundary points.
Let $R_I$ be the $U(1)$ generators normalized so that $R_I=\pm\frac{1}{2}$ for supercharges
and $R_I=0$ or $R_I=1$ for scalar component fields in chiral multiplets.
The action of $R_I$ on the toric fibers of $X$ is given by the vector $V_I=(x_I,y_I,1)$.
If $d>3$ $V_I$ are not linearly independent, and $d-3$ linear combinations of $V_I$ vanish.
Let $B_a$ ($a=1,\ldots,d-3$) be the corresponding linear combinations of $R_I$.
$B_a$ generate non-geometric symmetries,
which are often called baryonic symmetries.

The superconformal index of the $\N=1$ toric quiver gauge theory is defined by
\begin{align}
I=\tr\left[(-1)^Fq^{J_1}p^{J_2}\prod_{I=1}^dv_I^{R_I}\right],
\quad qp=\prod_{I=1}^dv_I^{R_I}.
\end{align}

We are interested in the spectrum of eigenvalues of $R_I$.
Let us denote the $R_I$ charge of an operator ${\cal O}$ by $R_I[{\cal O}]$.
For a BPS operator ${\cal O}$ contributing to the superconformal index
$R_I[{\cal O}]$ are expressed as a lattice point in the $d$-dimensional lattice,
and the Taylor expansion of the index can be regarded as
the formal sum of weighted lattice points.
The charges of an operator ${\cal O}$ is specified by giving the corresponding fugacity $v_{\cal O}$,
and we also use the notation $R_I[v_{\cal O}]=R_I[{\cal O}]$.
For example, the elementary fugacities $v_I$ satisfy $R_I[v_J]=\delta_{IJ}$,
and we can regard them as the dual basis of the R-charges $R_I$.

Supersymmetric three-cycles on which D3-brane can wrap and which contribute to the multiple-sum giant graviton expansion
can be defined for each $I$ as the fixed locus of $R_I$,
and we denote them by $S_I$.
If a boundary point $I$ of the toric diagram is an internal point of a side,
the fixed locus $S_I$ is a one-dimensional circle, which can be interpreted as a shrinking three-cycle,
and the cycle corresponding to a corner point has finite size.
For this reason the distinction between the corner points and the others is important.
We define the set of all boundary points $\II=\{1,\ldots,d\}$ and the subset $\II_C\subset\II$
corresponding to the corner points. The complement of $\II_C$ in $\II$ is denoted by $\II_C^*=\II\bs\II_C$.
As we will see below, wrapping numbers $m_I$ for $I\in\II_C^*$
are related to holonomies on branes wrapped around finite-size three-cycles.

For a toric diagram with perimeter $d$,
the giant graviton expansion is $d$-ple sum:
\begin{align}
\frac{I_N}{I_\infty}=\sum_{m_\II}v_\II^{m_\II}F_{m_\II} ,
\label{dplesum}
\end{align}
where we used short-hand notation $m_\II=\{m_1,\ldots,m_d\}$,
$v_\II^{m_\II}=\prod_{I\in\II}v_I^{m_I}$, and $F_{m_\II}=F_{m_1,\ldots,m_d}$.
However, because the sum over $m_{\II_C^*}$ is the holonomy sum,
it is natural to divide $m_\II$ into the genuine wrapping numbers $m_{\II_C}$ and
holonomy variables $m_{\II_C^*}$,
and
we rewrite (\ref{dplesum}) as
\begin{align}
\frac{I_N}{I_\infty}=\sum_{m_{\II_C}}v_{\II_C}^{m_{\II_C}}
\left(\sum_{m_{\II_C^*}}v_{\II_C^*}^{m_{\II_C^*}}
F_{m_\II}\right).
\label{nplesum}
\end{align}
We can regard the sum in the parentheses as the index of the theory on the
GG system with wrapping numbers $m_{\II_C}$.
In this sense the GG expansion is $|\II_C|$-ple sum,
and the GG expansion of the $\ZZ_k$ orbifold theory studied in Section \ref{orb.sec}
is triple-sum for generic degree assignment in this sense.
We want to reduce the sum by an appropriate choice of the degrees.

The index $F_{m_\II}$ is given by
\begin{align}
F_{m_\II}
=\int dg
\Pexp\left(
\sum_{I\in\II_C}P_I[f_I\chi_{\rm adj}^{U(m_I)}]+\sum_{I,I'}f_{II'}\chi^{(m_I,m_{I'})}
\right).
\label{genletter}
\end{align}
$\chi^{(m_I,m_{I'})}$ are the bi-fundamental characters in (\ref{chbifund}).
$P_I$ and $f_I$ in the first term of the letter index
are the projection and the letter index
associated with the finite-size cycle $S_I$.
Note that $P_I$ is the projection operator acting on both $f_I$ and $\chi_{\rm adj}^{U(m_I)}$,
and the action on $\chi_{\rm adj}^{U(m_I)}$ depends on the holonomy variables $m_{\II_C^*}$.
$\sum_{I,I'}$ in the second term is the summation over pairs of adjacent corners $(I,I')$,
and $f_{I,I'}$ is the letter index associated with the intersection $S_I\cap S_{I'}$.
In the second term we should also take account of the holonomy dependence,
which is omitted in the following for simplicity.

$f_I$ and $f_{I,I'}$ can be obtained
according to the toric structure encoded in the toric diagram as follows.

As we mentioned above, the Taylor expansion of the index can be expressed as the formal sum of a set of weighted lattice points
in the $d$-dimensional lattice.
In the large $N$ limit, operators with baryonic charges do not contribute, and $B_a[u]=0$ are satisfied
for all terms $u$ in the Taylor expansion of the index.
This define the $3$-dimensional sublattice in the $d$-dimensional lattice.
In addition, BPS bounds guarantee $R_\II[u]\geq0$. (Namely, $R_I[u]\geq0$ for all $I\in\II$.)
These inequalities define a cone in the three-dimensional lattice.

The large $N$ index $I_\infty$ is given by \cite{Eager:2012hx,Agarwal:2013pba}\footnote{This gives the large $N$ index
of quiver gauge theory with $SU$ gauge groups.
This is different from (\ref{largenorb1}) by the contribution from the IR-free $U(1)$ vector multiplets included in (\ref{largenorb1}).}
\begin{align}
I_\infty=\sum_{I=1}^d\frac{w_I}{1-w_I} ,
\label{toriclargen}
\end{align}
where $w_I$ for a specific $I$ is the primitive fugacity satisfying
\begin{align}
B_a[w_I]=R_I[w_I]=R_{I+1}[w_I]=0,\quad
R_\II[w_I]\geq0.
\end{align}
The primitive fugacity is the one such that all fugacities satisfying the same conditions
are given as positive powers of it.

For each corner point $I$ let $(I_{-1},I_0=I,I_{+1})$ be the three consecutive corner points.
We define the three non-baryonic fugacities $w_{I,\alpha}$ ($\alpha=0,\pm1$) for each $I$ as the dual basis of $R_{I_\alpha}$
by the conditions
\begin{align}
B_a[w_{I,\alpha}]=0,\quad
R_{I_\alpha}[w_{I,\beta}]=\delta_{\alpha\beta}\quad
(\alpha,\beta=0,\pm1).
\label{wiadef}
\end{align}
We can show $w_{I,-1}w_{I,0}w_{I,+1}=\prod_{I=1}^dv_I=qp$.

In the case of the orbifold $S^5/\Gamma$ the toric diagram is a triangle,
and there are three corner points corresponding to $R_x$, $R_y$, and $R_z$.
If $R_I=R_x$, $R_{I_{+1}}=R_y$, and $R_{I_{-1}}=R_z$,
the corresponding fugacities are $w_{I,0}=x$, $w_{I,+1}=y$, and $w_{I,-1}=z$.
In fact, the letter index $f_I$ for a giant graviton
wrapped around a finite-size cycle $S_I$ is obtained simply by
replacing $x$, $y$, and $z$ in $\sigma_xf_{\rm vec}$
by $w_{I,0}$, $w_{I,+1}$, and $w_{I,-1}$, respectively.
\begin{align}
f_I=(\sigma_xf_{\rm vec})|_{x\rightarrow w_{I,0},y\rightarrow w_{I,+1},z\rightarrow w_{I,-1}}
=1-\frac{(1-w_{I,0}^{-1})(1-q)(1-p)}{(1-w_{I,+1})(1-w_{I,-1})}.
\label{sigmaif}
\end{align}
If the lattice generated by the three vectors $V_{I_\alpha}$ is not the whole three-dimensional lattice
but its sub-lattice, then some of $R_J[w_{I,\alpha}]$ are fractional.
Such terms must be removed, and this is realized by the projection $P_I$ in
(\ref{genletter}).

The letter index of bi-fundamental fields
can also be written with these fugacities.
Let $I$ and $I'$ ($=I_{+1}$) be two consecutive corners.
We introduce fugacities $w_{I,\alpha}$ and $w_{I',\alpha}$
according to (\ref{wiadef}).
If one or both of the intersecting cycles $S_I$ and $S_{I'}$
is orbifolded, we need to take account of the coupling
with holonomies.
Let us consider the simple case with neither $S_I$ nor $S_{I'}$
being orbifolded.
This is the case when both $V_{I_\alpha}$ and $V_{I'_\alpha}$
span the whole 3d lattice.
Then, fugacities $w_{I,-1}$ and $w_{I',+1}$ are the same.
Let us denote them by $w$.
This is the fugacity associated with the $U(1)$ symmetry
shifting the intersection.
The letter index of the bi-fundamental fields
along the intersection $S_I\cap S_{I'}$ is obtained
from (\ref{hypers5}) by replacing $z$ with $w$.
\begin{align}
f_{II'}=\left(\frac{w}{qp}\right)^{\frac{1}{2}}
\frac{(1-q)(1-p)}{1-w}.
\label{fiip}
\end{align}

Now, let us consider degree assignment.
We assign $d_q$, $d_p$, and $d_I$ to
$q$, $p$, and $v_I$, respectively.
This means that we introduce auxiliary variable $t$ by
\begin{align}
(q,p,v_I)\rightarrow (t^{d_q}q,t^{d_p}p,t^{d_I}v_I)
\quad
d_q+d_p=\sum_{I=1}^dd_I,
\end{align}
and perform the $t$-expansion first.
$t$ is the fugacity for the charge
\begin{align}
Q_t\equiv d_qJ_1+d_pJ_2+R_t,\quad
R_t\equiv \sum_{I=1}^dd_IR_I ,
\end{align}
and the degree of a fugacity $u$ is
nothing but $\deg(u)=Q_t[u]$.
$R_t$ is a linear combination of $R_I$.
Because we are not interested in the baryonic charges,
it is enough to know its action on the toric fibers,
which is expressed as the linear combination of $V_I$.
With the normalization $\sum_{I=1}^dd_I=1$, it can be expressed
as a point $D=\sum_{I=1}^dd_I(x_I,y_I)$
in the toric diagram, and if the position of
$D$ is specified, degrees of non-baryonic fugacities are fixed.
In particular, the degrees of the fugacities $w_{I,\alpha}$
are determined by the relative positions of $D$ and three corners $I_{\alpha}$.
If $D$ is inside the triangle made by the three corners $I_\alpha$, $\deg(w_{I,\alpha})>0$,
while if $D$ and a corner $I_\alpha$ are on the opposite sides of the line passing through the other two corners,
then $\deg(w_{I,\alpha})<0$.
Remark that $D$ is always in the $|\II_C|$-gon of the toric diagram (including its boundary), and
$\deg(w_{I,\pm})$ are always
non-negative while $\deg(w_{I,0})$ may be negative.

Let us apply the decoupling criterion.
For decoupling of a finite-size cycle $S_I$,
the letter index $f_I$ with the projection $P_I$ applied
needs to include infinitely many terms with $Q_t<0$.
Because $J_1$ and $J_2$ (the exponents of $q$ and $p$) are non-negative for all terms in
the expansion of (\ref{sigmaif}), there must be infinitely many terms with
$R_t<0$.
This can be true if the following conditions hold:
\begin{align}
\deg(w_{I,0})>0
\quad\text{and}\quad
(\deg(w_{I,+1})=0
\quad\text{or}\quad
\deg(w_{I,-1})=0).
\end{align}
This means
\begin{align}
D\in
(I_{-1},I_0]\cup[I_0,I_{+1}),
\label{dinsegs}
\end{align}
where 
$(I_{-1},I_0]$ is the segment between $I_{-1}$ and $I_0$ with
$I_{-1}$ excluded and $I_0$ included.
$[I_0,I_{+1})$ is similarly defined.
Note that (\ref{dinsegs}) is a necessary condition,
and to show the decoupling of a cycle $S_I$ we need to confirm that infinitely many negative-degree terms remain
after the projection $P_I$.

Obviously, the condition (\ref{dinsegs}) is satisfied at most for two corners $I$,
and the $|\II_C|$-ple sum giant graviton expansion at best reduces to ($|\II_C|-2$)-ple sum expansion.
For an orbifold $S^5/\Gamma$, the toric diagram is a triangle with $|\II_C|=3$,
and as we discussed in Section \ref{orb.sec} it may be possible to obtain simple-sum expansion.
Unfortunately, this is not possible for the case with $|\II_C|\geq4$.

\subsection{Klebanov-Witten theory}
\begin{figure}[htb]
\centering
\includegraphics{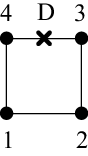}
\caption{The toric diagram of the conifold.}\label{t11toric.eps}
\end{figure}

As a simplest example of non-orbifold toric Calabi-Yau,
let us consider the conifold.
The toric diagram is shown in Figure \ref{t11toric.eps}.
It is a square, and the four corners are labeled by $I=1,2,3,4$.

There are a set of prescriptions to read off information
of the corresponding quiver gauge theory from the toric diagram \cite{Hanany:2005ve,Franco:2005rj}.
It is convenient to use the bipartite graph associated with the toric diagram.
The graph is drawn on the torus, and is called the brane tiling.
In the brane tiling, faces, edges, and vertices correspond
to gauge groups, bi-fundamental chiral multiplets, and terms in the superpotential, respectively.

\begin{figure}[htb]
\centering
\includegraphics{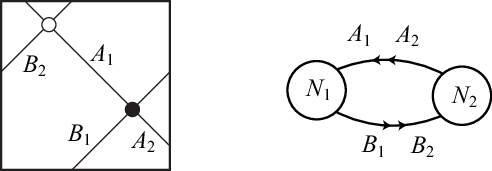}
\caption{The brane tiling and the quiver diagram of the Klebanov-Witten theory.}\label{t11tiling.eps}
\end{figure}

The brane tiling and the quiver diagram for the conifold are shown in
Figure \ref{t11tiling.eps}.
The quiver gauge theory is called the Klebanov-Witten theory \cite{Klebanov:1998hh}, which we denote by ${\rm KW}[N]$.

\begin{figure}[htb]
\centering
\includegraphics{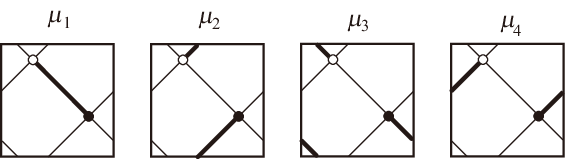}
\caption{The perfect matchings in the brane tiling for the conifold.}\label{conipm.eps}
\end{figure}

We can also read off charge assignment for $R_I$ from the brane tiling by using perfect matchings.
All perfect matchings in the tiling is
shown in Figure \ref{conipm.eps}.
Each perfect matching is associated with an internal or boundary lattice point of the toric diagram.
Let $\mu_I$ be a perfect matching associated with a boundary point $I$.
$\mu_I$ is a subset of edges, and defines a subset of
the chiral multiplets.
We assign $R_I=+1$ to them, and $R_I=0$ to the others.
There is the unique perfect matching for each $I\in\II_C$,
and charge assignment for $R_I$ ($I=1,2,3,4$) is uniquely determined as shown in Table \ref{kwfields}.
\begin{table}[htb]
\caption{Charge assignment for the chiral multiplets in the Klebanov-Witten theory are shown.
$\chi_\Phi$ is the contribution of the scalar component of the chiral multiplet to the letter index.
$\chi_{12}$ and $\chi_{21}$ are the characters for the bi-fundamental representations $(N_1,\ol N_2)$ and $(N_2,\ol N_1)$, respectively.}
\label{kwfields}
\centering
\begin{tabular}{cccccccc}
\hline
\hline
$\Phi$& $F_A$ & $F_B$ & $R_1$& $R_2$ & $R_3$ & $R_4$ & $\chi_\Phi$ \\
\hline
$A_1$ & $1$   & $0$   & $1$  & $0$   & $0$   & $0$   & $v_1\chi_{12}$ \\
$A_2$ & $-1$  & $0$   & $0$  & $0$   & $1$   & $0$   & $v_3\chi_{12}$ \\
$B_1$ & $0$   & $1$   & $0$  & $1$   & $0$   & $0$   & $v_2\chi_{21}$ \\
$B_2$ & $0$   & $-1$  & $0$  & $0$   & $0$   & $1$   & $v_4\chi_{21}$ \\
\hline
\end{tabular}
\end{table}

The global symmetry of the theory is
\begin{align}
G_{{\rm KW}_N}=U(1)_r\times SU(2)_A\times SU(2)_B\times U(1)_B.
\end{align}
and the $U(1)_r$ charge $r$, the $SU(2)_A$ Cartan generator $F_A$,
and $SU(2)_B$ Cartan generators $F_B$
are given by
\begin{align}
r=\frac{1}{2}(R_1+R_2+R_3+R_4),\quad
F_A=R_1-R_3,\quad
F_B=R_2-R_4,\quad
\end{align}
$U(1)_B$ is the baryonic symmetry.
Corresponding to the linear relation of four vectors $V_1-V_2+V_3-V_4=0$,
the $U(1)_B$ charge $B$ is given by
\begin{align}
NB=R_1-R_2+R_3-R_4.
\end{align}
The normalization of $B$ is chosen so that $\det A_1$ carries $B=+1$.

The degrees for the point $D$ shown in Figure \ref{t11toric.eps} are
\begin{align}
\deg(v_1,v_2,v_3,v_4)=(0,0,\tfrac{1}{2},\tfrac{1}{2})
\label{degreekw}
\end{align}
up to the ambiguity for baryonic charge.
With these degrees
the cycles $S_3$ and $S_4$ decouple,
and the expansion becomes double-sum
associated with $S_1$ and $S_2$:
\begin{align}
\frac{I_{{\rm KW}[N]}}{I_{{\rm KW}[\infty]}}
=\sum_{m_1,m_2}v_1^{m_1N}v_2^{m_2N}F_{m_1,m_2,0,0} .
\label{ggkw}
\end{align}

The fugacities $w_{I,\alpha}$ are given by
\begin{align}
w_{I,0}=v_Iv_{I+2}^{-1},\quad
w_{I,\pm1}=v_{I\pm1}v_{I+2}.
\end{align}
The letter index $f_I$ for a giant graviton wrapped on $S_I$ is
\begin{align}
f_I=1-\frac{(1-v_{I+2}/v_I)(1-q)(1-p)}{(1-v_{I+1}v_{I+2})(1-v_{I-1}v_{I+2})},
\label{t11adj}
\end{align}
and the letter index for the intersection modes on $S_I\cap S_{I+1}$ is
\begin{align}
f_{I,I+1}=\left(\frac{v_{I+2}v_{I+3}}{qp}\right)^{\frac{1}{2}}\frac{(1-q)(1-p)}{1-v_{I+2}v_{I+3}}.
\label{t11hyp}
\end{align}

\subsection{$\ZZ_2$ orbifold}
\begin{figure}[htb]
\centering
\includegraphics{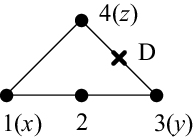}
\caption{The toric diagram of $\CC^3/\ZZ_2$.
There are four boundary points $\II=\{1,2,3,4\}$,
and three of them, $\II_C=\{1,3,4\}$, are corners.
$D$ is the point that specifies the degrees used for the $X$-expansion.}\label{z2toric.eps}
\end{figure}

\begin{figure}[htb]
\centering
\includegraphics{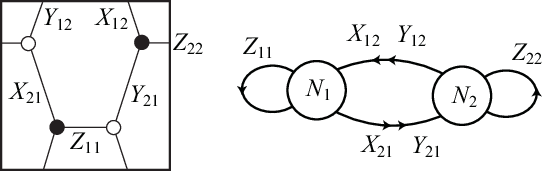}
\caption{The brane tiling and the quiver diagram of the $\ZZ_2$ orbifold theory  are shown.}\label{s5z2tiling.eps}
\end{figure}

Let us re-consider the $\ZZ_2$ orbifold theory $T_{N,N}$ discussed in \ref{zkbdr.sec} as
an example of toric quiver gauge theories.
The toric diagram is shown in Figure \ref{z2toric.eps}.
Corresponding to the three corners $\II_C=\{1,3,4\}$,
there are three finite-size cycles $S_1$, $S_3$, and $S_4$,
which are respectively referred to as $X=0$, $Y=0$, and $Z=0$ cycles in Section \ref{orb.sec},
and $R_I$ ($I\in\II_C$) are related to the $R$-charges used in Section \ref{orb.sec} by
\begin{align}
R_x=R_1,\quad
R_y=R_3,\quad
R_z=R_4.
\end{align}
The brane tiling and the quiver diagram for the $\ZZ_2$ orbifold theory are shown in
Figure \ref{s5z2tiling.eps}.
In terms of ${\cal N}=1$ multiplets,
$T_{N,N}$ consists of
six chiral multiplets shown in Table \ref{z2chiral.tbl}
and two $U(N)$ vector multiplets $V_{11}$ and $V_{22}$.

\begin{figure}[htb]
\centering
\includegraphics{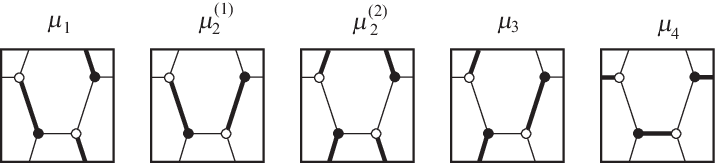}
\caption{Perfect matchings in the brane tiling of the $\ZZ_2$ orbifold theory are shown.}\label{s5z2matchings.eps}
\end{figure}

As in the previous example we can determine charge assignment of $R_I$ by using perfect matchings.
All perfect matchings for the $\ZZ_2$ theory are shown in Figure \ref{s5z2matchings.eps}.
Important difference from the previous example is that we have the vertex $I=2$ which
is not a corner.
In general, there are more than one perfect matchings associated with a point $I\in\II_C^*$,
and we should choose one of them to define the corresponding charge $R_I$.
This ambiguity affects the definition of the baryonic charges.
In the case of the $\ZZ_2$ orbifold theory $T_{N,N}$,
we have two perfect matchings $\mu_2^{(1)}$ and $\mu_2^{(2)}$ for
vertex $I=2$, and here we choose $\mu_2^{(1)}$.
The resulting charges are shown in Table \ref{z2chiral.tbl}.
\begin{table}[htb]
\caption{Chiral multiplets in the $\ZZ_2$ orbifold theory $T_{N,N}$ and their charges.}\label{z2chiral.tbl}
\centering
\begin{tabular}{ccccccc}
\hline
\hline
$\Phi$ & $R_x=R_1$ & $R_y=R_3$ & $R_z=R_4$ & $NB$ & $R_2$ & $\chi_\Phi$ \\
\hline
$X_{12}$ & $1$ & $0$ & $0$ & $1$  & $0$ & $xb^{\frac{1}{N}}\chi_{12}=v_1\chi_{12}$ \\
$X_{21}$ & $1$ & $0$ & $0$ & $-1$ & $1$ & $xb^{-\frac{1}{N}}\chi_{21}=v_1v_2\chi_{21}$ \\
$Y_{12}$ & $0$ & $1$ & $0$ & $1$  & $0$ & $yb^{\frac{1}{N}}\chi_{12}=v_3\chi_{12}$ \\
$Y_{21}$ & $0$ & $1$ & $0$ & $-1$ & $1$ & $yb^{-\frac{1}{N}}\chi_{21}=v_3v_2\chi_{21}$ \\
$Z_{11}$ & $0$ & $0$ & $1$ & $0$  & $0$ & $z\chi_{11}=v_4\chi_{11}$ \\
$Z_{22}$ & $0$ & $0$ & $1$ & $0$  & $0$ & $z\chi_{22}=v_4\chi_{22}$ \\
\hline
\end{tabular}
\end{table}

The global symmetry of the $T_{N,N}$ theory is
\begin{align}
G_{T_{N,N}}=SU(2)_R\times U(1)_R\times SU(2)_F\times U(1)_B ,
\end{align}
where $SU(2)_R\times U(1)_R$ is the $R$-symmetry of the ${\cal N}=2$ superconformal algebra,
$SU(2)_F$ is the flavor symmetry, and $U(1)_B$ is the baryonic symmetry.
$SU(2)_R\times U(1)_R\times SU(2)_F$ is the geometric symmetry in the sense
that it is realized as the isometry of the background geometry.
The $SU(2)_R$ Cartan generator $R_{SU(2)}$, the $U(1)_R$ charge $R_{U(1)}$,
and the $SU(2)_F$ Cartan generator $F$ are given by
\begin{align}
R_{SU(2)}=R_1+R_3,\quad
R_{U(1)}=R_4,\quad
F=R_1-R_3.
\end{align}
The baryonic charge is determined from the linear dependence $V_1-2V_2+V_3=0$ as
\begin{align}
NB=R_1-2R_2+R_3.
\label{nbdef}
\end{align}
The normalization of $B$ is chosen so that $\det X_{12}$ carries $B=+1$.

The cross in the toric diagram in Figure \ref{z2toric.eps}
shows the point $D$ that gives the degrees
\begin{align}
\deg(v_1,v_2,v_3,v_4)=(0,0,\tfrac{1}{2},\tfrac{1}{2})
\label{orbdegree}
\end{align}
(up to ambiguity for the baryonic charge).
These are the same as the degrees for the $X$-expansion used in Section \ref{orb.sec} up to
normalization.
According to the rule (\ref{dinsegs})
two cycles $S_3$ and $S_4$ decouple,
and the giant graviton expansion becomes the double sum
\begin{align}
\frac{I_{T_{N,N}}}{I_{T_\infty}}=\sum_{m_1=0}^\infty\sum_{m_2}v_1^{m_1}v_2^{m_2}F_{m_1,m_2,0,0}.
\label{tnndoubleexp}
\end{align}
$m_1$ is the wrapping number for the finite-size cycle $S_1$, and runs over all non-negative integers.
$m_2$, the wrapping number for the shrining cycle $S_2$, can be regarded as the holonomy variable,
as we will explain below.

In the analysis in Section \ref{orb.sec} we consider the
index $I_{T_{N_1,\ldots,N_k}}^{[B_1,\ldots,B_k]}$ for sectors with specific baryonic charges.
For the comparison to
(\ref{tnndoubleexp})
we should introduce the baryonic fugacities and define the index as the sum over all baryonic sectors.
For $\ZZ_2$ orbifold theory $T_{N,N}$ we define the baryonic charge $B=B_1-B_2$ and introduce a
single fugacity $b$ for $B$.
Then, the index including all baryonic sectors is
\begin{align}
I_{T_{N,N}}
=\sum_{B=-\infty}^\infty b^BI_{T_{N,N}}^{[B,0]}.
\end{align}
By substituting (\ref{ggexpz20}) into $I_{T_{N,N}}^{[B,0]}$ we obtain
\begin{align}
\frac{I_{T_{N,N}}}{I_{T_\infty}}
=\sum_{m',m''=0}^\infty
b^{m'-m''}
x^{(m'+m'')N}\sigma_xI^0_{\wt T_{m',m''}}.
\label{itnnbggx}
\end{align}

Let us confirm that (\ref{tnndoubleexp}) and (\ref{itnnbggx}) are the same.
The relation between two sets of fugacities $(x,y,z,b)$ and $(v_1,v_2,v_3,v_4)$ can be
read off from $x^{R_x}y^{R_y}z^{R_z}b^B=v_1^{R_1}v_2^{R_2}v_3^{R_3}v_4^{R_4}$ as follows
\begin{align}
v_1=xb^{\frac{1}{N}},\quad
v_2=b^{-\frac{2}{N}},\quad
v_3=yb^{\frac{1}{N}},\quad
v_4=z.
\label{s5z2fugs}
\end{align}
With these relations
we can rewrite (\ref{itnnbggx}) as follows.
\begin{align}
\frac{I_{T_{N,N}}}{I_{T_\infty}}
=\sum_{m',m''=0}^\infty
v_1^{(m'+m'')N}
v_2^{m''N}
\sigma_xI^0_{\wt T_{m',m''}}.
\label{itnnbggx2}
\end{align}
By comparing the prefactors in (\ref{tnndoubleexp}) and (\ref{itnnbggx2}) we obtain
\begin{align}
m'=m_1-m_2,\quad
m''=m_2.
\label{m1m2}
\end{align}
These relations imply that $m_2$ is the parameter specifying the symmetry breaking pattern
$U(m_1)\rightarrow U(m_1-m_2)\times U(m_2)$.
In other words, $m_2$ is the parameter specifying the holonomy
\begin{align}
\{\wt h_a\}=\{1^{m_1-m_2},2^{m_2}\}
\label{holom2}
\end{align}
on the $m_1$ coincident giant gravitons.
This interpretation require $0\leq m_2\leq m_1$.
The upper bound $m_2\leq m_1$
would be interpreted as a kind of $s$-rules.

We can also confirm that $F_{m_1,m_2,0,0}$ in (\ref{tnndoubleexp})
reproduces the result in Section \ref{orb.sec}.
Because the cycles $S_3$ and $S_4$ decouple
(\ref{genletter}) gives the letter index
\begin{align}
P_1f_1\chi_{\rm adj}^{U(m_1)}
\label{p1f1}
\end{align}
The fugacities $w_{1,\alpha}$ are given by
\begin{align}
w_{1,0}=v_1v_2^{\frac{1}{2}}=x,\quad
w_{1,+1}=v_3v_2^{\frac{1}{2}}=y,\quad
w_{1,-1}=v_4=z.
\end{align}
With these fugacities we can see
$f_1=\sigma_xf_{\rm vec}$
and (\ref{p1f1}) agrees with $i_x[m_x]$ in (\ref{impk})
with $m_x=m_1$ and the holonomy (\ref{holom2}).

\subsection{RG flow}
The orbifold theory $T_{N,N}$ flows to
the Klebanov-Witten theory ${\rm KW}[N]$
by the deformation with the superpotential
\begin{align}
W=m(\tr Z_{11}^2-\tr Z_{22}^2).
\label{massterms}
\end{align}
Because the RG flow does not change the superconformal index
we can relate $I_{T_{N,N}}$ and $I_{{\rm KW}[N]}$ by the RG flow.

The deformation breaks the global symmetry $G_{T_{N,N}}$ of the UV theory
to $G_{\rm flow}=U(1)_r\times SU(2)_F\times U(1)_B$,
and at the IR fixed point $G_{\rm flow}$ is enhanced to $G_{\rm KW}$.
Therefore, $I_{T_{N,N}}=I_{{\rm KW}[N]}$
holds only for restricted values of fugacities corresponding to $G_{\rm flow}$.
Let $r$, $f$, and $b$ be the fugacities for $U(1)_r$, $SU(2)_F$, and $U(1)_B$ in $G_{\rm flow}$, respectively.
The restriction on the fugacities of the UV theory is given by
\begin{align}
x=fr^{\frac{1}{2}},\quad
y=f^{-1}r^{\frac{1}{2}},\quad
z=r,\quad
b=b,\quad(r^2=qp) .
\end{align}
For the restricted values the fugacity for the mass terms (\ref{massterms}),
which carry $(R_1,R_2,R_3,R_4)=(-1,-1,-1,1)$, is
\begin{align}
\frac{v_4}{v_1v_2v_3}=\frac{z}{xy}=1.
\end{align}

Concerning the symmetry $G_{\rm KW}$,
only the diagonal subgroup of $SU(2)_A\times SU(2)_B$ is preserved in $G_{\rm flow}$.
The restricted fugacities $v_I'$ of the IR theory are
\begin{align}
v_1'=fr^{\frac{1}{2}}b^{\frac{1}{N}},\quad
v_2'=fr^{\frac{1}{2}}b^{-\frac{1}{N}},\quad
v_3'=f^{-1}r^{\frac{1}{2}}b^{\frac{1}{N}},\quad
v_4'=f^{-1}r^{\frac{1}{2}}b^{-\frac{1}{N}}.
\end{align}
(In this subsection we use the primed variables $v_I'$ for ${\rm KW}[N]$ for distinction
from $v_I$ for $T_{N,N}$.)

With the localization formula on the gauge theory side
we can easily check that the two indices agree for the restricted values of fugacities.
The letter index of the $\ZZ_2$ orbifold theory
is given by
\begin{align}
f_{\rm v}(\chi_{11})+f_{\rm v}(\chi_{22})
+\sum_\Phi f_{\rm c}(\chi_\Phi)
\label{quiverletter}
\end{align}
with $\chi_\Phi$ shown in Table \ref{z2chiral.tbl},
where
$f_{\rm v}$ and $f_{\rm c}$ are the letter index for the vector multiplet and the chiral multiplet:
\begin{align}
f_{\rm c}(\chi_R)=\frac{\chi_R-qp\ol\chi_R}{(1-q)(1-p)},\quad
f_{\rm v}(\chi_{\rm adj})=\left(1-\frac{1-qp}{(1-q)(1-p)}\right)\chi_{\rm adj}.
\end{align}
The letter index of the Klebanov-Witten theory is also given by
(\ref{quiverletter}) with $\chi_\Phi$ shown in Table \ref{kwfields}.
Indeed, we can easily check
\begin{align}
f_{X_{12}}=f_{A_1},\quad
f_{Y_{12}}=f_{A_2},\quad
f_{X_{21}}=f_{B_1},\quad
f_{Y_{21}}=f_{B_2},\quad
f_{Z_{11}}=f_{Z_{22}}=0 ,
\end{align}
and the two indices agree at the level of the letter indices
in the localization formula.

Finally, let us confirm the GG expansions (\ref{itnnbggx2}) and (\ref{ggkw}) are the same.
The two sets of degrees
(\ref{orbdegree}) and (\ref{degreekw}) are both
consistent with the degrees
\begin{align}
\deg(f,r,b)=(-\tfrac{1}{2},1,0).
\end{align}
We can easily confirm that the prefactor
of the orbifold theory in (\ref{tnndoubleexp})
and the prefactor of the Klebanov-Witten theory in (\ref{ggkw})
agree
provided we identify $(m',m'')$ on the $T_{N,N}$ side
with $(m_1,m_2)$ on the ${\rm KW}[N]$ side:
\begin{align}
v_1^{(m'+m'')N}v_2^{m''N}=x^{(m'+m'')N}b^{m'-m''}=(fr^{\frac{1}{2}})^{(m'+m'')N}b^{m'-m''}=v_1'^{m'N}v_2'^{m''N}.
\end{align}
We can also confirm the agreement of the letter index for giant gravitons.
In the GG expansion of $T_{N,N}$
the letter index for $m_1$ coincident giant gravitons on the $X=0$ cycle
with the holonomy (\ref{holom2}) is
\begin{align}
i_x[m]
&=P_2[\sigma_x f_{\rm vec}\chi^{U(m)}_{\rm adj}]
\nonumber\\
&=\left(1-\frac{(1-x^{-1}y)(1-q)(1-p)}{(1-y^2)(1-z)}\right)(\chi_{\rm adj}^{U(m')}+\chi_{\rm adj}^{U(m'')})
\nonumber\\
&\quad +\frac{(x^{-1}-y)(1-q)(1-p)}{(1-y^2)(1-z)}\chi^{(m',m'')}.
\label{tnnix}
\end{align}
For the Klebanov-Witten theory, 
the letter index is given by
\begin{align}
f_1\chi^{U(m_1)}_{\rm adj}+f_2\chi^{U(m_2)}_{\rm adj}
+f_{12}\chi^{(m_1,m_2)} ,
\label{ixmkw}
\end{align}
where $f_1$ and $f_2$ are given in (\ref{t11adj})
and $f_{12}$ is given in (\ref{t11hyp}).
Two letter indices (\ref{tnnix}) and (\ref{ixmkw}) agree for the restricted values of fugacities.

\section{Discussion}\label{disc.sec}
In this paper we discussed the reduction of the multiple-sum giant graviton expansions
to the simple-sum expansions.
For orbifold and orientifold examples the triple-sum expansion for generic degrees
can be reduced to the simple-sum expansion by assigning appropriate degrees.

All theories studied in this paper are Lagrangian theories,
and we can calculate the superconformal index directly by using the localization formula.
This is in general not the case.
For example,
M2-brane theory (ABJM theory \cite{Aharony:2008ug}) and M5-brane theory (six-dimentional $(2,0)$ theory)
are related by the simple-sum giant graviton expansions \cite{Imamura:2022aua}.
In this case, the calculation of the index of $(2,0)$ theory
is only possible with indirect methods \cite{Beem:2015aoa,Lockhart:2012vp,Kim:2012qf,Kim:2013nva,Beem:2014kka}.
We can use the giant graviton expansion as another convenient method to calculate
the index of $(2,0)$ theory by using the ABJM index \cite{Kim:2009wb}.
The orbifold version of the relation between M2-theory and M5-theory is also interesting because
it may enable us to calculate 
the index of six-dimensional $(1,0)$ theories
by using orbifolds of ABJM theory.

Another interesting class of non-Lagrangian theories
includes Argyres Douglas and Minahan Nemeschanski theories
realized on D3-branes probing $7$-brane backgrounds with constant axiodilation.
They are labeled by the type of the $7$-brane $G=H_0,H_1,H_2,D_4,E_6,E_7,E_8$ and the rank $N$.
The AdS/CFT correspondence of these theories in the large $N$ limit was studied in \cite{Fayyazuddin:1998fb,Aharony:1998xz},
and it was confirmed in \cite{Imamura:2021dya}
that the multiple-sum expansion works well at least for the leading giant graviton contributions.
If we can apply the simple-sum expansion
to this class of theories it gives interesting relations
among the indices.
Actually, the $D_4[N]$ theory studied in Section \ref{o7.sec}
is a special case of the general theories $G[N]$.
We confirmed that the $Z$-expansion of the $D_4[N]$ theory works and is self-dual.
If this is also the case for general $G[N]$, the following relation should hold
\begin{align}
\frac{I_{G[N]}}{I_{G[\infty]}}
=\sum_{m=0}^\infty z^{\Delta_GmN}\sigma_zI_{G[m]},
\label{gnexpansion}
\end{align}
with $\Delta_G$ being the dimension of the Coulomb branch operator of
the $G[1]$ theory.
It would be interesting to confirm whether this relation holds,
and if so, to what extent we can bootstrap the index $I_{G[N]}$
with the relation (\ref{gnexpansion}).

In the orbifold and orientifold cases, we can consider three expansions:
$X$, $Y$, and $Z$-expansions.
Although not always the decoupling occurs, for some examples
we can perform the expansion in more than one ways,
and we can consider ``the web of GG expansions'',
by applying GG expansions repeatedly.
In the case of ${\cal N}=4$ $U(N)$ SYM
all the three expansions work.
In this case the giant graviton expansion is ``self-dual'' in the sense that
the theory on the giant graviton is also ${\cal N}=4$ $U(N)$ SYM,
and application of the giant graviton expansions gives just different frames.
Each step of giant graviton expansions is specified by one of the variable changes
$\sigma_x$, $\sigma_y$, and $\sigma_z$,
and a frame is specified by the composition of the variable changes.
Combining three variable changes $\sigma_x$, $\sigma_y$, $\sigma_z$ and
permutations in $(x,y,z)$ and $(q,p)$
associated with the Weyl groups of $SU(4)_R$ and $SO(4)_{\text{spin}}$ symmetries
we can generate many frames.
All these frames gives ${\cal N}=4$ $U(N)$ SYM.
However, in more general orbifolds and orientifolds,
it may be possible to generate the web including different theories.
By applying a projection $P$, we can obtain an orbifold theory,
and in a frame specified by the variable change $\sigma$ the projection is replaced by
$\sigma P\sigma^{-1}$, and in general this is different from the original one.
Therefore, the web consists of theories with different orbifold actions.
Although in the orbifold case all theories in the web are Lagrangian theories,
in more general cases (like $G[N]$ and orbifolds of M-brane theories) the web may contain
both Lagrangian and non-Lagrangian theories, and then it will be convenient to analyze
theories that are difficult to analyze directly.
This is analogous to the S-duality of type IIB string theory.
The duality takes the type IIB theory to the same theory (with the different coupling constant).
Let us take $\ZZ_2$ projection with the worldsheet parity $\Omega$, which gives type I theory.
Then the other side of the duality is also $\ZZ_2$ projected theory.
However, this $\ZZ_2$ is not $\Omega$ but $(-1)^{F_R}$, which gives the $SO(32)$ heterotic string.
In this way, by applying a projection, we can generate new duality from the duality for the
theories before the projection.
We can consider a similar situation for the web of giant graviton expansions.

As we have emphasized, the simple-sum expansion is much more easy to calculate
because it does not have issues of integration contours.
This is because the contribution from single cycle is obtained
by a simple variable change from the standard index, for which we can adopt
the standard choice of the contours.
In Section \ref{toric.sec} we saw that the giant graviton expansions of the orbifold theory $T_{N,N}$ and the Klebanov-Witten theory
${\rm KW}[N]$ are related by the RG flow.
In this relation, the simple sum expansion of $T_{N,N}$ generates
the double-sum expansion of ${\rm KW}[N]$.
This gives information about rules for pole selection in ${\rm KW}[N]$.
This kind of relation may be useful to find general rules for the pole selection
for general multiple-sum giant graviton expansions.

In all the examples in this work the simple-sum expansions are invertible.
Namely, if expansion of the index of $T_N$ gives the index of $\wt T_m$,
then the expansion of $\wt T_m$ gives $T_N$.
It would be interesting if we could analytically prove the invertibility.
Suppose that we have functions $F_N$ ($N\in\ZZ$) which vanishes for negative $N$,
and they have giant graviton expansion of the form
\begin{align}
F_N=F_\infty\sum_{m\in\ZZ}x^{mN}\sigma_x\wt F_m.
\label{fnexp}
\end{align}
The invertibility means that a similar expansion of $\wt F_m$ gives the
original functions $F_N$.
Namely, if the functions $F'_N$ appearing in the expansion of $\wt F_m$
\begin{align}
\wt F_m=\wt F_\infty\sum_{N\in\ZZ}x^{mN}\sigma_x F_N'
\label{wtfnexp}
\end{align}
are the same as $F_N$, then we can say the expansion is invertible.
Unfortunately, we have not yet succeeded in proving (or disproving) this fact.
We will only comment on what happen if we naively substitute (\ref{wtfnexp}) to (\ref{fnexp}).
It gives
\begin{align}
F_N=(F_\infty\sigma_x\wt F_\infty)\sum_{m\in\ZZ}\sum_{N'\in\ZZ}x^{m(N-N')}F_{N'}' .
\end{align}
Let us suppose $x$ is a generic phase factor $x=e^{2\pi i\alpha}$ with
irrational $\alpha$.
Then with an appropriate regularization we obtain
\begin{align}
\sum_{m\in\ZZ}x^{m(N-N')}={\cal N}\delta_{N,N'} ,
\end{align}
where ${\cal N}$ is a large number depending on the regularization.
Then we obtain
\begin{align}
F_N={\cal N}(F_\infty\sigma_x\wt F_\infty)F_{N}' .
\end{align}
If the product of $F_\infty$ and $\sigma_x\wt F_\infty$ gave
${\cal N}^{-1}$ we would obtain expected result $F_N=F_N'$.
Because the large $N$ index $F_\infty$ and $\wt F_\infty$
are given as the plethystic exponential of the letter indices
$f_\infty$ and $\wt f_\infty$ of
the massless fields in AdS,
this can be formally calculated as follows.
\begin{align}
F_\infty\sigma_x\wt F_\infty
=\Pexp(f_\infty+\sigma_x \wt f_\infty).
\label{fxf}
\end{align}
Interestingly, in all examples of pairs of theories
related by the $X$-expansion, we find the relation
\begin{align}
f_\infty+\sigma_x \wt f_\infty=-1
\end{align}
holds.
Therefore, naively, (\ref{fxf}) becomes $\Pexp(-1)=0$.
This is nice because we want to obtain ${\cal N}^{-1}$ as the result of the regularization.
Of course the above manipulation is so naive and formal that
it does not make sense as it is.
It would be nice if we can improve the derivation.

\section*{Acknowledgments}
The authors are  grateful to Keita Kuwabara for collaboration on the early stages of this work.
The work of Y.~I. and D.~Y. was
partially supported by Grand-in-Aid for Scientific Research (C) (No.21K03569),
Ministry of Education, Science and Culture, Japan.
S.~F. is supported by the South African Research Chairs Initiative of the Department of Science and Innovation
and the National Research Foundation grant 78554.
S.~M. is supported by JST, the establishment of university fellowships towards the creation of science technology innovation, Grant Number JPMJFS2112.

\appendix
\section{Characters and Haar measures}\label{haar.sec}
In general, the Haar measure is given by the plethystic exponential of the
negative of the adjoint character with constant terms removed.

For $G=U(N)$ the adjoint character is given by
\begin{align}
\chi^{U(N)}_{\rm adj}(\zeta_a)=\chi^{U(N)}_{\rm fund}\chi^{U(N)}_{\ol{\rm fund}}=\sum_{1\leq a,b\leq N}\frac{\zeta_a}{\zeta_b} ,
\label{unadj}
\end{align}
where $\chi^{U(N)}_{\rm fund}$ and $\chi^{U(N)}_{\ol{\rm fund}}$
are the characters of the fundamental and the anti-fundamental representations, respectively:
\begin{align}
\chi_{\rm fund}^{U(N)}=\sum_{a=1}^N\zeta_a,\quad
\chi_{\ol{\rm fund}}^{U(N)}=\sum_{a=1}^N\frac{1}{\zeta_a}.
\label{chifaf}
\end{align}
In the sum in (\ref{unadj}) each term with $a=b$ gives $1$, and in total the constant term in $\chi^{U(N)}_{\rm adj}$ is $N$.
The Haar measure is given by
\begin{align}
\int_{U(N)}d\mu
&=\frac{1}{N!}\int\prod_{a=1}^N\frac{d\zeta_a}{2\pi i\zeta_a}\Pexp(-\chi_{\rm adj}^{U(N)}(\zeta_a)+N)
\nonumber\\
&=\frac{1}{N!}\int\prod_{a=1}^N\frac{d\zeta_a}{2\pi i\zeta_a}\prod_{a\neq b}\left(1-\frac{\zeta_a}{\zeta_b}\right) .
\end{align}
The overall factor $1/N!$ can be fixed by the normalization condition $\int_Gd\mu=1$.

The characters for the symplectic and orthogonal groups are obtained from $U(N)$ characters
by appropriate restrictions and projections.
To obtain $Sp(N)$ and $SO(2N)$ characters we start from $U(2N)$, and impose
the following constraints on the $2N$ fugacities $\zeta_a$ ($a=1,\ldots,2N$):
\begin{align}
\zeta_{a+N}=\frac{1}{\zeta_a}\quad(a=1,\ldots,N).
\end{align}
Then the fundamental and the anti-fundamental characters in (\ref{chifaf}) (with $N$ replaced by $2N$)
become the same.
The adjoint characters for $Sp(N)$ and $SO(2N)$ are given by
taking the symmetric and anti-symmetric products of two copies of the fundamental character:
\begin{align}
\chi_{\rm adj}^{Sp(N)}
&=\sum_{1\leq a\leq b\leq 2N}\zeta_a\zeta_b
=\sum_{1\leq a\leq b\leq N}\left(\zeta_a\zeta_b+\frac{1}{\zeta_a\zeta_b}\right)
+\sum_{1\leq a,b\leq N}\frac{\zeta_a}{\zeta_b} ,\nonumber\\
\chi_{\rm adj}^{SO(2N)}
&=\sum_{1\leq a<b\leq 2N}\zeta_a\zeta_b
=\sum_{1\leq a< b\leq N}\left(\zeta_a\zeta_b+\frac{1}{\zeta_a\zeta_b}\right)
+\sum_{1\leq a,b\leq N}\frac{\zeta_a}{\zeta_b} .
\end{align}
Both these characters include constant term $N$, and the Haar measures are given by
\begin{align}
\int_{Sp(N)}d\mu
&=\frac{1}{N!2^N}\int\prod_{a=1}^N\frac{d\zeta_a}{2\pi i\zeta_a}\Pexp(-\chi_{\rm adj}^{Sp(N)}(\zeta_a)+N) ,
\nonumber\\
\int_{SO(2N)}d\mu
&=\frac{1}{N!2^{N-1}}\int\prod_{a=1}^N\frac{d\zeta_a}{2\pi i\zeta_a}\Pexp(-\chi_{\rm adj}^{SO(2N)}(\zeta_a)+N) .
\end{align}

The adjoint character and the Haar measure of $SO(2N+1)$ are obtained in a similar way
starting from $U(2N+1)$ and imposing the constraints
\begin{align}
\zeta_{a+N}=\frac{1}{\zeta_a}\quad(a=1,\ldots,N),\quad
\zeta_{2N+1}=1.
\end{align}
The adjoint character is
\begin{align}
\chi^{SO(2N+1)}_{\rm adj}=\sum_{1\leq a < b\leq 2N+1}\zeta_a\zeta_b
=\chi_{\rm adj}^{SO(2N)}+\sum_{a=1}^N\left(\zeta_a+\frac{1}{\zeta_a}\right) ,
\end{align}
and the normalized Haar measure is
\begin{align}
\int_{SO(2N+1)}d\mu
&=\frac{1}{N!2^N}\int\prod_{a=1}^N\frac{d\zeta_a}{2\pi i\zeta_a}\Pexp(-\chi_{\rm adj}^{SO(2N+1)}(\zeta_a)+N) .
\end{align}

The orthogonal group $O(2N)$ consists of two disconnected components:
the component $G_+=SO(2N)\subset O(2N)$
containing the identity and the other component which we denote by $G_-\subset O(2N)$.
We normalize the Haar measure for each component by $\int_{G_\pm}d\mu=1$,
and the normalized measure for $O(2N)$ is given by
\begin{align}
\int_{O(2N)}d\mu=\frac{1}{2}\int_{G_+}d\mu+\frac{1}{2}\int_{G_-}d\mu.
\end{align}
To obtain the adjoint character and the Haar measure for $G_-$ we
start from $U(2N)$ and impose the constraints
\begin{align}
\zeta_{a+N}=\frac{1}{\zeta}_a\quad(a=1,\ldots,N-1),\quad
\zeta_N=1,\quad
\zeta_{2N}=-1.
\end{align}
Remark that the value $\zeta_{2N}=-1$ should be set after the calculation of the plethystic exponential.
For example, $\Pexp \zeta_{2N}=1/(1-\zeta_{2N})=1/2$.
To avoid the confusion with fermionic terms with negative coefficients, we introduce an auxiliary variable $\theta$
which is set to be $-1$ after the calculation of the plethystic exponential.
The adjoint character is
\begin{align}
\chi^{G_-}_{\rm adj}
=\sum_{1\leq a<b\leq 2N}\zeta_a\zeta_b
=\chi^{SO(2N-1)}_{\rm adj}+\theta\left[\sum_{a=1}^N\left(\zeta_a+\frac{1}{\zeta_a}\right)+1\right] ,
\end{align}
and the normalized Haar measure is
\begin{align}
\int_{G_-}d\mu
&=\frac{1}{(N-1)!2^N}\int\prod_{a=1}^{N-1}\frac{d\zeta_a}{2\pi i\zeta_a}\left.
\left[\Pexp(N-1+\theta-\chi_{\rm adj}^{G_-}(\zeta_a))\right]\right|_{\theta=-1} .
\end{align}

\section{Contributions from fixed loci}
\subsection{Tensor multiplet on $AdS_5\times S^1$}\label{tensor51}
The fixed locus of the $\ZZ_k$ generator (\ref{s5zk})
\begin{align}
U_k=\exp\left(\frac{2\pi i}{k}(R_x-R_y)\right)
\end{align}
is $AdS_5\times S^1$, and tensor multiplets live on it.
All the component fields are $U_k$ neutral,
that is, $R_x-R_y=0$ for all fields.
Let us express the other quantum numbers by the notation
\begin{align}
[\tfrac{J_1+J_2}{2},\tfrac{J_1-J_2}{2}]_H^{\frac{R_x+R_y}{2},R_z}.
\end{align}
The modes of a single tensor multiplet on the fixed locus belong to
${\cal E}_{\ell(0,0)}$ representations $(\ell\geq1)$, and
the conformal representations with the following primaries appear
\begin{align}
\underline{[0,0]_\ell^{0,+\ell}},\quad
\underline{[0,\tfrac{1}{2}]_{\ell+\frac{1}{2}}^{\frac{1}{2},+\ell-\frac{1}{2}}},\quad
[0,1]_{\ell+1}^{0,+\ell-1}
\oplus\underline{[0,0]_{\ell+1}^{1,+\ell-1}},\quad
[0,\tfrac{1}{2}]_{\ell+\frac{3}{2}}^{\frac{1}{2},+\ell-\frac{3}{2}},\quad
[0,0]_{\ell+2}^{0,+\ell-2}.
\end{align}
The primary states of the underlined representations contribute to the index
\begin{align}
i\left([0,0]_\ell^{0,+\ell}\right)
&=\frac{z^\ell}{(1-q)(1-p)} &(\ell=1,2,\ldots),\nonumber\\
i\left([0,\tfrac{1}{2}]_{\ell+\frac{1}{2}}^{+\frac{1}{2},+\ell-\frac{1}{2}}\right)
&=-\frac{z^{\ell-1}(q+p)}{(1-q)(1-p)}
+\delta_{\ell,1}\frac{qp}{(1-q)(1-p)}
&(\ell=1,2,\ldots),\nonumber\\
i\left([0,0]_{\ell+1}^{+1,+\ell-1}\right)
&=\frac{z^{\ell-2}qp}{(1-q)(1-p)}&(\ell=2,3,\ldots) .
\end{align}
The term including $\delta_{\ell,1}$ is the contribution from the equation of motion of the fermion.
By summing up all contributions we obtain
\begin{align}
f_{\rm tensor}^{AdS_5\times S^1}=\frac{z}{1-z}-\frac{q}{1-q}-\frac{p}{1-p}
\label{ftads5s1}
\end{align}
for a single tensor multiplet.

\subsection{Tensor multiplet on $AdS_3\times S^3$}\label{tensor33}

The fixed locus of (\ref{ggtwist})
\begin{align}
\wt U_k=\exp\left(-\frac{2\pi i}{k}(J_1+R_x)\right)
\end{align}
is $AdS_3\times S^3$, and $(2,0)$ tensor multiplets live on the six-dimensional locus.
All component fields are $\wt U_k$ neutral, that is, $J_1+R_x=0$.
Let us express other quantum numbers in the form
\begin{align}
[J_2]_H^{\frac{R_y+R_z}{2},\frac{R_y-R_z}{2},\frac{R_x-J_1}{2}}.
\end{align}
The mode expansion of a single tensor multiplet on the fixed locus gives the conformal representations
with the following primaries.
\begin{align}
\begin{array}{ccccc}
&& [0]_{\ell+2}^{\frac{\ell-2}{2},\frac{\ell-2}{2},0} \\
& [+\frac{1}{2}]_{\ell+\frac{3}{2}}^{\frac{\ell-1}{2},\frac{\ell-2}{2},\frac{1}{2}}
&& [-\frac{1}{2}]_{\ell+\frac{3}{2}}^{\frac{\ell-2}{2},\frac{\ell-1}{2},\frac{1}{2}} \\{}
\underline{[+1]_{\ell+1}^{\frac{\ell}{2},\frac{\ell-2}{2},0}}
&&[0]_{\ell+1}^{\frac{\ell-1}{2},\frac{\ell-1}{2},0\oplus1}
&&[-1]_{\ell+1}^{\frac{\ell-2}{2},\frac{\ell}{2},0} \\{}
& \underline{[+\frac{1}{2}]_{\ell+\frac{1}{2}}^{\frac{\ell}{2},\frac{\ell-1}{2},\frac{1}{2}}}
&& [-\frac{1}{2}]_{\ell+\frac{1}{2}}^{\frac{\ell-1}{2},\frac{\ell}{2},\frac{1}{2}} \\{}
&& \underline{[0]_\ell^{\frac{\ell}{2},\frac{\ell}{2},0}}
\end{array}
\end{align}
The underlined components include BPS states contributing to the index.
The contributions from these conformal representations
are
\begin{align}
i\left([0]_\ell^{+\frac{\ell}{2},\frac{\ell}{2},0}\right)
&=\frac{(y^\ell+\cdots+z^\ell)}{1-p}&(\ell=1,2,\ldots),\nonumber\\
i\left([+\tfrac{1}{2}]_{\ell+\frac{1}{2}}^{+\frac{\ell}{2},\frac{\ell-1}{2},\frac{1}{2}}\right)
&=-\frac{(y^{\ell-1}+\cdots+z^{\ell-1})(yz+p)}{1-p}&(\ell=1,2,\ldots),\nonumber\\
i\left([+1]_{\ell+1}^{+\frac{\ell}{2},\frac{\ell-2}{2},0}\right)
&=\frac{(x^{\ell-2}+\cdots+y^{\ell-2})pyz}{1-p}&(\ell=2,3,\ldots).
\end{align}
Summing up all contributions with $\ell=1,2,\ldots$ we obtain
\begin{align}
f_{\rm tensor}^{AdS_3\times S^3}=\frac{y}{1-y}+\frac{z}{1-z}-\frac{p}{1-p}.
\label{ftads3s3}
\end{align}

\subsection{Vector multiplet on $AdS_5\times S^3$}\label{vector53}

The fixed locus of (\ref{o7flip})
\begin{align}
U_{O7}=e^{i\pi(R_x-R_y-S)}
\end{align}
is $AdS_5\times S^3$.

All components are $U_{O7}$ neutral, that is, $R_x-R_y=S$.
Because $S$ commute with all superconformal generators,
the eigenvalue of $S$ is common for all component fields in
an irreducible superconformal representation.
So is the generator $\frac{R_x-R_y}{2}$.
Namely, this is the Cartan generator of the flavor symmetry $SU(2)_F$.
Let us use the notation
\begin{align}
[\tfrac{J_1+J_2}{2},\tfrac{J_1-J_2}{2}]_H^{\frac{R_x+R_y}{2},\frac{R_x-R_y}{2},R_z}
\end{align}
to represent the quantum numbers.
We obtain the conformal representations with the following primaries.
\begin{align}
\begin{array}{ccccc}
&&[0,0]_{\ell+4}^{\frac{\ell-2}{2},\frac{\ell}{2},0} \\
&[0,\frac{1}{2}]_{\ell+\frac{7}{2}}^{\frac{\ell-1}{2},\frac{\ell}{2},+\frac{1}{2}}
&&[\frac{1}{2},0]_{\ell+\frac{7}{2}}^{\frac{\ell-1}{2},\frac{\ell}{2},-\frac{1}{2}} \\{}
[0,0]_{\ell+3}^{\frac{\ell}{2},\frac{\ell}{2},+1}
&&[\frac{1}{2},\frac{1}{2}]_{\ell+3}^{\frac{\ell}{2},\frac{\ell}{2},0}
&&[0,0]_{\ell+3}^{\frac{\ell}{2},\frac{\ell}{2},-1} \\{}
&\underline{[\frac{1}{2},0]_{\ell+\frac{5}{2}}^{\frac{\ell+1}{2},\frac{\ell}{2},+\frac{1}{2}}}
&&[0,\frac{1}{2}]_{\ell+\frac{5}{2}}^{\frac{\ell+1}{2},\frac{\ell}{2},-\frac{1}{2}} \\{}
&&\underline{[0,0]_{\ell+2}^{\frac{\ell+2}{2},\frac{\ell}{2},0}}
\end{array}
\end{align}
This is the representation $\wh{\cal B}_{\frac{\ell+2}{2}}\otimes[\frac{\ell}{2}]_{SU(2)_F}$.
Two conformal representations with underlines contribute to the index.
The contributions are given by
\begin{align}
i([0,0]_{\ell+2}^{\frac{\ell+2}{2},\frac{\ell}{2},0})
&=\frac{xy(x^\ell+\cdots+y^\ell)}{(1-q)(1-p)}&(\ell=0,1,2,\ldots),\nonumber\\
i([\tfrac{1}{2},0]_{\ell+\frac{5}{2}}^{\frac{\ell+1}{2},\frac{\ell}{2},+\frac{1}{2}})
&=-\frac{qp(x^\ell+\cdots+y^\ell)}{(1-q)(1-p)}&(\ell=0,1,2,\ldots).
\end{align}
By summing up all contributions,
we obtain
\begin{align}
f_{\rm vector}^{AdS_5\times S^3}
=\frac{xy(1-z)}{(1-q)(1-p)(1-x)(1-y)}.
\label{fvads5s3}
\end{align}

\subsection{Vector multiplet on $AdS_3\times S^5$}\label{vector35}

The fixed locus of O7-flip (\ref{o7wt})
\begin{align}
\wt U_{O7}=e^{\pi i(J_2+A)}
\end{align}
is $AdS_3\times S^5$.
Let us use the notation
\begin{align}
[J_1]_H^{R_{SU(4)},J_2}
\end{align}
to represent the quantum numbers.
$R_{SU(4)}$ is given by the Dynkin labels.
The mode expansion in $AdS_3\times S^5$ gives
the conformal representations with the following primaries.
\begin{align}
\begin{array}{ccccc}
&& [-1]_{\ell+3}^{[0,\ell-2,0],0} \\{}
& [-\frac{1}{2}]_{\ell+\frac{5}{2}}^{[0,\ell-2,1],+\frac{1}{2}}
&& [-\frac{1}{2}]_{\ell+\frac{5}{2}}^{[1,\ell-2,0],-\frac{1}{2}} \\{}
[0]_{\ell+2}^{[0,\ell-1,0],+1}
&& [0]_{\ell+2}^{[1,\ell-2,1],0}
&& [0]_{\ell+2}^{[0,\ell-1,0],-1} \\{}
& \underline{[+\frac{1}{2}]_{\ell+\frac{3}{2}}^{[1,\ell-1,0],+\frac{1}{2}}}
&& [+\frac{1}{2}]_{\ell+\frac{3}{2}}^{[0,\ell-1,1],-\frac{1}{2}} \\{}
&& \underline{[+1]_{\ell+1}^{[0,\ell,0],0}}
\end{array}
\end{align}
Dynkin labels $[R_1,R_2,R_3]$ correspond to the following $R$-charges.
\begin{align}
(R_x,R_y,R_z)
=R_1(\tfrac{1}{2},\tfrac{1}{2},\tfrac{1}{2})
+R_2(1,0,0)
=R_3(\tfrac{1}{2},\tfrac{1}{2},-\tfrac{1}{2}).
\end{align}
Underlined conformal representations contribute to the index.
\begin{align}
i([+1]_{\ell+1}^{[0,\ell,0],0})
&=\frac{q(x^\ell+\cdots+y^\ell+\cdots+z^\ell)}{1-q}&(\ell=0,1,2,\ldots),\nonumber\\
i([+\tfrac{1}{2}]_{\ell+\frac{3}{2}}^{[1,\ell-1,0],+\frac{1}{2}})
&=-\frac{pq(x^{\ell-1}+\cdots+y^{\ell-1}+\ldots+z^{\ell-1})}{(1-q)}&(\ell=1,2,3,\ldots).
\end{align}
Summing up all contributions,
we obtain
\begin{align}
f_{\rm vector}^{AdS_3\times S^5}
=\frac{q(1-p)}{(1-q)(1-x)(1-y)(1-z)}.
\label{fvads3s5}
\end{align}



\begin{thebibliography}{99}

\bibitem{Maldacena:1997re} 
  J.~M.~Maldacena,
  ``The Large N limit of superconformal field theories and supergravity,''
  Int.\ J.\ Theor.\ Phys.\  {\bf 38}, 1113 (1999)
  [Adv.\ Theor.\ Math.\ Phys.\  {\bf 2}, 231 (1998)]
  doi:10.1023/A:1026654312961, 10.4310/ATMP.1998.v2.n2.a1
  [hep-th/9711200].


\bibitem{Gubser:1998bc}
  S.~S.~Gubser, I.~R.~Klebanov and A.~M.~Polyakov,
  ``Gauge theory correlators from noncritical string theory,''
  Phys.\ Lett.\ B {\bf 428}, 105 (1998)
  doi:10.1016/S0370-2693(98)00377-3
  [hep-th/9802109].


\bibitem{Witten:1998qj}
  E.~Witten,
  ``Anti-de Sitter space and holography,''
  Adv.\ Theor.\ Math.\ Phys.\  {\bf 2}, 253 (1998)
  doi:10.4310/ATMP.1998.v2.n2.a2
  [hep-th/9802150].

\bibitem{Romelsberger:2005eg}
C.~Romelsberger,
``Counting chiral primaries in N = 1, d=4 superconformal field theories,''
Nucl. Phys. B \textbf{747}, 329-353 (2006)
doi:10.1016/j.nuclphysb.2006.03.037
[arXiv:hep-th/0510060 [hep-th]].

\bibitem{Kinney:2005ej} 
  J.~Kinney, J.~M.~Maldacena, S.~Minwalla and S.~Raju,
  ``An Index for 4 dimensional super conformal theories,''
  Commun.\ Math.\ Phys.\  {\bf 275}, 209 (2007)
  doi:10.1007/s00220-007-0258-7
  [hep-th/0510251].

\bibitem{Gadde:2011uv} 
  A.~Gadde, L.~Rastelli, S.~S.~Razamat and W.~Yan,
  ``Gauge Theories and Macdonald Polynomials,''
  Commun.\ Math.\ Phys.\  {\bf 319}, 147 (2013)
  doi:10.1007/s00220-012-1607-8
  [arXiv:1110.3740 [hep-th]].


\bibitem{Hosseini:2017mds}
S.~M.~Hosseini, K.~Hristov and A.~Zaffaroni,
``An extremization principle for the entropy of rotating BPS black holes in AdS$_{5}$,''
JHEP \textbf{07}, 106 (2017)
doi:10.1007/JHEP07(2017)106
[arXiv:1705.05383 [hep-th]].

\bibitem{Cabo-Bizet:2018ehj}
A.~Cabo-Bizet, D.~Cassani, D.~Martelli and S.~Murthy,
``Microscopic origin of the Bekenstein-Hawking entropy of supersymmetric AdS$_{5}$ black holes,''
JHEP \textbf{10}, 062 (2019)
doi:10.1007/JHEP10(2019)062
[arXiv:1810.11442 [hep-th]].

\bibitem{Choi:2018hmj}
  S.~Choi, J.~Kim, S.~Kim and J.~Nahmgoong,
  ``Large AdS black holes from QFT,''
  [arXiv:1810.12067 [hep-th]].


\bibitem{Witten:1998xy}
  E.~Witten,
  ``Baryons and branes in anti-de Sitter space,''
  JHEP {\bf 9807}, 006 (1998)
  doi:10.1088/1126-6708/1998/07/006
  [hep-th/9805112].


\bibitem{McGreevy:2000cw} 
  J.~McGreevy, L.~Susskind and N.~Toumbas,
  ``Invasion of the giant gravitons from Anti-de Sitter space,''
  JHEP {\bf 0006}, 008 (2000)
  doi:10.1088/1126-6708/2000/06/008
  [hep-th/0003075].

\bibitem{Grisaru:2000zn} 
  M.~T.~Grisaru, R.~C.~Myers and O.~Tafjord,
  ``SUSY and goliath,''
  JHEP {\bf 0008}, 040 (2000)
  doi:10.1088/1126-6708/2000/08/040
  [hep-th/0008015].
\bibitem{Hashimoto:2000zp} 
  A.~Hashimoto, S.~Hirano and N.~Itzhaki,
  ``Large branes in AdS and their field theory dual,''
  JHEP {\bf 0008}, 051 (2000)
  doi:10.1088/1126-6708/2000/08/051
  [hep-th/0008016].


\bibitem{Mikhailov:2000ya} 
  A.~Mikhailov,
  ``Giant gravitons from holomorphic surfaces,''
  JHEP {\bf 0011}, 027 (2000)
  doi:10.1088/1126-6708/2000/11/027
  [hep-th/0010206].




\bibitem{Arai:2019xmp}
R.~Arai and Y.~Imamura,
``Finite $N$ Corrections to the Superconformal Index of S-fold Theories,''
PTEP \textbf{2019}, no.8, 083B04 (2019)
doi:10.1093/ptep/ptz088
[arXiv:1904.09776 [hep-th]].


\bibitem{Arai:2020qaj}
R.~Arai, S.~Fujiwara, Y.~Imamura and T.~Mori,
``Schur index of the ${\cal N}=4$ $U(N)$ supersymmetric Yang-Mills theory via the AdS/CFT correspondence,''
Phys. Rev. D \textbf{101}, no.8, 086017 (2020)
doi:10.1103/PhysRevD.101.086017
[arXiv:2001.11667 [hep-th]].

\bibitem{Imamura:2021ytr}
Y.~Imamura,
``Finite-N superconformal index via the AdS/CFT correspondence,''
PTEP \textbf{2021}, no.12, 123B05 (2021)
doi:10.1093/ptep/ptab141
[arXiv:2108.12090 [hep-th]].


\bibitem{Arai:2019wgv}
  R.~Arai, S.~Fujiwara, Y.~Imamura and T.~Mori,
  ``Finite $N$ corrections to the superconformal index of orbifold quiver gauge theories,''
  JHEP {\bf 1910}, 243 (2019)
  doi:10.1007/JHEP10(2019)243
  [arXiv:1907.05660 [hep-th]].

\bibitem{Arai:2019aou}
R.~Arai, S.~Fujiwara, Y.~Imamura and T.~Mori,
``Finite $N$ corrections to the superconformal index of toric quiver gauge theories,''
PTEP \textbf{2020}, no.4, 043B09 (2020)
doi:10.1093/ptep/ptaa023
[arXiv:1911.10794 [hep-th]].

\bibitem{Arai:2020uwd}
  R.~Arai, S.~Fujiwara, Y.~Imamura, T.~Mori and D.~Yokoyama,
    ``Finite-$N$ corrections to the M-brane indices,''
    JHEP \textbf{11}, 093 (2020)
    doi:10.1007/JHEP11(2020)093
    [arXiv:2007.05213 [hep-th]].

\bibitem{Fujiwara:2021xgu}
    S.~Fujiwara, Y.~Imamura and T.~Mori,
    ``Flavor symmetries of six-dimensional ${\cal N}=(1,0)$ theories from AdS/CFT correspondence,''
    JHEP \textbf{05}, 221 (2021)
    doi:10.1007/JHEP05(2021)221
    [arXiv:2103.16094 [hep-th]].

\bibitem{Imamura:2021dya}
Y.~Imamura and S.~Murayama,
``Holographic index calculation for Argyres-Douglas and Minahan-Nemeschansky theories,''
[arXiv:2110.14897 [hep-th]].

\bibitem{Fujiwara:2023azx}
S.~Fujiwara,
``Schur-like index of the Klebanov-Witten theory via the AdS/CFT correspondence,''
[arXiv:2302.04697 [hep-th]].



\bibitem{Gaiotto:2021xce}
D.~Gaiotto and J.~H.~Lee,
``The Giant Graviton Expansion,''
[arXiv:2109.02545 [hep-th]].

\bibitem{Lee:2022vig}
J.~H.~Lee,
``Exact Stringy Microstates from Gauge Theories,''
[arXiv:2204.09286 [hep-th]].




\bibitem{Murthy:2022ien}
S.~Murthy,
Pure Appl. Math. Quart. \textbf{19}, no.1, 299-340 (2023)
doi:10.4310/PAMQ.2023.v19.n1.a12
[arXiv:2202.06897 [hep-th]].

\bibitem{Liu:2022olj}
J.~T.~Liu and N.~J.~Rajappa,
``Finite N indices and the giant graviton expansion,''
JHEP \textbf{04}, 078 (2023)
doi:10.1007/JHEP04(2023)078
[arXiv:2212.05408 [hep-th]].

\bibitem{Eniceicu:2023uvd}
D.~S.~Eniceicu,
``Comments on the Giant-Graviton Expansion of the Superconformal Index,''
[arXiv:2302.04887 [hep-th]].

\bibitem{Choi:2022ovw}
S.~Choi, S.~Kim, E.~Lee and J.~Lee,
``From giant gravitons to black holes,''
[arXiv:2207.05172 [hep-th]].


\bibitem{Beccaria:2023hip}
M.~Beccaria and A.~Cabo-Bizet,
``Large black hole entropy from the giant brane expansion,''
[arXiv:2308.05191 [hep-th]].





\bibitem{Biswas:2006tj}
  I.~Biswas, D.~Gaiotto, S.~Lahiri and S.~Minwalla,
  ``Supersymmetric states of N=4 Yang-Mills from giant gravitons,''
  JHEP {\bf 0712}, 006 (2007)
  doi:10.1088/1126-6708/2007/12/006
  [hep-th/0606087].


\bibitem{Mandal:2006tk} 
  G.~Mandal and N.~V.~Suryanarayana,
  ``Counting 1/8-BPS dual-giants,''
  JHEP {\bf 0703}, 031 (2007)
  doi:10.1088/1126-6708/2007/03/031
  [hep-th/0606088].

\bibitem{Bourdier:2015wda}
  J.~Bourdier, N.~Drukker and J.~Felix,
  ``The exact Schur index of $\mathcal{N}=4$ SYM,''
  JHEP {\bf 1511}, 210 (2015)
  doi:10.1007/JHEP11(2015)210
  [arXiv:1507.08659 [hep-th]].
\bibitem{Bourdier:2015sga}
  J.~Bourdier, N.~Drukker and J.~Felix,
  ``The $\mathcal{N}=2$ Schur index from free fermions,''
  JHEP {\bf 1601}, 167 (2016)
  doi:10.1007/JHEP01(2016)167
  [arXiv:1510.07041 [hep-th]].


\bibitem{Dolan:2002zh}
F.~A.~Dolan and H.~Osborn,
``On short and semi-short representations for four-dimensional superconformal symmetry,''
Annals Phys. \textbf{307}, 41-89 (2003)
doi:10.1016/S0003-4916(03)00074-5
[arXiv:hep-th/0209056 [hep-th]].




\bibitem{Kim:1985ez}
  H.~J.~Kim, L.~J.~Romans and P.~van Nieuwenhuizen,
  ``The Mass Spectrum of Chiral N=2 D=10 Supergravity on S**5,''
  Phys.\ Rev.\ D {\bf 32}, 389 (1985).
  doi:10.1103/PhysRevD.32.389


\bibitem{Gunaydin:1984fk}
  M.~Gunaydin and N.~Marcus,
  ``The Spectrum of the s**5 Compactification of the Chiral N=2, D=10 Supergravity and the Unitary Supermultiplets of U(2, 2/4),''
  Class.\ Quant.\ Grav.\  {\bf 2}, L11 (1985).
  doi:10.1088/0264-9381/2/2/001


\bibitem{Beccaria:2023zjw}
M.~Beccaria and A.~Cabo-Bizet,
``On the brane expansion of the Schur index,''
JHEP \textbf{08}, 073 (2023)
doi:10.1007/JHEP08(2023)073
[arXiv:2305.17730 [hep-th]].

\bibitem{Imamura:2022aua}
Y.~Imamura,
``Analytic continuation for giant gravitons,''
PTEP \textbf{2022}, no.10, 103B02 (2022)
doi:10.1093/ptep/ptac127
[arXiv:2205.14615 [hep-th]].



\bibitem{Kachru:1998ys}
  S.~Kachru and E.~Silverstein,
  ``4-D conformal theories and strings on orbifolds,''
  Phys.\ Rev.\ Lett.\  {\bf 80}, 4855 (1998)
  doi:10.1103/PhysRevLett.80.4855
  [hep-th/9802183].


\bibitem{Lawrence:1998ja}
  A.~E.~Lawrence, N.~Nekrasov and C.~Vafa,
  ``On conformal field theories in four-dimensions,''
  Nucl.\ Phys.\ B {\bf 533}, 199 (1998)
  doi:10.1016/S0550-3213(98)00495-7
  [hep-th/9803015].


\bibitem{Nakayama:2005mf}
  Y.~Nakayama,
  ``Index for orbifold quiver gauge theories,''
  Phys.\ Lett.\ B {\bf 636}, 132 (2006)
  doi:10.1016/j.physletb.2006.03.045
  [hep-th/0512280].

\bibitem{Douglas:1996sw}
M.~R.~Douglas and G.~W.~Moore,
``D-branes, quivers, and ALE instantons,''
[arXiv:hep-th/9603167 [hep-th]].



\bibitem{Sen:1996vd}
A.~Sen,
``F theory and orientifolds,''
Nucl. Phys. B \textbf{475}, 562-578 (1996)
doi:10.1016/0550-3213(96)00347-1
[arXiv:hep-th/9605150 [hep-th]].


\bibitem{Fayyazuddin:1998fb}
A.~Fayyazuddin and M.~Spalinski,
``Large N superconformal gauge theories and supergravity orientifolds,''
Nucl. Phys. B \textbf{535}, 219-232 (1998)
doi:10.1016/S0550-3213(98)00545-8
[arXiv:hep-th/9805096 [hep-th]].


\bibitem{Aharony:1998xz}
O.~Aharony, A.~Fayyazuddin and J.~M.~Maldacena,
``The Large N limit of N=2, N=1 field theories from three-branes in F theory,''
JHEP \textbf{07}, 013 (1998)
doi:10.1088/1126-6708/1998/07/013
[arXiv:hep-th/9806159 [hep-th]].




\bibitem{Feng:2000mi}
  B.~Feng, A.~Hanany and Y.~H.~He,
  ``D-brane gauge theories from toric singularities and toric duality,''
  Nucl.\ Phys.\ B {\bf 595}, 165 (2001)
  doi:10.1016/S0550-3213(00)00699-4
  [hep-th/0003085].
\bibitem{Feng:2002zw}
  B.~Feng, S.~Franco, A.~Hanany and Y.~H.~He,
  ``Symmetries of toric duality,''
  JHEP {\bf 0212}, 076 (2002)
  doi:10.1088/1126-6708/2002/12/076
  [hep-th/0205144].


\bibitem{Hanany:2005ve}
A.~Hanany and K.~D.~Kennaway,
``Dimer models and toric diagrams,''
[arXiv:hep-th/0503149 [hep-th]].

\bibitem{Franco:2005rj}
S.~Franco, A.~Hanany, K.~D.~Kennaway, D.~Vegh and B.~Wecht,
``Brane dimers and quiver gauge theories,''
JHEP \textbf{01}, 096 (2006)
doi:10.1088/1126-6708/2006/01/096
[arXiv:hep-th/0504110 [hep-th]].

\bibitem{Franco:2005sm}
S.~Franco, A.~Hanany, D.~Martelli, J.~Sparks, D.~Vegh and B.~Wecht,
``Gauge theories from toric geometry and brane tilings,''
JHEP \textbf{01}, 128 (2006)
doi:10.1088/1126-6708/2006/01/128
[arXiv:hep-th/0505211 [hep-th]].

\bibitem{Hanany:2005ss}
A.~Hanany and D.~Vegh,
``Quivers, tilings, branes and rhombi,''
JHEP \textbf{10}, 029 (2007)
doi:10.1088/1126-6708/2007/10/029
[arXiv:hep-th/0511063 [hep-th]].

\bibitem{Feng:2005gw}
B.~Feng, Y.~H.~He, K.~D.~Kennaway and C.~Vafa,
``Dimer models from mirror symmetry and quivering amoebae,''
Adv. Theor. Math. Phys. \textbf{12}, no.3, 489-545 (2008)
doi:10.4310/ATMP.2008.v12.n3.a2
[arXiv:hep-th/0511287 [hep-th]].

\bibitem{Franco:2006gc}
S.~Franco and D.~Vegh,
``Moduli spaces of gauge theories from dimer models: Proof of the correspondence,''
JHEP \textbf{11}, 054 (2006)
doi:10.1088/1126-6708/2006/11/054
[arXiv:hep-th/0601063 [hep-th]].








\bibitem{Eager:2012hx}
  R.~Eager, J.~Schmude and Y.~Tachikawa,
  ``Superconformal Indices, Sasaki-Einstein Manifolds, and Cyclic Homologies,''
  Adv.\ Theor.\ Math.\ Phys.\  {\bf 18}, no. 1, 129 (2014)
  doi:10.4310/ATMP.2014.v18.n1.a3
  [arXiv:1207.0573 [hep-th]].
\bibitem{Agarwal:2013pba}
  P.~Agarwal, A.~Amariti and A.~Mariotti,
  ``A Zig-Zag Index,''
  arXiv:1304.6733 [hep-th].





\bibitem{Klebanov:1998hh}
I.~R.~Klebanov and E.~Witten,
``Superconformal field theory on three-branes at a Calabi-Yau singularity,''
Nucl. Phys. B \textbf{536}, 199-218 (1998)
doi:10.1016/S0550-3213(98)00654-3
[arXiv:hep-th/9807080 [hep-th]].




\bibitem{Aharony:2008ug}
O.~Aharony, O.~Bergman, D.~L.~Jafferis and J.~Maldacena,
``N=6 superconformal Chern-Simons-matter theories, M2-branes and their gravity duals,''
JHEP \textbf{10}, 091 (2008)
doi:10.1088/1126-6708/2008/10/091
[arXiv:0806.1218 [hep-th]].




\bibitem{Beem:2015aoa}
C.~Beem, M.~Lemos, L.~Rastelli and B.~C.~van Rees,
``The (2, 0) superconformal bootstrap,''
Phys. Rev. D \textbf{93}, no.2, 025016 (2016)
doi:10.1103/PhysRevD.93.025016
[arXiv:1507.05637 [hep-th]].



\bibitem{Lockhart:2012vp}
G.~Lockhart and C.~Vafa,
``Superconformal Partition Functions and Non-perturbative Topological Strings,''
JHEP \textbf{10}, 051 (2018)
doi:10.1007/JHEP10(2018)051
[arXiv:1210.5909 [hep-th]].

\bibitem{Kim:2012qf}
H.~C.~Kim, J.~Kim and S.~Kim,
``Instantons on the 5-sphere and M5-branes,''
[arXiv:1211.0144 [hep-th]].


\bibitem{Kim:2013nva}
H.~C.~Kim, S.~Kim, S.~S.~Kim and K.~Lee,
``The general M5-brane superconformal index,''
[arXiv:1307.7660 [hep-th]].

\bibitem{Beem:2014kka}
C.~Beem, L.~Rastelli and B.~C.~van Rees,
``$ \mathcal{W} $ symmetry in six dimensions,''
JHEP \textbf{05}, 017 (2015)
doi:10.1007/JHEP05(2015)017
[arXiv:1404.1079 [hep-th]].

\bibitem{Kim:2009wb}
S.~Kim,
``The Complete superconformal index for N=6 Chern-Simons theory,''
Nucl. Phys. B \textbf{821}, 241-284 (2009)
doi:10.1016/j.nuclphysb.2009.06.025
[arXiv:0903.4172 [hep-th]].



\end{thebibliography}
\end{document}